\documentclass{elsart}
\usepackage{epsfig}
\newcommand{\be}{\begin{equation}}
\newcommand{\ee}{\end{equation}}

\begin{document}

\begin{frontmatter}

\title{Multi-peaked localized states of DNLS in one and two dimensions}

\author{G. Kalosakas \corauthref{cor1}}
\corauth[cor1]{Fax: +49-351-871-1999. E-mail: georgek@pks.mpg.de}
\address{Max Planck Institute for the Physics of Complex
Systems, \\ N\"othnitzer Str. 38, Dresden, 01187, Germany}


\begin{abstract}

Multi-peaked localized stationary solutions of the discrete
nonlinear Schr\"odinger (DNLS) equation are presented in
one (1D) and two (2D) dimensions. These are excited states
of the discrete spectrum and correspond to multi-breather
solutions. A simple, very fast, and efficient numerical method,
suggested by Aubry, has been used for their calculation.
The method involves no diagonalization, but just iterations of a map,
starting from trivial solutions of the anti-continuous limit.
Approximate analytical expressions are presented and compared with
the numerical results. The linear stability of the calculated stationary
states is discussed and the structure of the linear stability spectrum
is analytically obtained for relatively large values of nonlinearity.

\end{abstract}

\end{frontmatter}


\section{Introduction}

The DNLS equation, Eq. (\ref{dnls}), has been extensively used
as a generic model for studying nonlinear effects (breathers,
for example) in a discrete system
\cite{ELS,eilb,MA,JA,FKM,HT,RABT,JA2,BBJ,BKRK,VMK,floria}. In addition,
specific applications have been proposed for the description of:
{\it i)} local intramolecular stretching vibrations in symmetric
polyatomic molecules \cite{lm},
{\it ii)} arrays of coupled nonlinear optical waveguides \cite{nopt},
{\it iii)} interacting electron-lattice models \cite{hol,KAT}
(or, equivalently, intramolecular excitation-phonon coupled systems
\cite{dav,acn,zolo}) in solid state physics,
where localized solutions of DNLS correspond to
polarons (vibrational polarons, respectively), and recently
{\it iv)} Bose-Einstein condensates \cite{TS,KRB}.

In DNLS the evolution of a complex probability amplitude $\Psi_n$
at the site $n$ of a $d$-dimensional lattice is given by
\be   \label{dnls}
 i \frac{d \Psi_n}{dt} = -V \sum_{\delta} \Psi_{n+\delta}
  + \chi |\Psi_n|^2 \Psi_n,
\ee
where $V$ represents nearest neighbor coupling,
$\chi$ is the strength of nonlinearity, and the sum over $\delta$
contains the nearest neighbors of the lattice site~$n$.
For example, in 1D is $\sum_{\delta} \Psi_{n+\delta}= \Psi_{n+1}+\Psi_{n-1}$,
while in a 2D square lattice, if each lattice site $n$ is represented
by a pair ($n_x,n_y$), is $\sum_{\delta} \Psi_{n+\delta}=
\Psi_{n_x+1,n_y}+\Psi_{n_x-1,n_y}+\Psi_{n_x,n_y+1}+\Psi_{n_x,n_y-1}$.

The DNLS equation is derived from the Hamiltonian (assuming an
infinite lattice, or periodic boundary conditions)
\be  \label{ham}
H = - \; V \; \sum_n \sum_{\delta} \Psi_n^\star \Psi_{n+\delta}
      \; + \; \frac{\chi}{2} \sum_n |\Psi_n|^4.
\ee

There exist two conserved quantities during the DNLS dynamics:
the Hamiltonian (\ref{ham}) and the norm
\be
N = \sum_n |\Psi_n|^2.
\ee

A trivial transformation ($\Psi_n \rightarrow \Psi_n / \sqrt{N}$)
connects the solutions of DNLS
for an arbitrary norm $N$ with those normalized to unity, through a
rescalling of the nonlinearity ($\chi \rightarrow \chi \cdot N$).
Therefore, any solution $\Phi_n$ of Eq.(\ref{dnls}) with norm $N$
is obtained from a solution $\Psi_n$ with norm 1, through
\be  \label{metN}
\Phi_n(t;V,\chi) = \sqrt{N} \cdot \Psi_n(t;V,\chi N)
\ee

Furthermore, the magnitude of the hopping integral
$V$ can be considered unity by appropriate rescalling of time and $\chi$.
Changing sign in $V$ is equivalent to the transformation
$\Psi_n \rightarrow (-1)^n\Psi_n $ (or
$\Psi_{n_x,n_y} \rightarrow (-1)^{n_x+n_y} \Psi_{n_x,n_y}$ in 2D, etc.).
Therefore, in the following we
consider $N=1$ and $V=1$ without loss of the generality, while the
nonlinearity $\chi$ remains the only free parameter of the system.

\subsection{Stationary solutions of DNLS}

The stationary states of DNLS are characterized by a simple harmonic
evolution with frequency $\omega$:
\be  \label{statsol}
\Psi_n = \psi_n \cdot e^{-i \omega t}.
\ee

The time independent amplitudes $\psi_n$ satisfy the nonlinear
eigenvalue problem
\be  \label{sdnls}
\omega \psi_n = - \sum_{\delta} \psi_{n+\delta} + \chi
   |\psi_n|^2 \psi_n
\ee
(since we consider $V=1$).
The energy $E_s$ of a stationary state is related to its
frequency $\omega$ through
\be
E_s = \omega - \frac{\chi}{2} \sum_n |\psi_n|^4.
\ee

There are two kinds of stationary solutions of Eq.~(\ref{sdnls}):
extended (like Bloch states and standing waves \cite{MJKA}) and
localized states. The Bloch states, $\psi_n^q \sim \exp(iqn)$, form
a band of frequencies or energies from $-2dV$ to $2dV$ (for an infinite
$d$-dimensional lattice). Localized states have discrete
frequencies outside of the Bloch band.

The most obvious and well studied localized state is the single-peaked
solution, which has its maximum amplitude on one lattice site. This state
has extreme (minimal for negative $\chi$ and maximal for positive $\chi$)
energy and frequency, compared to the other localized states.
However, in general there are infinite (in an infinite lattice)
multi-peaked stationary solutions of DNLS (which, for example, can
be continued from trivial multi-peaked solutions of the anti-continuous
limit), resulting in a very rich discrete spectrum with many
quasi-degenerate levels (see below). In spite of this complexity, a part of
the spectrum can be understood by classifying the stationary states from the
anti-continuous limit. For the general concept of the anti-continuous,
or anti-integrable, limit in nonlinear lattices see Ref.~\cite{sergeAC}.
Here, multi-peaked and real stationary solutions of high symmetry, which
have direct counterparts at the anti-continuous limit, are presented.
Their location on the DNLS spectrum, their stability, and approximate
analytical expressions are discussed.

\section{Aubry's method for the numerical calculation of localized
stationary states of DNLS}

Localized stationary states can be obtained as attractors of the map:
\be  \label{map}
\overline{\psi} \; \longrightarrow \; \overline{\psi'} =
\; sgn\chi \; \cdot \; \frac{ \aleph \{ \overline{\psi} \} }
{|| \aleph \{ \overline{\psi} \} ||}.
\ee
In this equation, $ \overline{\psi} = (\psi_1,\ldots,\psi_L)$,
where $L$ is the total number of lattice sites,
$sgn\chi$ denotes the sign of $\chi$, $\aleph \{ \overline{\psi} \}$
is defined through the right-hand-side of the stationary equation
(\ref{sdnls}), with its $n$-th component given by
\be
\aleph \{ \overline{\psi} \}_n = - \sum_{\delta} \psi_{n+\delta} + \chi
   |\psi_n|^2 \psi_n,
\ee
and $|| \aleph \{ \overline{\psi} \} ||$ represents its
norm $\sqrt{\sum_{n=1}^{L} |\aleph \{ \overline{\psi} \}_n|^2}$.

Starting the iterations from appropriate initial states, obtained through
trivial solutions of the anti-continuous limit, i.e.
\be
\psi_n^{(r=0)} = \delta_{n,n_0}, \hspace{0.5cm} \mbox{or} \hspace{0.5cm}
\psi_n^{(r=0)} =\frac{1}{\sqrt{2}}(\delta_{n,n_0} \pm \delta_{n,n_1}),
\hspace{0.5cm} \mbox{etc.,}
\ee
depending on the desired localized stationary solution (single-peaked,
double-peaked with interpeak separation $|n_0-n_1|$, etc.),
and iteratively applying the map (\ref{map})
\be
\psi_n^{(r+1)} = \; sgn\chi \; \cdot \; \frac{\aleph
\{\overline{\psi}^{(r)}\}_n }{|| \aleph \{\overline{\psi}^{(r)}\} ||},
\ee
the procedure can rapidly converge to the corresponding localized state.

Up to now this method has been successfully applied for calculating
single-peaked stationary states of DNLS in 1D, 2D, and 3D \cite{KAT},
as well as for the single-peaked ground states
in other similar systems \cite{voul,VKBT,dirk,KH,HSAP}.
The method has been invented by S. Aubry for finding polarons
in the adiabatic Holstein model \cite{AQ,AAR,serge1}, a problem
which reduces to the stationary solutions of DNLS.

In the following, after briefly recalling some results obtained in
Ref~\cite{KAT} regarding the stable single-peaked
stationary states of DNLS (section 3), this method is applied for
calculating multi-peaked stationary solutions in 1D (section 4) and
2D (section 5). Stationary states are presented in sections 4 and 5
for negative values of $\chi$; the obtained solutions in this
case can be directly transformed to the corresponding ones at $-\chi$
(i.e. at positive nonlinearities) by changing the sign of the frequency
($\omega \rightarrow -\omega$) and energy ($E_s \rightarrow -E_s$),
and making the transformation $\psi_n \rightarrow (-1)^n \psi_n$ (or
$\Psi_{n_x,n_y} \rightarrow (-1)^{n_x+n_y} \Psi_{n_x,n_y}$ in 2D, etc.)
in the wavefunction. However, analytical results and the general discussion
concern both signs of $\chi$.

\section{Single-peaked (SP) stationary states}

For negative (positive) nonlinearity the single-peaked solution of
DNLS corresponds to the lowest (highest) frequency stationary state.
In 1D there is always a SP state with extreme frequency
and energy, for any nonzero value of $\chi$. As $\chi$ is approaching
zero from negative (positive) values, the frequency and the energy
of the SP solution tends to the bottom (top) of the Bloch band.
The branch of SP solutions, as obtained by varying $\chi$, has
two well-known limits: for large $|\chi|$ (in the anti-continuous limit,
where the first term of the right-hand-side of Eq. (\ref{sdnls})
can be neglected) tends to a single-site localized state
($\psi_n = \delta_{n,n_0}$), while for $|\chi| \rightarrow 0$ tends to
a solution obtained by the soliton of the continuous nonlinear
Schr\"odinger equation (see Eq. ~(\ref{nlsw}) below).

The picture is qualitatively different in 2D and 3D. In these cases
there is a critical value of nonlinearity $\chi_1 >0$, such that
for $|\chi|<\chi_1$ does not exist a single-peaked, or any other
localized, stationary state. At $|\chi|=\chi_1$ a pair of SP
states appears, through a saddle-node bifurcation; a narrow stable
solution of high amplitude and an unstable one of relatively large
extent and small amplitude. Due to the simultaneous existence of two
SP states and the proximity of the unstable one with the extended
Bloch states of the band edges,
it is not paid any attention to the unstable solution and from now on
we exclusively refer to the stable one wherever a single-peaked state
is mentioned in 2D or 3D. A second nonlinearity threshold $\chi_2 >\chi_1$
exists, such that for $\chi_1<|\chi|<\chi_2$ the SP state has
extreme frequency, but not extreme energy, since its energy $E_s$ lies
inside the Bloch band. Only for $|\chi|>\chi_2$ the single-peaked stationary
solution provides an extreme of the energy (i.e. it is the ground state
for negative $\chi$). In 2D the values of these thresholds are
$\chi_1 \approx 5.701$ and $\chi_2 \approx 6.679$, while in 3D are
$\chi_1 \approx 7.852$ and $\chi_2 \approx 10.816$ \cite{KAT}\footnote{
Note that the nonlinear parameter $\chi$ used in this work
corresponds to $-k^2$ of Ref.~\cite{KAT}.}.

Analytical approximate expressions have been presented in Ref. \cite{KAT},
which accurately describe the SP stationary states.
In particular, for the whole branch of SP solutions in 2D and
3D, as well as for values of $|\chi|$ larger than about 3 in 1D,
the exponentially decaying function
\be  \label{vap}
\psi_{n_x,n_y,n_z}^{SP} = \left( \frac{1-\zeta^2}{1+\zeta^2} \right) ^{d/2}
\cdot \zeta^{|n_x|+|n_y|+|n_z|}, \hspace{1.0cm} \mbox{with} \hspace{0.2cm}
\zeta = -\frac{1}{\chi} - \frac{4d-2}{\chi^3}
\ee
(where one has to disregard the index $n_z$ in 2D and both $n_z$ and
$n_y$ in 1D), can be used to describe the exact SP solution.
The corresponding frequencies and energies are given by
\be
\omega = \chi - \frac{2d(4d-3)}{x^3} \hspace{1.0cm} \mbox{and}
\hspace{1.0cm} E_s = \frac{\chi}{2}+ \frac{2d}{\chi}+ \frac{d(4d-3)}{\chi^3}.
\ee
The expression (\ref{vap}) is derived from a variational method, by
employing a perturbative expansion of $\zeta$ in powers of $1/\chi$
in the condition providing the minima of the variational energy.
The next non-zero correction of $\zeta$ in Eq.(\ref{vap}) is of the order
of $1/\chi^5$.

The above analytical results describe accurately (the larger the $|\chi|$
the better the approximation) the SP solutions of DNLS in all
cases, except for relatively small values of $|\chi|$ in 1D. Then
a smooth transition occurs for $|\chi|$ in the region $2.5-3$, from the
above expressions to the static soliton of the continuous nonlinear
Schr\"odinger equation. Therefore, for $|\chi|$ smaller than about 2.5
the SP solutions in 1D can be approximated by
\be   \label{nlsw}
\psi_n^{SP} = (-sgn \chi )^n \; \sqrt{\frac{|\chi|}{8}}
\cdot \frac{1}{\cosh \frac{\chi n}{4} },
\ee
with corresponding frequency and energy
\be
\omega = sgn \chi \cdot \left( 2 + \frac{\chi^2}{16} \right)
\hspace{1.0cm} \mbox{and} \hspace{1.0cm}
 E_s = sgn \chi \cdot \left( 2 + \frac{\chi^2}{48} \right).
\ee
Note that the $(-sgn\chi)^n$ term in (\ref{nlsw}) provides the alteration
of signs in successive lattice sites, which characterizes the solution
at positive $\chi$. In Eq.~(\ref{vap}) this is obtained through the
negative sign of $\zeta$.

\section{Multi-peaked solutions in 1D}

\subsection{Frequency spectrum}

As it has been already mentioned in the introduction, the frequency (or
energy) spectrum of DNLS in 1D comprises many discrete levels, corresponding
to single-peaked and multi-peaked localized stationary solutions.

These discrete levels have well determined limits in the
anti-continuous regime. In this limit the discrete spectrum is given by
\be   \label{acsp}
\omega = \frac{\chi}{M}, \hspace{0.5cm} E_s = \frac{\chi}{2M},
\hspace{0.5cm} \mbox{where} \; M=1,2,3, \ldots \; \; .
\ee
Each (highly degenerate) level of Eq.(\ref{acsp}) corresponds to any
$M$-peaked stationary state, where all the $M$ peaks (at arbitrary sites)
have the same norm $1/\sqrt{M}$ (and arbitrary complex phases).
Such stationary solutions can be continued away from the anti-continuous
limit \cite{AAR} up to some value of $\chi$, depending on the
particular state.

\begin{figure}
\centerline{\hbox{\psfig{figure=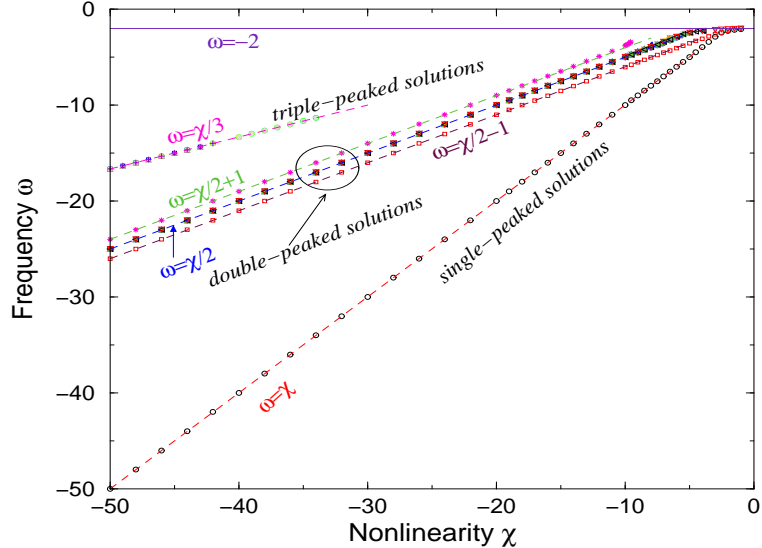,width=10cm,height=7.5cm}}}
\caption{Frequencies of single- double- and triple-peaked stationary
solutions of DNLS in 1D (points). Dashed lines show analytical
expressions obtained for large values of $|\chi|$.
The horizontal line at $\omega=-2$ indicates the lower edge of the
band of Bloch stationary states, which extends from $-2$ to 2.
The spectrum is antisymmetric on $\chi$; $\omega (-\chi)=-\omega (\chi)$.}
\label{figFS1}
\end{figure}

A stationary solution at a given value of $\chi$ is classified as a
single- double- or, in general, $M$-peaked state, depending on how
many sites (one, two, or in general $M$, respectively) are occupied
at the anti-continuous limit of the branch in which this state belongs.
Such a classification is always possible for stationary states of high
symmetry. Moreover, it can be used for {\it each} stationary solution,
providing a complete description of the discrete spectrum, up to $\omega
\approx -5.45$ \cite{ABK}.

Fig.~\ref{figFS1} shows a part of the frequency spectrum
which contains contributions from many double-peaked solutions, as well
as few branches (just shown for indication) of triple-peaked states.
Analytical expressions, obtained at large values of $|\chi|$ and shown
by dashed lines, describe well these branches (the next corrections are of
the order of inverse powers of $\chi$, see Appendix).
Bifurcations lead to the disappearance of some branches (or
merging with other branches) by decreasing the strength of nonlinearity.

What appears as a middle branch of the double-peaked (DP)
solutions in Fig.~\ref{figFS1} actually consists of
many branches of closely spaced levels (corresponding to DP
solutions with any interpeak separation larger than one lattice constant),
which merge to the level of Eq.(\ref{acsp}) for $M=2$ as $|\chi|$
increases. The lower and upper branch of the DP solutions are
single branches corresponding to symmetric and antisymmetric, respectively,
double-peaked states with their peaks at neighboring lattice sites
(examples are shown at the upper left plots of Figs. \ref{fDS1} and
\ref{fDA1}, respectively). The interpeak separation of a DP state
is determined by the distance $S$ of the sites where the peaks appear.
Symmetric and antisymmetric DP solutions in 1D, with various
interpeak distances, are presented in the following subsection.

\subsection{Double-peaked symmetric and antisymmetric solutions}

\subsubsection{Stationary states}

\begin{figure}
\begin{center}
\begin{tabular}{cc}
\epsfig{file=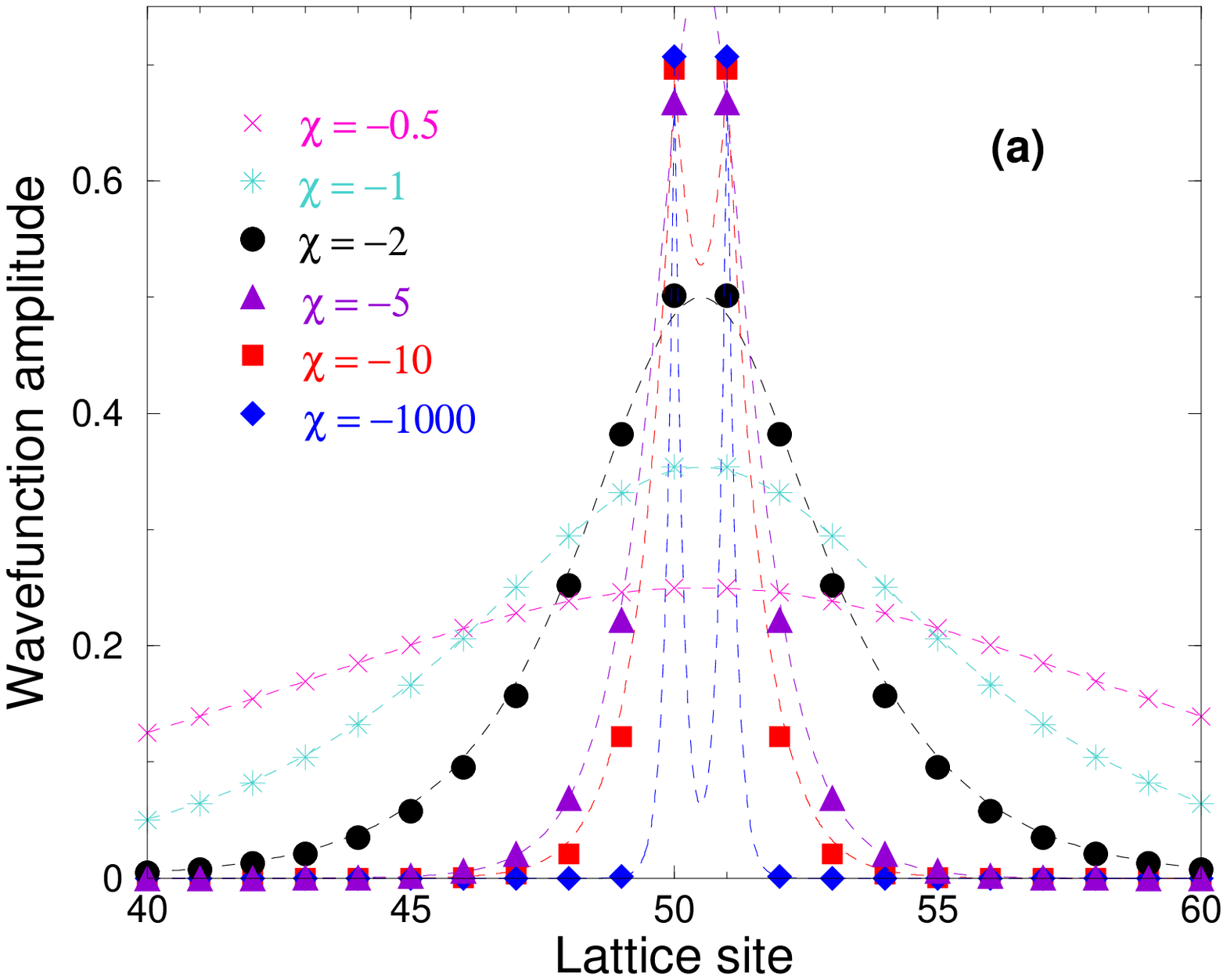,height=6.0cm,width=6.8cm} & \hspace{-0.3cm} 
\epsfig{file=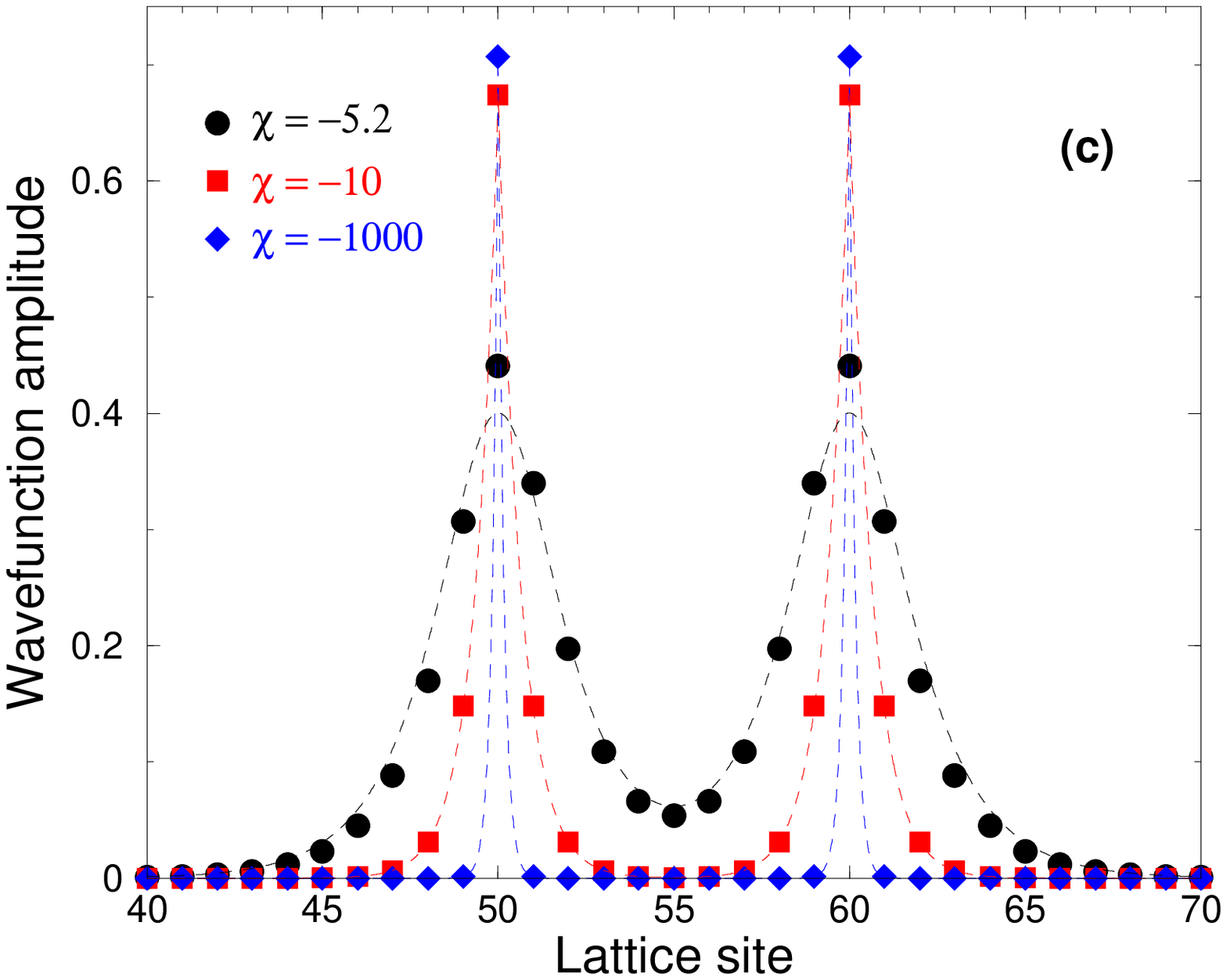,height=6.0cm,width=6.8cm}  \\
\epsfig{file=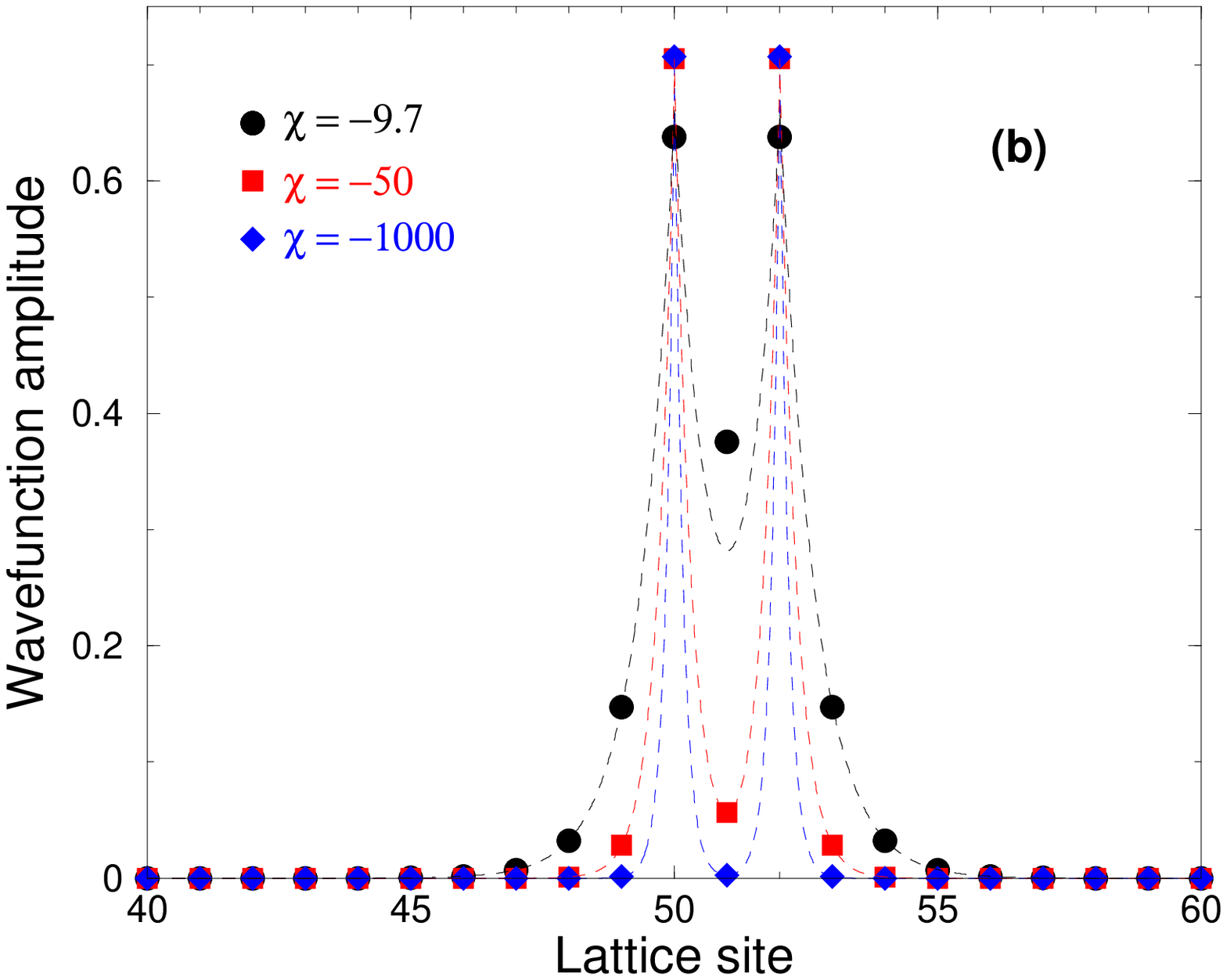,height=6.0cm,width=6.8cm} & \hspace{-0.3cm}
\epsfig{file=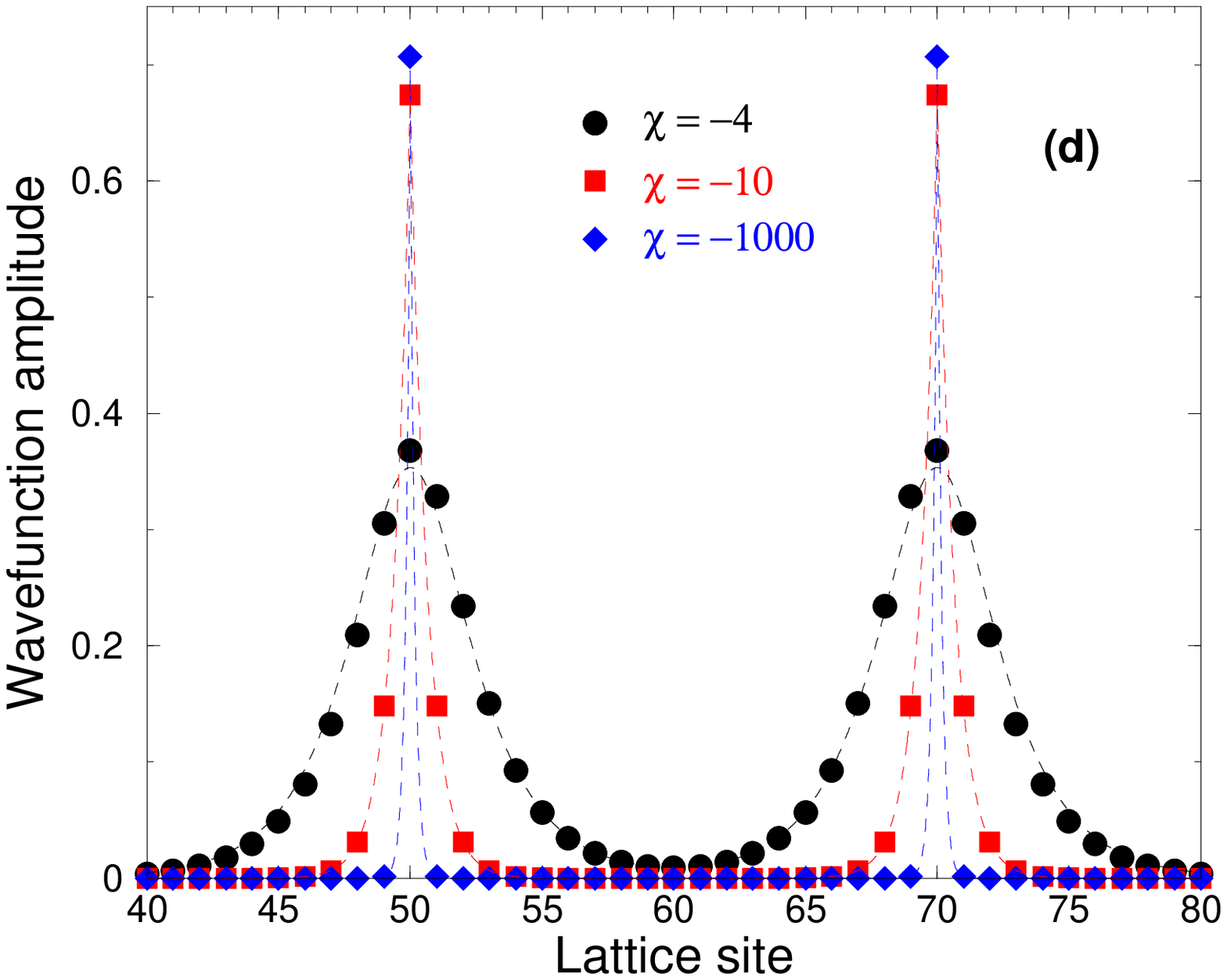,height=6.0cm,width=6.8cm}  \\
\end{tabular}
\end{center}
\caption{Double-peaked symmetric stationary solutions of DNLS in 1D
(points) for different values of the nonlinearity $\chi$ and various
interpeak separations $S$: {\bf (a)} $S=1$ lattice site, {\bf (b)} $S=2$
sites, {\bf (c)} $S=10$ sites, and {\bf (d)} $S=20$ sites. Dashed lines
show analytical approximations of the solutions using Eq.~(\ref{dpd}) for
the discrete cases where $|\chi|>6$, Eq.~(\ref{dpc}) for $\chi=-5.2$ in (c)
and $\chi=-4$ in (d), and Eq.~(\ref{nlsw}) centered in the middle between the
sites of maximum amplitude for $\chi=-0.5$ up to $-5$ in (a) (see text).  }
\label{fDS1}
\end{figure}

In order to find the branches of symmetric and antisymmetric
DP solutions, the method described in section 2 is applied
using as initial states
\be
\psi_n^{(r=0)}= \frac{1}{\sqrt{2}}(\delta_{n,n_1} + \delta_{n,n_2})
\hspace{1.0cm}  \mbox{and} \hspace{1.0cm}
\psi_n^{(r=0)}= \frac{1}{\sqrt{2}}(\delta_{n,n_1} - \delta_{n,n_2}),
\ee
respectively. The distance $|n_2-n_1|$ determines the interpeak separation
$S$ of the corresponding solution.

\begin{figure}
\begin{center}
\begin{tabular}{cc}
\epsfig{file=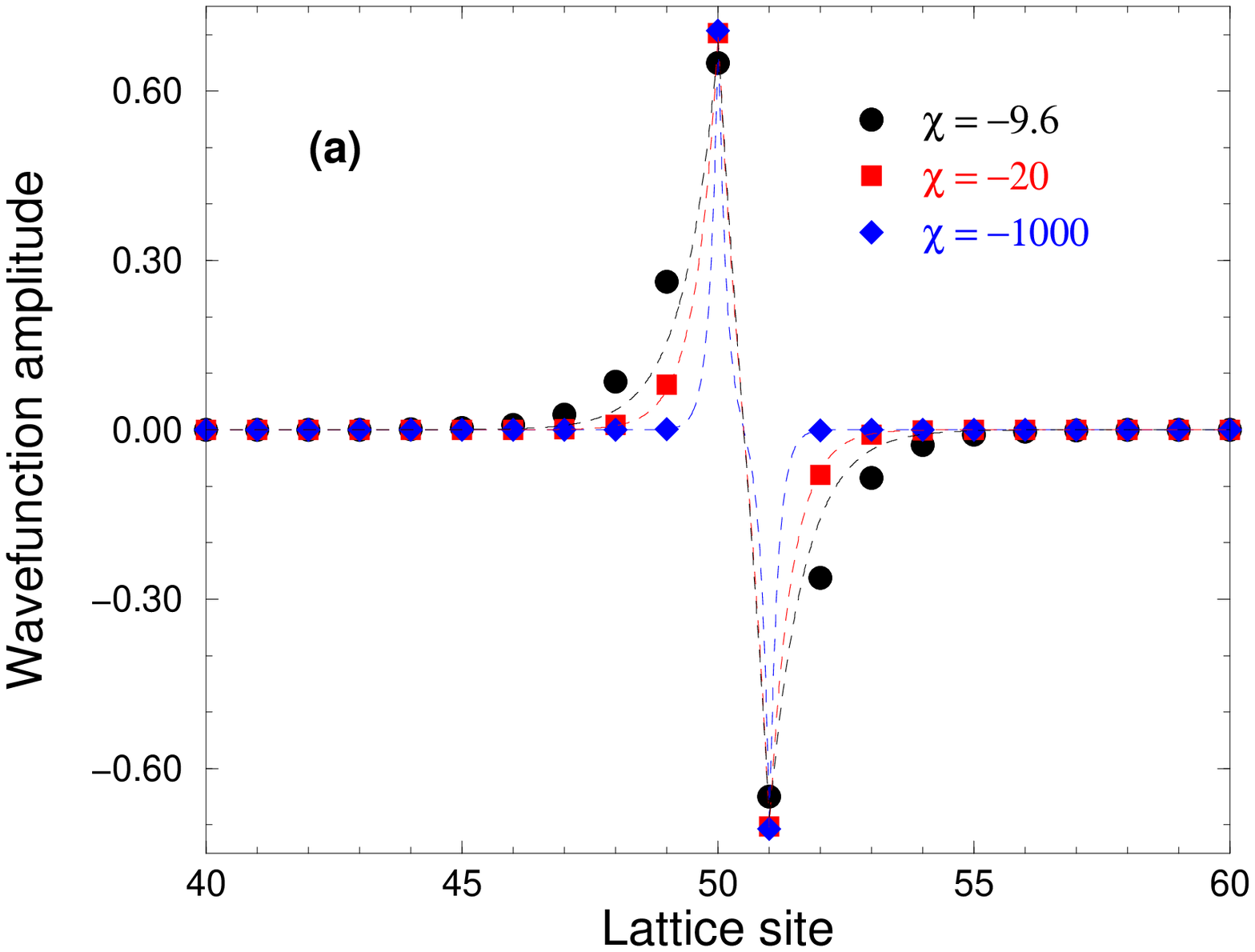,height=6.0cm,width=6.8cm} & \hspace{-0.3cm} 
\epsfig{file=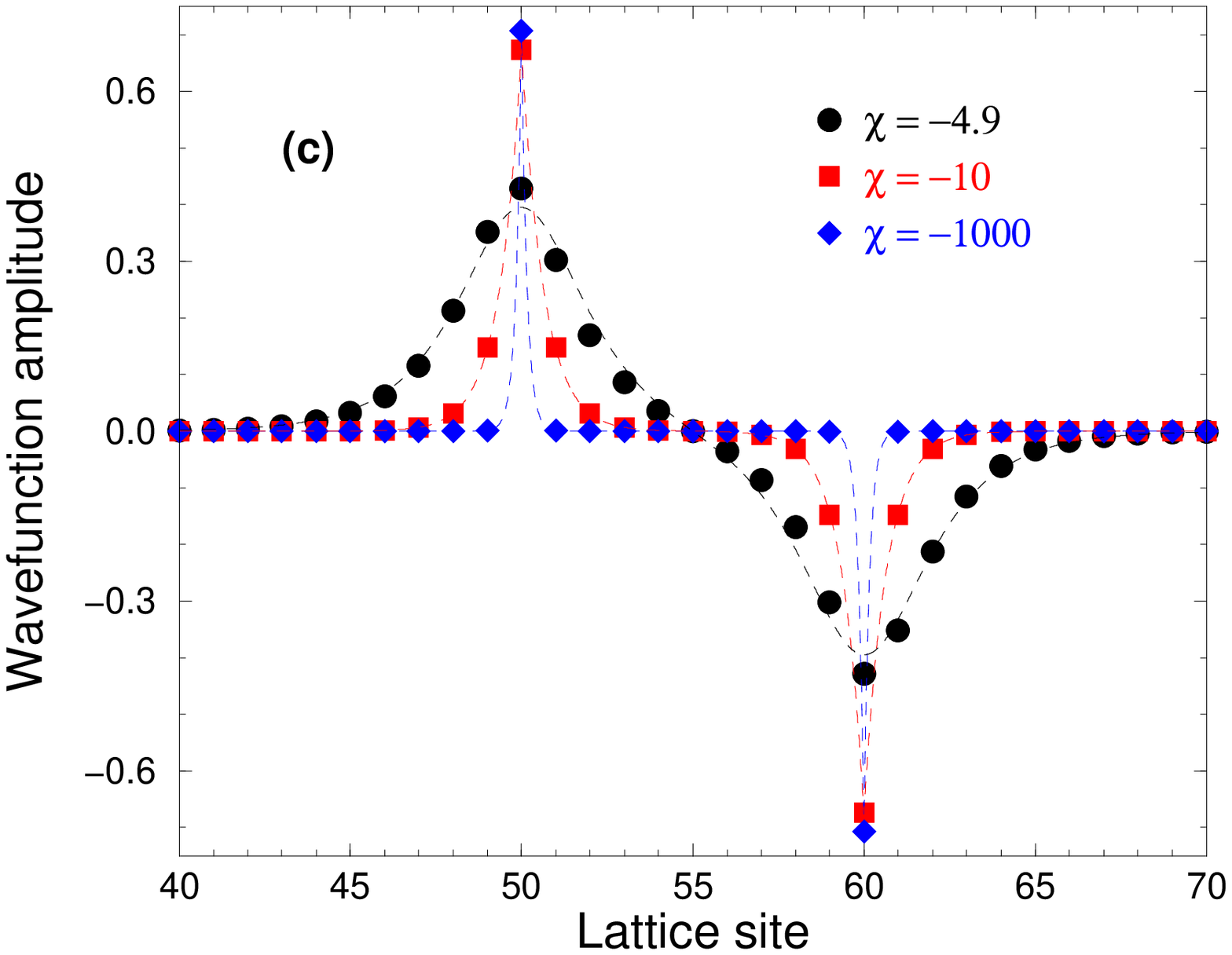,height=6.0cm,width=6.8cm}  \\
\epsfig{file=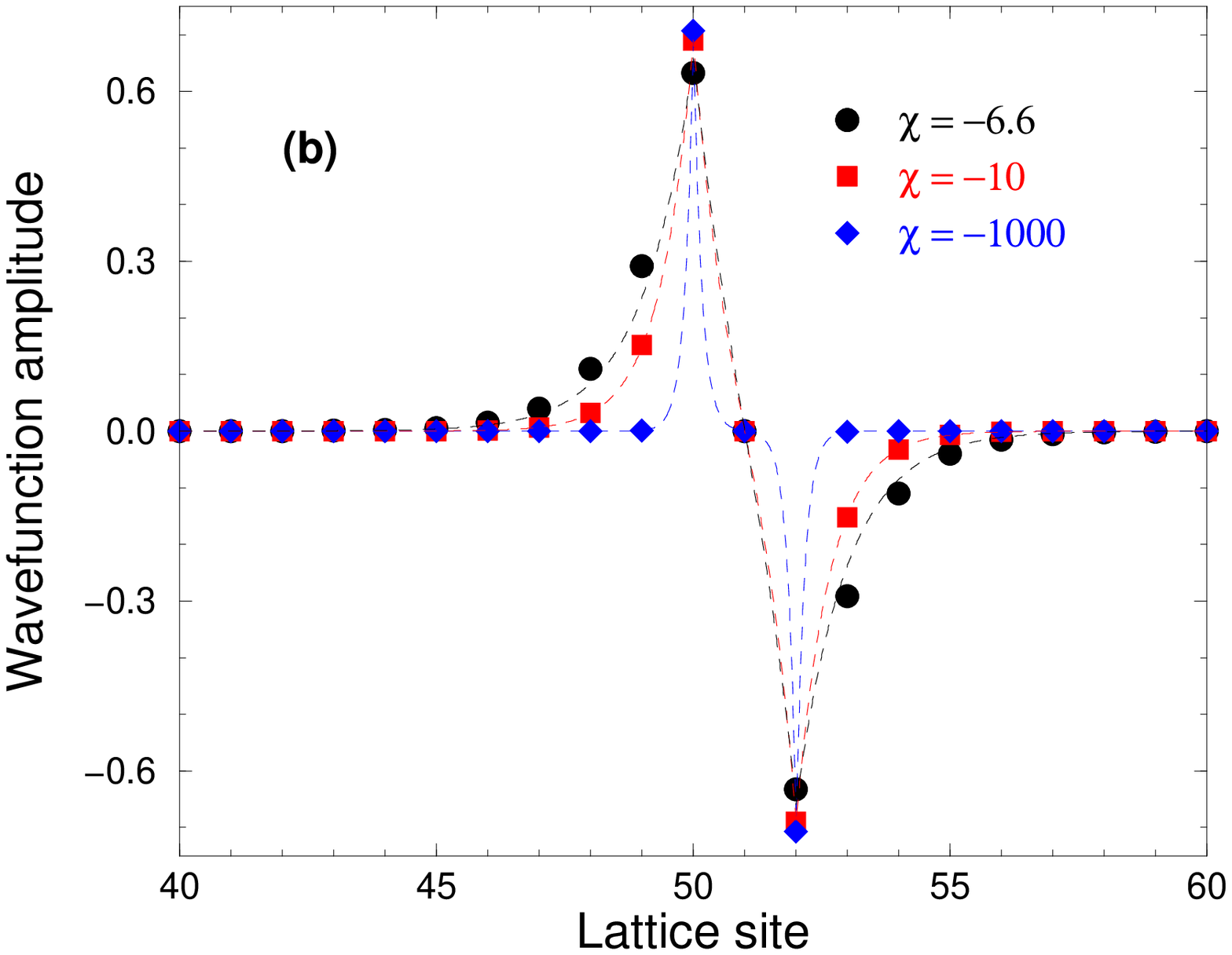,height=6.0cm,width=6.8cm} & \hspace{-0.3cm}
\epsfig{file=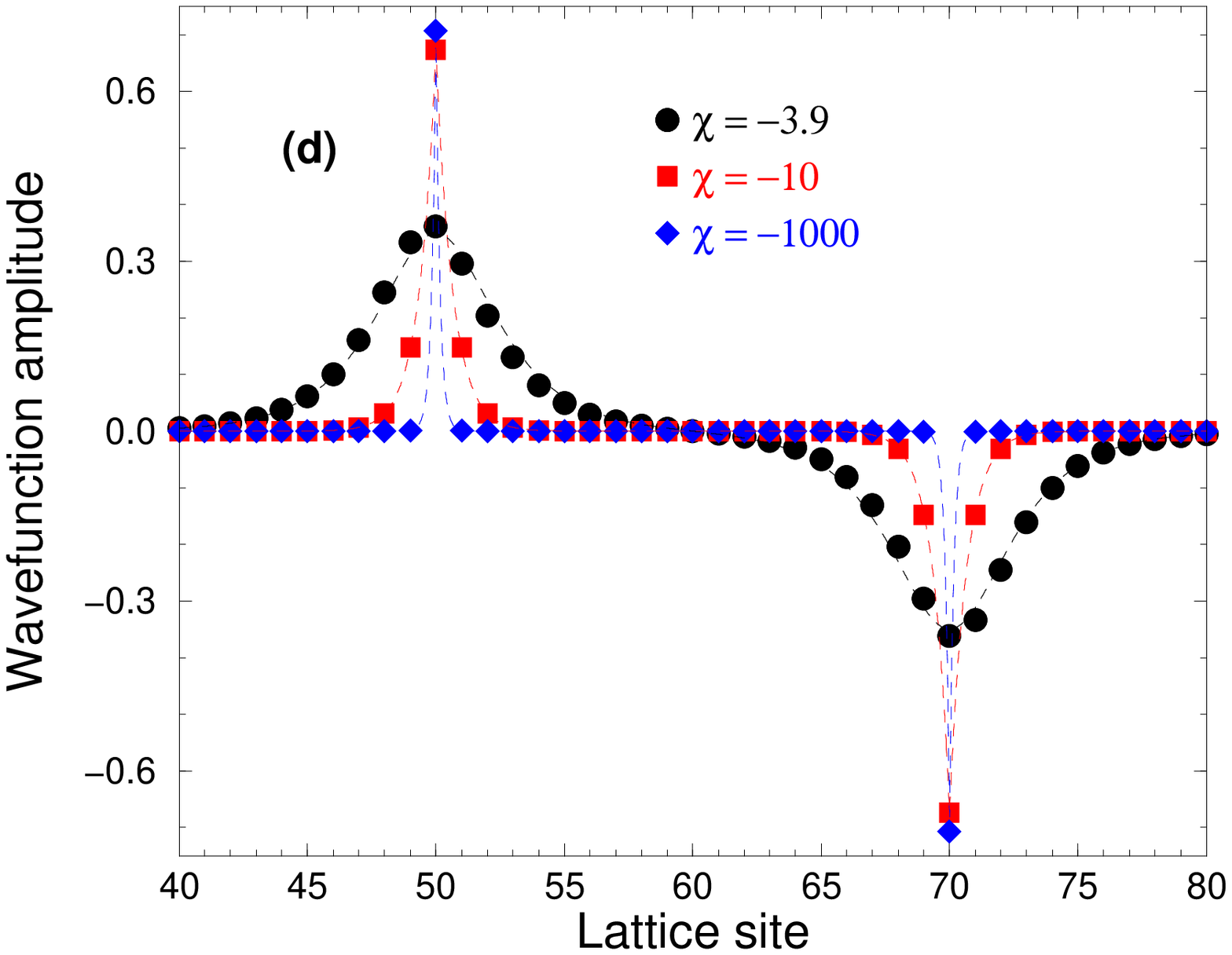,height=6.0cm,width=6.8cm}  \\
\end{tabular}
\end{center}
\caption{Double-peaked antisymmetric stationary solutions of DNLS in 1D
(points) for different values of the nonlinearity $\chi$ and various
interpeak separations $S$: {\bf (a)} $S=1$ lattice site, {\bf (b)} $S=2$
sites, {\bf (c)} $S=10$ sites, and {\bf (d)} $S=20$ sites. Dashed lines
show analytical approximations of the solutions using Eq.~(\ref{dpd}) in the
discrete cases where $|\chi|>6$ and Eq.~(\ref{dpc}) in the two more extended
cases of (c) and (d).}
\label{fDA1}
\end{figure}

Stationary solutions with interpeak separations 1, 2, 10, and 20 sites
are shown for different values of $\chi$ in Fig.~\ref{fDS1} for symmetric
and in Fig.~\ref{fDA1} for antisymmetric states.
As it is expected, by decreasing $|\chi|$ the solutions spread
more and more, until a value of $\chi$ where a bifurcation occurs
resulting in the disappearance of the corresponding branch.
The smaller values of $|\chi|$ shown at each plot of Fig.~\ref{fDS1}
(apart from the case of $S=1$, displayed in Fig.~\ref{fDS1}a) and
Fig.~\ref{fDA1} are close to the bifurcation point and therefore represent
one of the most extended solutions of the corresponding branch.
The case of symmetric states with interpeak distance equal to one lattice
constant (Fig.~\ref{fDS1}a) is exceptional in this respect, since the
corresponding branch survives until the limit $|\chi| \rightarrow 0$, as it
merges to the SP branch for sufficiently small values of $|\chi|$.
As the continuous limit is approached, the DP symmetric state with $S=1$
becomes practically indistinguishable from the
SP state, i.e. the solution given by Eq.~(\ref{nlsw}).

The symmetric (antisymmetric) states with $S=1$ provide the lower (upper)
single branch of the DP solutions of Fig.~\ref{figFS1},
with frequency equal to $\omega = \frac{\chi}{2}-1$
($\omega = \frac{\chi}{2}+1$), for relatively large $|\chi|$. All the
other symmetric and antisymmetric DP solutions with interpeak
separations $S>1$ are congested in the middle branch (with
frequency equal to $\omega = \frac{\chi}{2}$ for large $|\chi|$).
In particular, in Fig.~\ref{figFS1} the middle branch contains
solutions for $S=2$, 3, 4, 5, 10, 20, and 200.
As the interpeak separation $S$ increases, the two peaks start to
not overlap much, even for small values of $|\chi|$, and in this case
the corresponding branch survives longer (i.e. persists closer to $\chi=0$).
These branches, as $|\chi|$ decreases, eventually deviate from the line
$\omega = \frac{\chi}{2}$, and for even smaller values of $|\chi|$ they
can be described by double-peaked solutions of the continuous nonlinear
Schr\"odinger equation.

Approximate analytical expressions, which describe the DP
states in the most of the cases, can be obtained by appropriate
superpositions of the single-peaked states (\ref{vap}) and (\ref{nlsw}).
One has to take into account that when the two peaks do not significantly
overlap the norm of each peak is about half of the total norm. Then
from Eq.~(\ref{metN}) follows that the two individual wavefunctions
superimposed in a double-peaked solution should be provided by Eqs.
(\ref{vap}) or (\ref{nlsw}) corresponding to $\frac{\chi}{2}$. As a result
a symmetric or antisymmetric DP solution with interpeak separation $S$,
where the first peak is located at the site $n_1$ and the second at
$n_2=n_1+S$ ($S$ is assumed to be positive), can be approximated by
\be  \label{dpd}
\psi_n^{DP} = \frac{1}{\sqrt{2(1\pm P)}} \sqrt{\frac{1-\zeta^2}{1+\zeta^2}}
\left( \zeta^{|n-n_1|} \pm (-sgn \chi)^S \zeta^{|n-n_1-S|} \right),
\ee
\be  \label{ov1}
\mbox{where} \hspace{0.3cm}
P = (-sgn \chi)^S \; \frac{(1+S) \zeta^S - (S-1) \zeta^{S+2}}{1+\zeta^2}
\ee
\be
\mbox{and} \hspace{0.3cm} \zeta = -\frac{1}{\chi/2} - \frac{2}{(\chi/2)^3}
= -\frac{2}{\chi} - \frac{16}{\chi^3},
\ee
or
\be  \label{dpc}
\psi_n^{DP} = \frac{(-sgn \chi)^{n-n_1}}{\sqrt{2(1\pm P)}}
\sqrt{\frac{|\chi|}{16}} \left( \frac{1}{\cosh \frac{\chi (n-n_1)}{8}}
\pm \frac{(-sgn \chi)^S}{\cosh \frac{ \chi (n-n_1-S)}{8}} \right),
\ee
\be \label{ov1c}
\mbox{where} \hspace{0.3cm}
P= (-sgn \chi)^S \; \frac{ \frac{\chi S}{8}}{\sinh \frac{\chi S}{8}}.
\ee
The former (latter) solution can be used for a DP state
with discrete (rather extended) peaks, valid for relatively large
(small) values of $|\chi|$. Roughly speaking the transition
from one form to the other occurs for $|\chi|$ in the region $5-6$.
The normalization factors $P$ in Eqs. (\ref{ov1}) and (\ref{ov1c}) result
from the non-orthogonality of the superimposed wavefunctions, i.e.
$P=\sum_n \psi_n^{SP[n_1]}(\frac{\chi}{2}) \psi_n^{SP[n_1+S]}(\frac{\chi}{2})$,
where $\psi_n^{SP[m]}(\frac{\chi}{2})$ denotes the SP solution
centered at the site $m$ and calculated for the value $\frac{\chi}{2}$
of the nonlinearity parameter. The plus (minus) signs in these approximate
solutions correspond to symmetric (antisymmetric) states, except for the
case of positive $\chi$ and odd $S$, where they give the antisymmetric
(symmetric) DP state.

Eq.~(\ref{dpc}) is not applicable for small values of $|\chi|$ in the
exceptional case of symmetric states with $S=1$, due to the significant
overlap of the two non-distinguished peaks (Fig.~\ref{fDS1}a for $|\chi|$
approximately less than 5-6). In this case the
corresponding DP solution, whose branch is merging to the
single-peaked branch, can be well described by Eq.~(\ref{nlsw})
centered in the middle between the consecutive sites of the two peaks.
The analytical approximations discussed above have been plotted in
Figs. \ref{fDS1} and \ref{fDA1} (dashed lines) along with the numerical
solutions (points) for comparison. In particular, Eq.~(\ref{nlsw}) has been
used, as just explained, in
Fig.~\ref{fDS1}a for $\chi=-0.5$, $-1$, $-2$, and $-5$, Eq.~(\ref{dpc})
has been used only for the smallest values of $|\chi|$ in Figs.
\ref{fDS1}c, \ref{fDS1}d, \ref{fDA1}c, and \ref{fDA1}d (where
$|\chi|<6$), and Eq.~(\ref{dpd}) has been plotted in all the other
cases. The more distinguished the two peaks, i.e. the higher the
$S$, or the higher the $|\chi|$ even for smaller values of $S$, the better
the approximations (\ref{dpd}) and (\ref{dpc}) are.

\subsubsection{Linear stability}

Symmetric and antisymmetric DP states show qualitatively different
behavior regarding their stability. A detailed investigation of the
stability eigenvalues is presented below for negative values of $\chi$.
From this, the behavior at positive $\chi$ can be obtained as follows:
for even $S$ the situation, regarding the stability eigenvalues,
is exactly the same as that of $\chi<0$, while
for odd $S$ the symmetric (antisymmetric) states behave exactly like
the antisymmetric (symmetric) states of $\chi<0$. Note here that the
transformation $(-1)^n \psi_n$, connecting stationary solutions at opposite
values of $\chi$, turns a symmetric DP state to antisymmetric and
vis versa when $S$ is odd. Therefore, what is mentioned below for $\chi<0$,
also holds as it is for $\chi>0$ when $S$ is even, while it valids after
interchanging roles between symmetric and antisymmetric states when $S$
is odd.

The eigenvalues of the linear stability problem are always obtained as pairs
of opposite sign and there always exists a pair of eigenvalues at zero.
In our approach (see Appendix), if there is an eigenvalue with non-zero
imaginary part, then the stationary solution is unstable. All relevant
discussions in this and the following section
will refer to those eigenvalues with non-negative real part, without usually
mentioning the pinned pair of eigenvalues at zero. The spectrum of
eigenvalues is always symmetric with respect to the imaginary axis.
The complete structure of the linear stability spectrum close to the
anti-continuous limit is analytically derived in the Appendix.

For negative $\chi$, the nonlinear term of DNLS has the same sign as the
tunneling term. Then it is known that the symmetric DP states are unstable
\cite{LKS,KKM}, in contrary to the case of the antisymmetric ones
that may be linearly stable \cite{JA,PKF}. In fact, the latter
are linearly stable in the larger part of the corresponding branch.
When this branch becomes unstable, by decreasing $|\chi|$,
soon it disappears. The larger the interpeak distance, the smaller the
$|\chi|$'s at which the branch turns unstable and disappears. The scenario
for development of instability and disappearance seems to roughly be as
follows. For large $|\chi|$, linear stability analysis provides,
a discrete eigenvalue outside of the band.
For relatively large $|\chi|$ the band extends from $\frac{|\chi|}{2}-2$ to
$\frac{|\chi|}{2}+2$, apart from the case of symmetric (antisymmetric) states
with interpeak separation $S=1$, where it extends from $\frac{|\chi|}{2}-1$
to $\frac{|\chi|}{2}+3$ (from $\frac{|\chi|}{2}-3$ to $\frac{|\chi|}{2}+1$).
As $|\chi|$ decreases, the band moves towards zero, while the discrete
eigenvalue keeps off zero for $S>2$, remains almost constant for $S=2$, and
goes to zero (but slower than the band) for $S=1$. After their collision (or
the collision of the discrete eigenvalue with another eigenvalue splitted off
the band) the instability develops.
As $|\chi|$ decreases more, the band is approaching zero and the
branch disappears when an eigenvalue splitted from the band collides
with the pinned eigenvalue at zero. Table~1 shows for antisymmetric
DP solutions of various interpeak separations (first column)
the nonlinearity regime where the corresponding branch becomes unstable
(second column; for larger values of $|\chi|$ the solutions are linearly
stable) and the nonlinearity regime where the branch disappears
(third column).

\begin{table}
\caption{Nonlinearity strength $\chi$ for the development of instability and
disappearance of antisymmetric and the disappearance of symmetric double-peaked
stationary solutions of DNLS at different interpeak separations. Values in
parentheses in the second column show rough analytical estimates using
Eq.~(\ref{antun}).}
\centerline{
\begin{tabular}{|c|c|c|c|} \hline 
Interpeak & Regime of $\chi$ at which & Regime of $\chi$ at which
& Regime of $\chi$ at which   \\   separation & the antisymmetric branch &
the antisymmetric & the symmetric   \\
 & becomes unstable & branch disappears & branch disappears   \\  \hline
$S=1$  &  $[-19.73,-19.72]$ \hspace{0.05cm} ($-18$) &  $[-9.58,-9.57]$  &  $-$
\\ $S=2$  & $[-8.8,-8.7]$ \hspace{0.5cm} ($-8$)  &  $[-6.54,-6.53]$  &
$[-9.63,-9.62]$  \\   $S=3$  &  $[-7.1,-7.0]$ \hspace{0.5cm} ($-6.3$)  & 
$[-5.9,-5.8]$ &  $[-7.9,-7.8]$   \\   $S=4$  & $[-6.5,-6.4]$ \hspace{0.5cm}
($-5.5$) &  $[-5.7,-5.6]$  &  $[-7.0,-6.9]$   \\   $S=5$ &  $[-6.2,-6.1]$
\hspace{0.5cm} ($-5.0$) & $[-5.5,-5.4]$  &  $[-6.5,-6.4]$   \\   $S=10$  &
$[-5.0,-4.9]$ \hspace{0.5cm} ($-4.2$) &  $[-4.9,-4.8]$ &  $[-5.2,-5.1]$  \\
$S=20$  &  $[-3.90,-3.89]$ \hspace{0.15cm} ($-4.0$)  &  $[-3.90,-3.89]$  &
$[-4.0,-3.9]$
\\    \hline
\end{tabular}  }
\end{table}

\begin{figure}
\centerline{\hbox{\psfig{figure=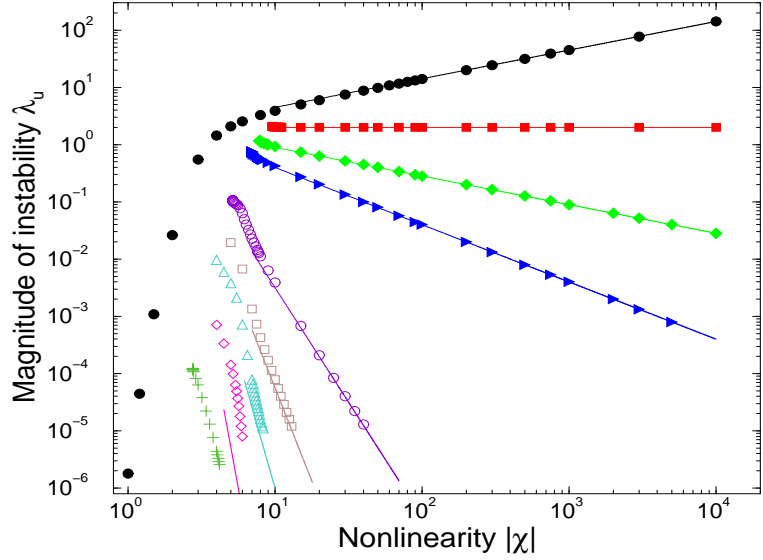,width=10cm,height=7.5cm}}}
\caption{Magnitude of instability $\lambda_u$ of symmetric double-peaked
solutions of DNLS for $\chi<0$ in 1D (points) as a function of the strength
of nonlinearity
$|\chi|$, for various interpeak separations: $S=1$ (filled circles), $S=2$
(filled squares), $S=3$ (filled diamonds), $S=4$ (filled triangles), $S=10$
(open circles), $S=15$ (open squares), $S=20$ (open triangles), $S=30$
(open diamonds), and $S=50$ (crosses). Lines show the power-law relation,
Eq.(\ref{inst1}), derived for relatively large values of $|\chi|$.}
\label{fstab}
\end{figure}

Regarding the unstable symmetric DP states, linear stability
analysis reveals that there is always one pair of purely imaginary unstable
eigenvalues. If the magnitude of instability is denoted by $\lambda_u$
(i.e. the unstable eigenvalues are $\pm i \lambda_u$),
then for fixed value of $\chi$, $\lambda_u$ decreases
as the interpeak separation increases. For fixed interpeak distance $S$,
but larger than two lattice sites, $\lambda_u$ decreases by increasing
$|\chi|$. For $S=2$, $\lambda_u$ tends to a constant value for large $|\chi|$,
while for $S=1$, $\lambda_u$ decreases with decreasing $|\chi|$, in accordance
with the fact that this branch merges to the stable single-peaked branch for
$|\chi| \rightarrow 0$. This behavior is shown in Fig.~\ref{fstab}.
In some cases, especially for large interpeak separations, the magnitude
of instability is so small that for any practical purpose the corresponding
solution can be considered as quasi-stable \cite{LKS}.
The log-log plot of $\lambda_u$ as a function of $|\chi|$ in Fig.~\ref{fstab}
demonstrates a power-law dependence $\lambda_u \sim |\chi|^a$ at large
values of $|\chi|$. Relating the stability eigenvalues with the energy
spectrum of a tight-binding problem in the presence of a deep and narrow
double-well potential (see Appendix), one obtains
\be   \label{inst1}
\lambda_u = 2^{S/2} \; |\chi|^{1-\frac{S}{2}}, \hspace{1.0cm} 
\mbox{for} \; \; |\chi| \gg 1.
\ee
This relation is plotted in Fig.~\ref{fstab} for $S=$ 1, 2, 3, 4, 10, 15, 20,
and 30 (solid lines) along with the corresponding numerical results (points).
The agreement is very good for $|\chi|$ larger than $10-20$.

The disappearance of the symmetric branches occurs for non-zero $\chi$
(apart from the case of $S=1$, where the corresponding branch merges with
the SP branch) when, as in the antisymmetric case, a real
eigenvalue splitted from the band collides with the pinned eigenvalues
at zero. Table~1 shows (fourth column) the nonlinearity regime where
different branches of symmetric DP solutions disappear. It seems that for
fixed $S$ ($S>1$), the antisymmetric branches survive until smaller
values of $|\chi|$.

As it is shown in the Appendix, close to the anti-continuous limit the
stable (real) discrete eigenvalue of an antisymmetric DP state has the same
dependence like in Eq.~(\ref{inst1}) (see Eq.~(\ref{steig})). Numerical
simulations confirm that for relative large $|\chi|$ the stable eigenvalues
of antisymmetric states have the same magnitude with the unstable of the
symmetric ones. This result can be used for a rough estimate of the
nonlinearity value, $\chi_{un}$, where an antisymmetric state becomes
unstable (through the collision of the discrete eigenvalue with the band), by
\be     \label{antun}
\frac{|\chi|}{2}-2= 2^{S/2} \; |\chi|^{1-\frac{S}{2}} \hspace{0.4cm} 
\mbox{for} \; \; S>1, \hspace{0.6cm} \mbox{or} \hspace{0.6cm}
\frac{|\chi|}{2}-3=\sqrt{2|\chi|}  \hspace{0.4cm} \mbox{for} \; \; S=1,
\ee
where $\frac{|\chi|}{2}-2$ ($\frac{|\chi|}{2}-3$) is the lower band edge.
The larger the $|\chi_{un}|$ the better the estimate, since Eq.~(\ref{inst1})
is valid for $|\chi| \gg 1$. Estimates of $\chi_{un}$, resulting from the
solution of Eq.~(\ref{antun}), are shown in the second column of Table~1
inside parentheses, next to the numerical results. The relative error is less
than $10\%$ for $S=1$ and $S=2$, but it increases for
larger $S$ where $|\chi_{un}|$ is getting smaller.

\section{Multi-peaked solutions in 2D}

\subsection{Frequency spectrum}

\begin{figure}
\centerline{\hbox{\psfig{figure=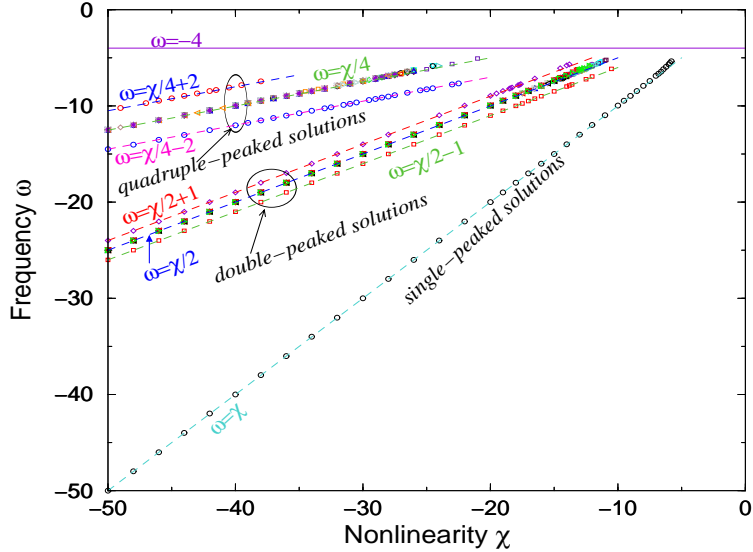,width=9.9cm,height=7.5cm}}}
\caption{Frequencies of single- double- and quadruple-peaked stationary
solutions of DNLS in 2D (points). Dashed lines show analytical expressions
obtained for large values of $|\chi|$. The horizontal line at $\omega=-4$
indicates the lower edge of the band of Bloch stationary states, which
extends from $-4$ to 4. The spectrum is antisymmetric on $\chi$;
$\omega (-\chi)=-\omega (\chi)$.}    \label{fFS2}
\end{figure}

Similarly to the 1D case, also in 2D there are many families of multi-peaked
localized stationary states corresponding to discrete levels in the
frequency spectrum. For large values of $|\chi|$ these levels tend
to the anti-continuous limit spectrum.
Fig.~\ref{fFS2} shows a part of the frequency spectrum in 2D, which,
apart from the single-peaked states of lowest frequency,
contains many branches of double-peaked and quadruple-peaked (QP)
stationary solutions at various interpeak distances.

All the calculated DP states (symmetric or antisymmetric, along the
lattice axes or along the diagonal, and with different interpeak separations)
have frequencies around $\omega = \frac{\chi}{2}$, apart from the
symmetric and antisymmetric states with their two peaks at neighboring sites
(examples are shown at the left columns of Fig.~\ref{fDS2} and
Fig.~\ref{fDA2}), which have frequencies around
$\omega = \frac{\chi}{2}-1$ and $\omega = \frac{\chi}{2}+1$, respectively.
Therefore, from the branches of the DP solutions of
Fig.~\ref{fFS2}, the middle one is highly crowded, tending to the
level of Eq.(\ref{acsp}) for $M=2$ at large $|\chi|$, while the two
external branches are single branches.

Similar considerations are valid for the branches of quadruplet stationary
states. What appears as a middle branch of the QP states
in Fig.~\ref{fFS2} actually contains many stationary solutions with
frequencies around $\omega = \frac{\chi}{4}$. Below and above these
highly congested branches there are
single branches corresponding to the symmetric (see left column of
Fig.~\ref{fQS2}) and antisymmetric (see left column of Fig.~\ref{fQA2})
solutions, respectively, with their four peaks on the corners of the
unit cell of the square lattice.

Multi-peaked stationary states, representative of some of the branches
shown in Fig.~\ref{fFS2}, are presented in the following two subsections
and their linear stability is discussed. Note that there also exist
triple-peaked solutions (with their positive or negative peaks along the
axes, or along the diagonal, or at random sites), which are not discussed
here. Only to mention that their corresponding branches tend to
$\omega=\frac{\chi}{3}$ for large values of $|\chi|$, i.e. are in between
the double-peaked and quadruple-peaked branches shown in Fig.~\ref{fFS2}.

\subsection{Double-peaked solutions}

\begin{figure}
\begin{center}
\begin{tabular}{ccc}
\epsfig{file=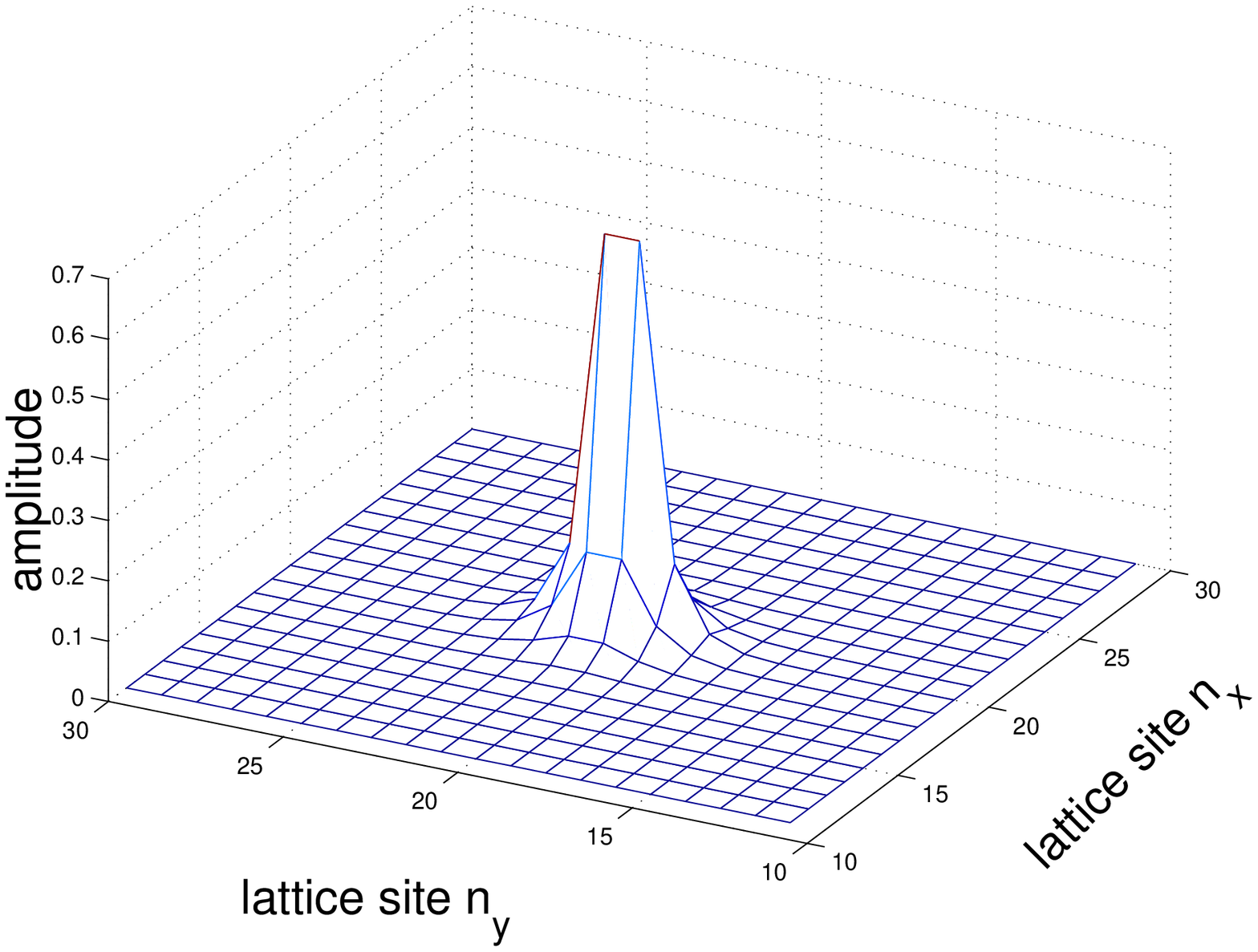,height=4.4cm,width=4.4cm} & \hspace{-0.2cm} 
\epsfig{file=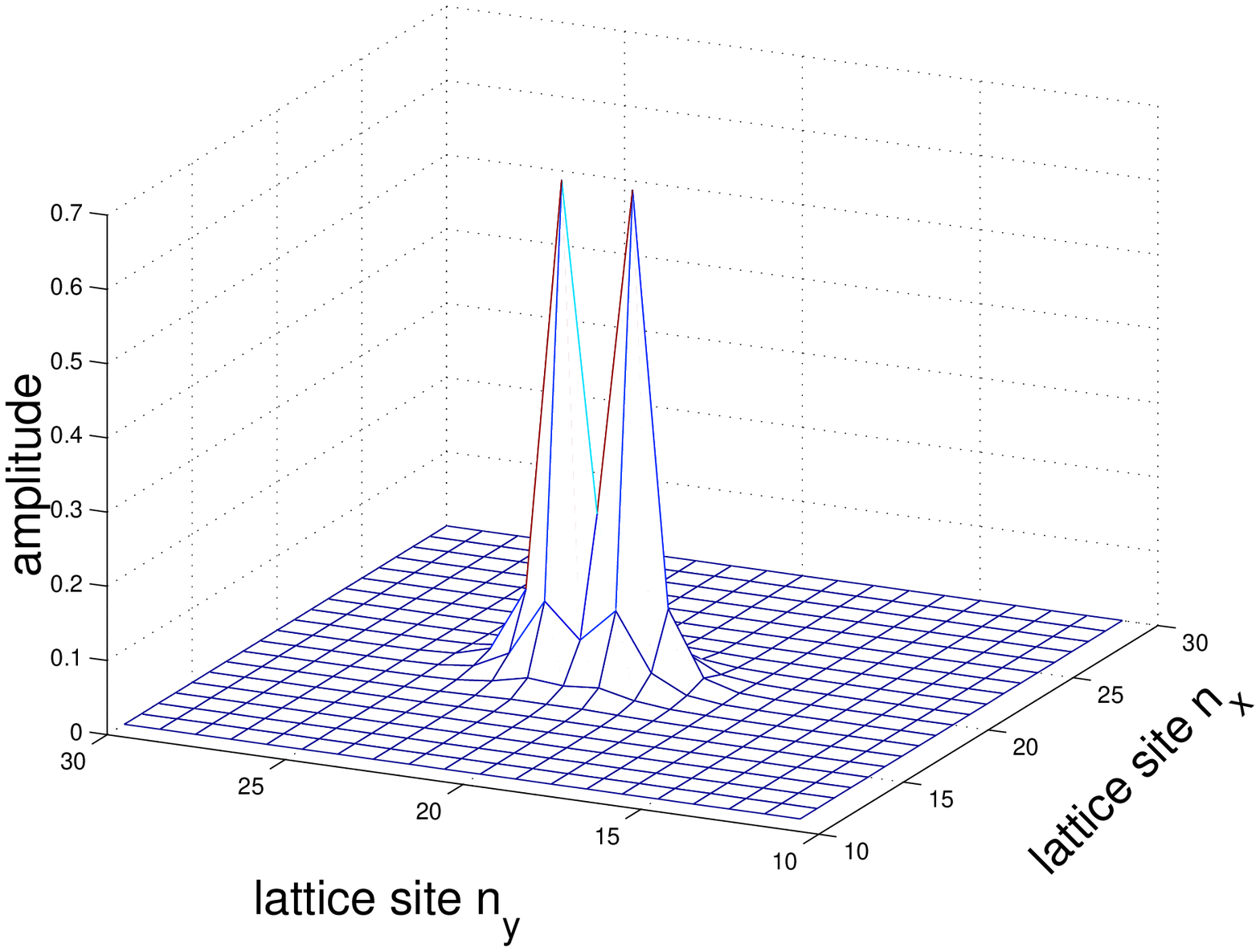,height=4.4cm,width=4.4cm} & \hspace{-0.2cm}
\epsfig{file=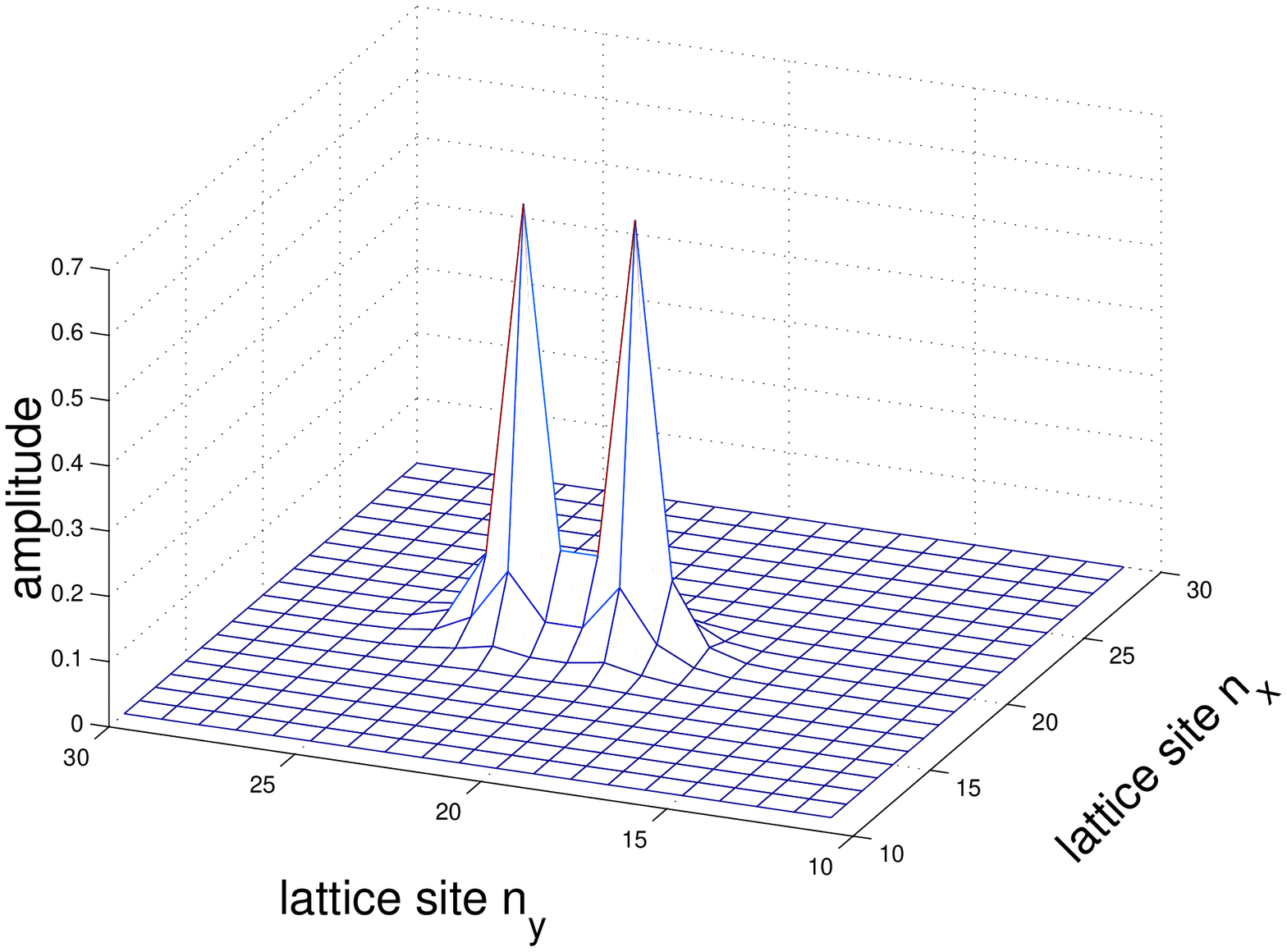,height=4.4cm,width=4.4cm}  \\
\epsfig{file=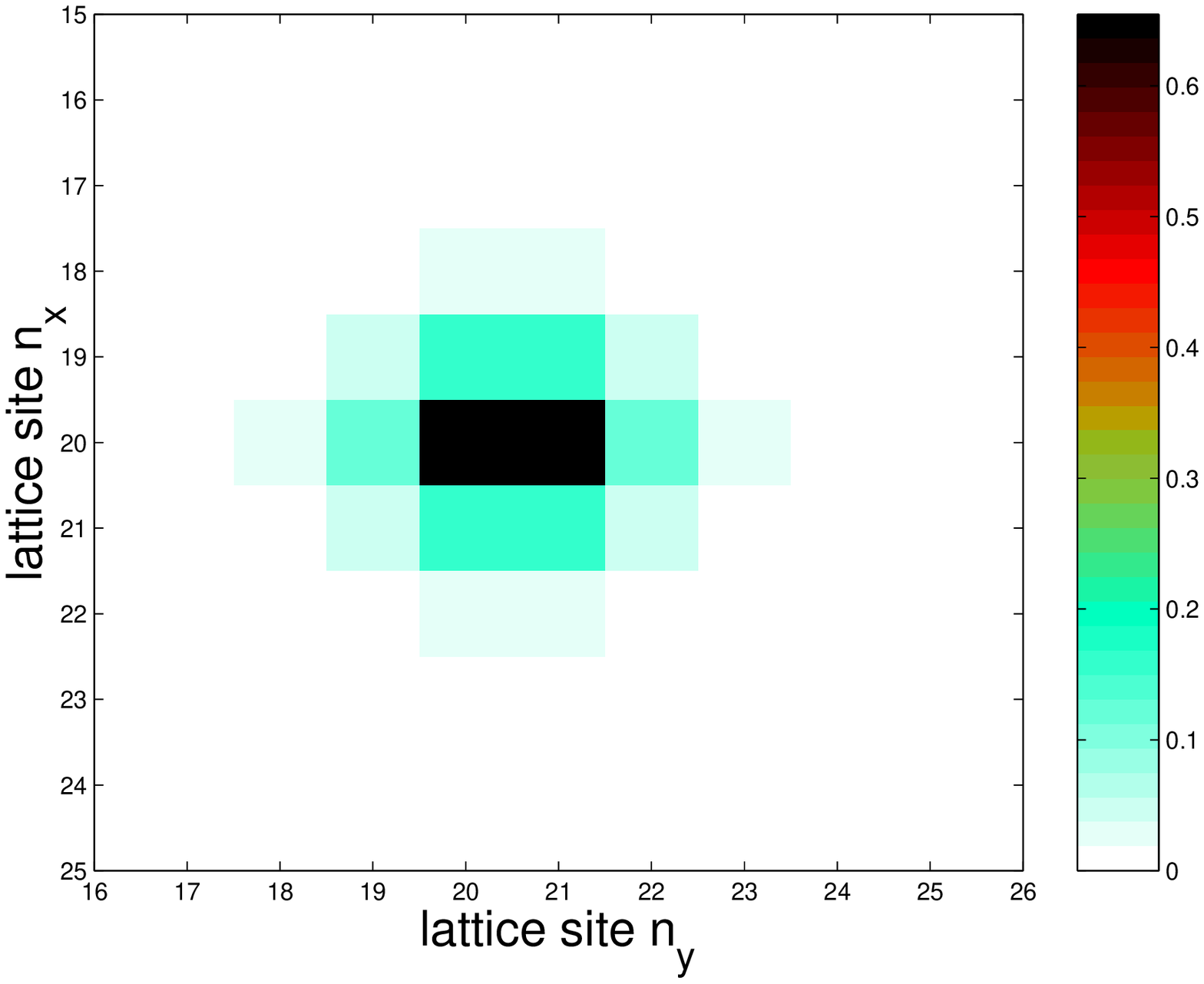,height=4.4cm,width=4.4cm} & \hspace{-0.2cm} 
\epsfig{file=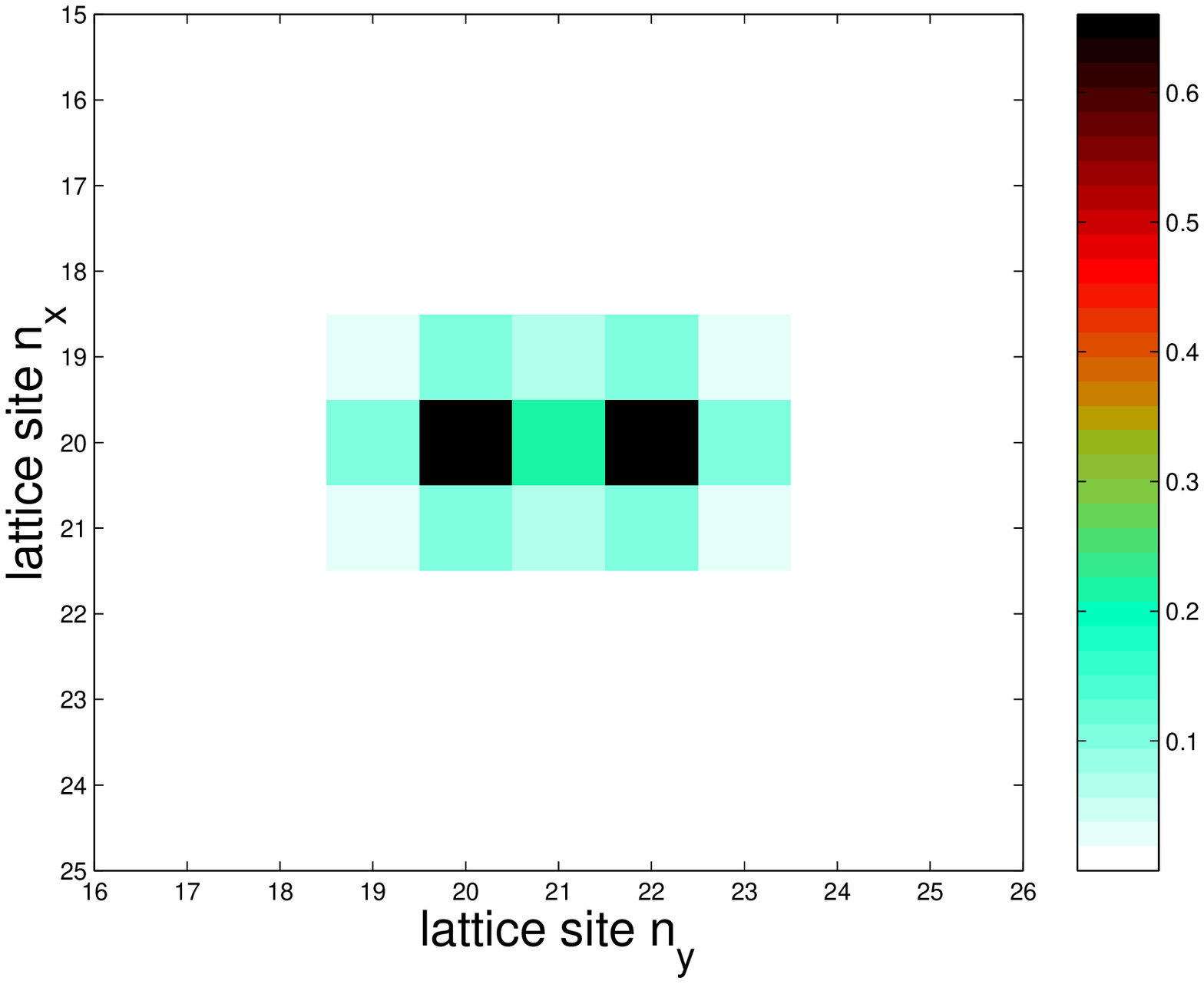,height=4.4cm,width=4.4cm} & \hspace{-0.2cm}
\epsfig{file=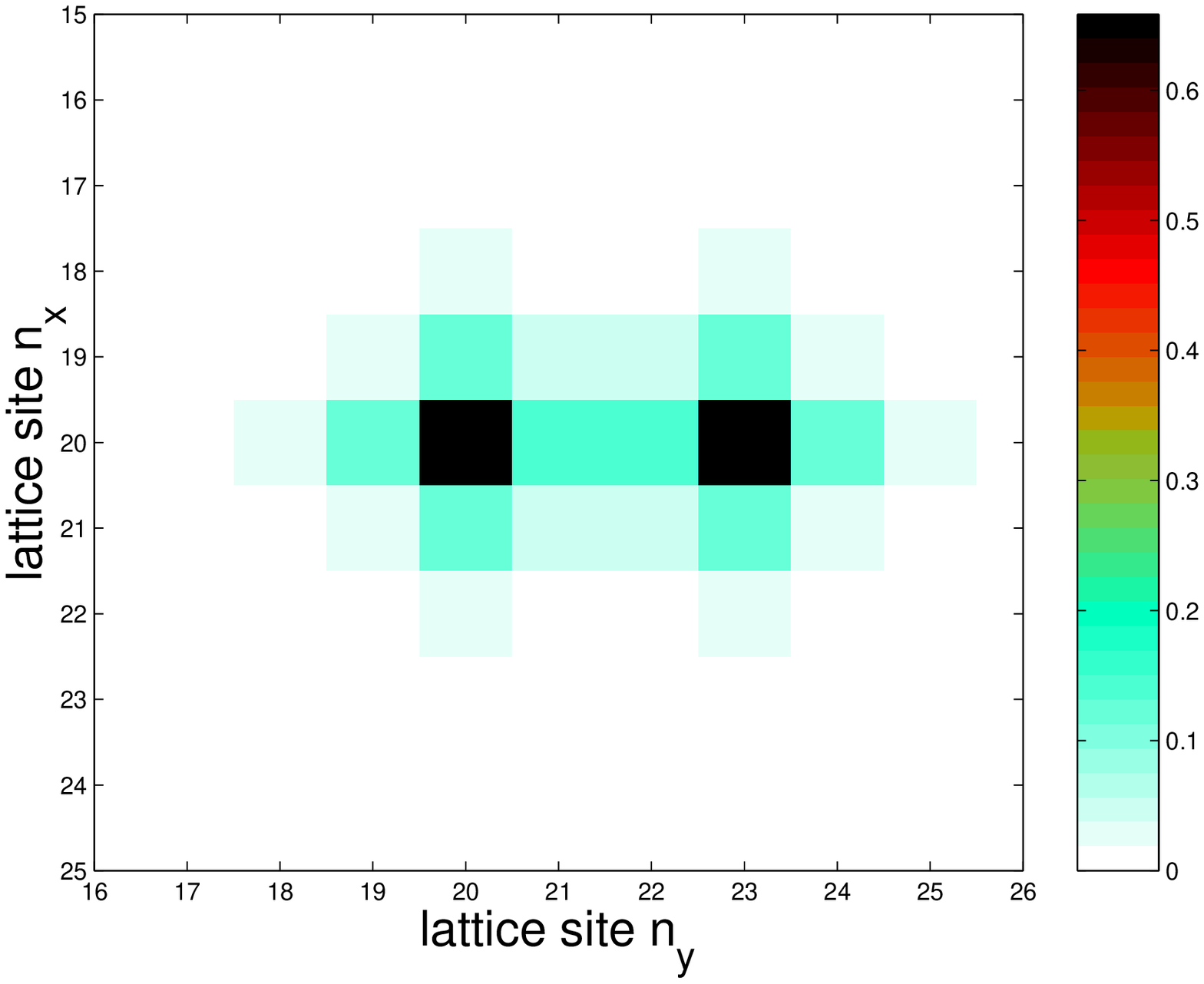,height=4.4cm,width=4.4cm}  \\
\epsfig{file=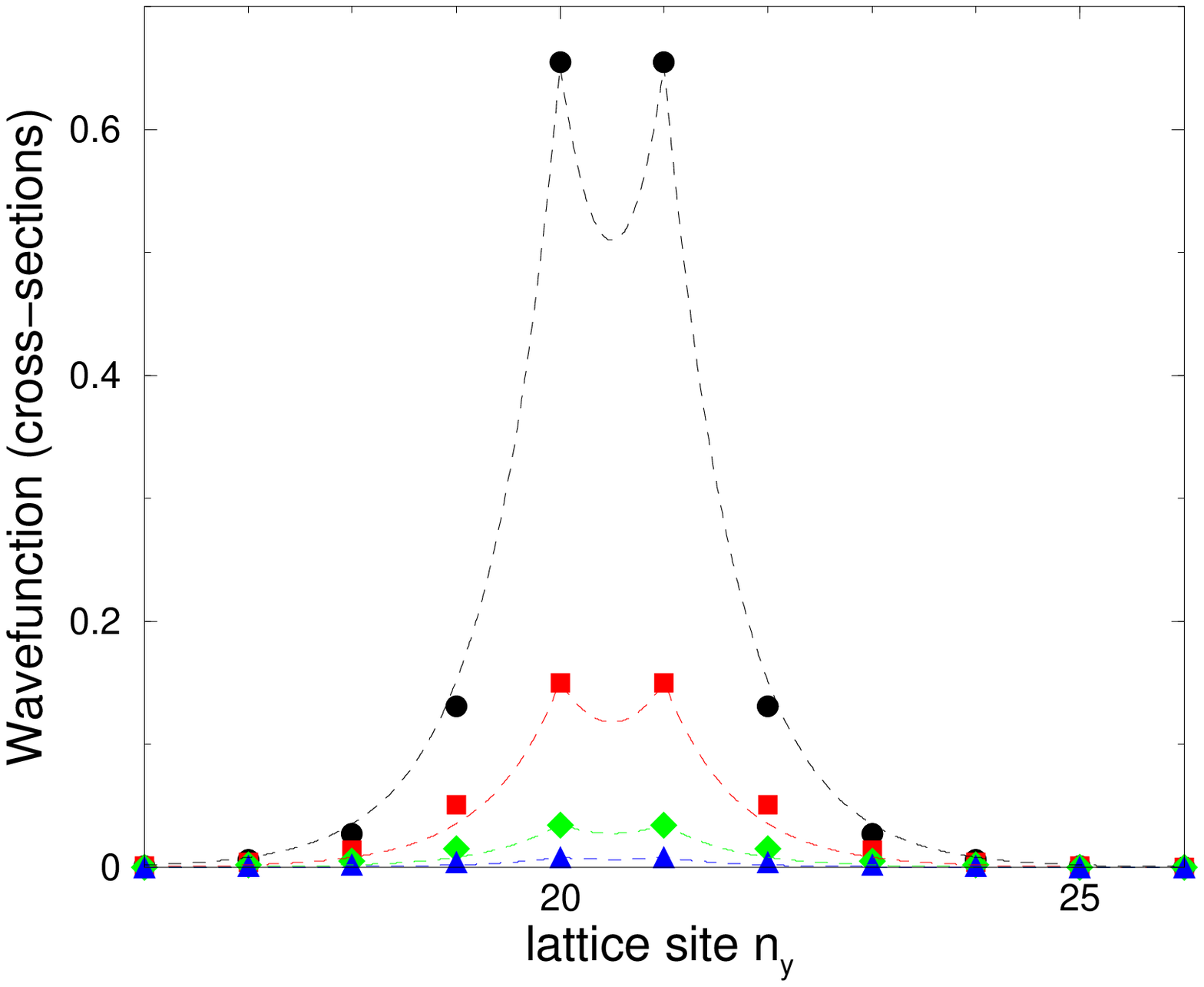,height=4.4cm,width=4.4cm} & \hspace{-0.2cm} 
\epsfig{file=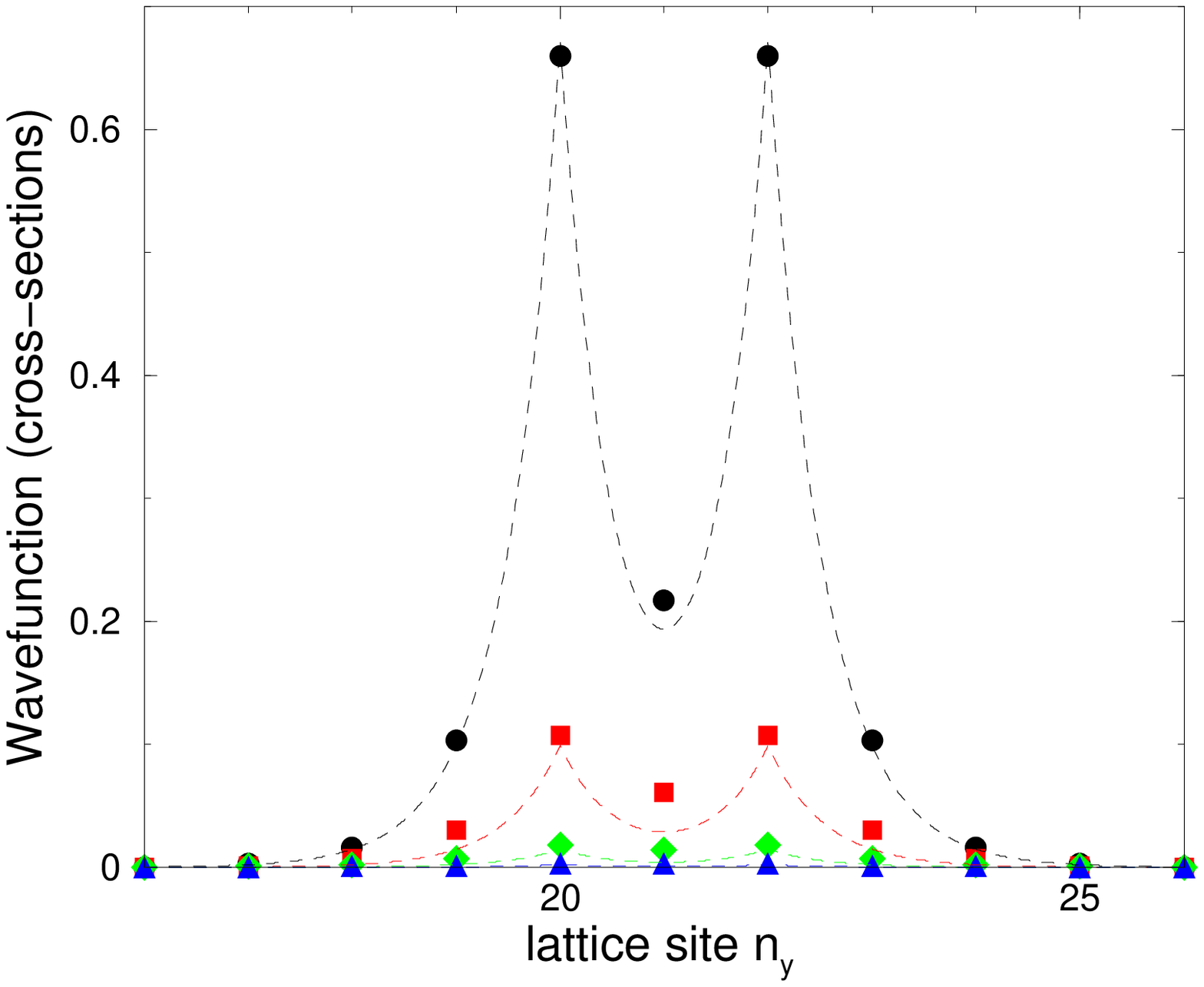,height=4.4cm,width=4.4cm} & \hspace{-0.2cm}
\epsfig{file=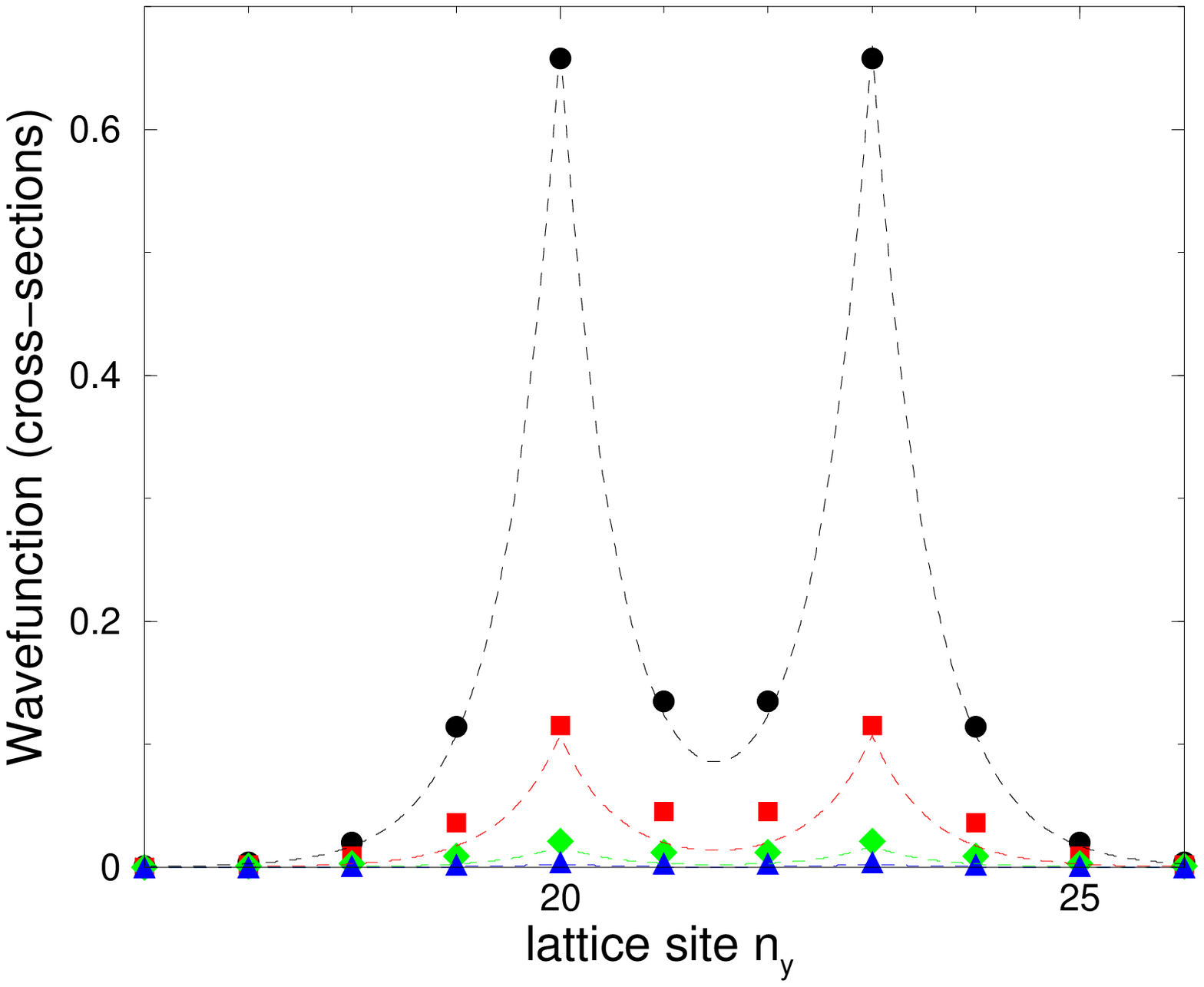,height=4.4cm,width=4.4cm}  \\
\end{tabular}
\end{center}
\caption{3D plots (first row) and density plots (second row) of
double-peaked symmetric solutions (with their peaks along a lattice axis)
of DNLS in 2D. {\it Left column:} interpeak separation $S=1$ lattice site,
$\chi=-10.5$. {\it Middle column:} interpeak separation $S=2$ sites,
$\chi=-15$. {\it Right column:} interpeak separation $S=3$ sites, $\chi=-14$.
Cross-sections of the wavefunctions are shown in the third row with points:
$\psi_{n_x=n_1,n_y}$ (filled circles), $\psi_{n_x=n_1+1,n_y}$ (filled
squares), $\psi_{n_x=n_1+2,n_y}$ (filled diamonds), $\psi_{n_x=n_1+3,n_y}$
(filled triangles), where $n_1=20$ is the x$-$coordinate of the two peaks.
Dashed lines show analytical approximations of the solutions using
Eq.~(\ref{dp2d}). }
\label{fDS2}
\end{figure}

\begin{figure}
\begin{center}
\begin{tabular}{ccc}
\epsfig{file=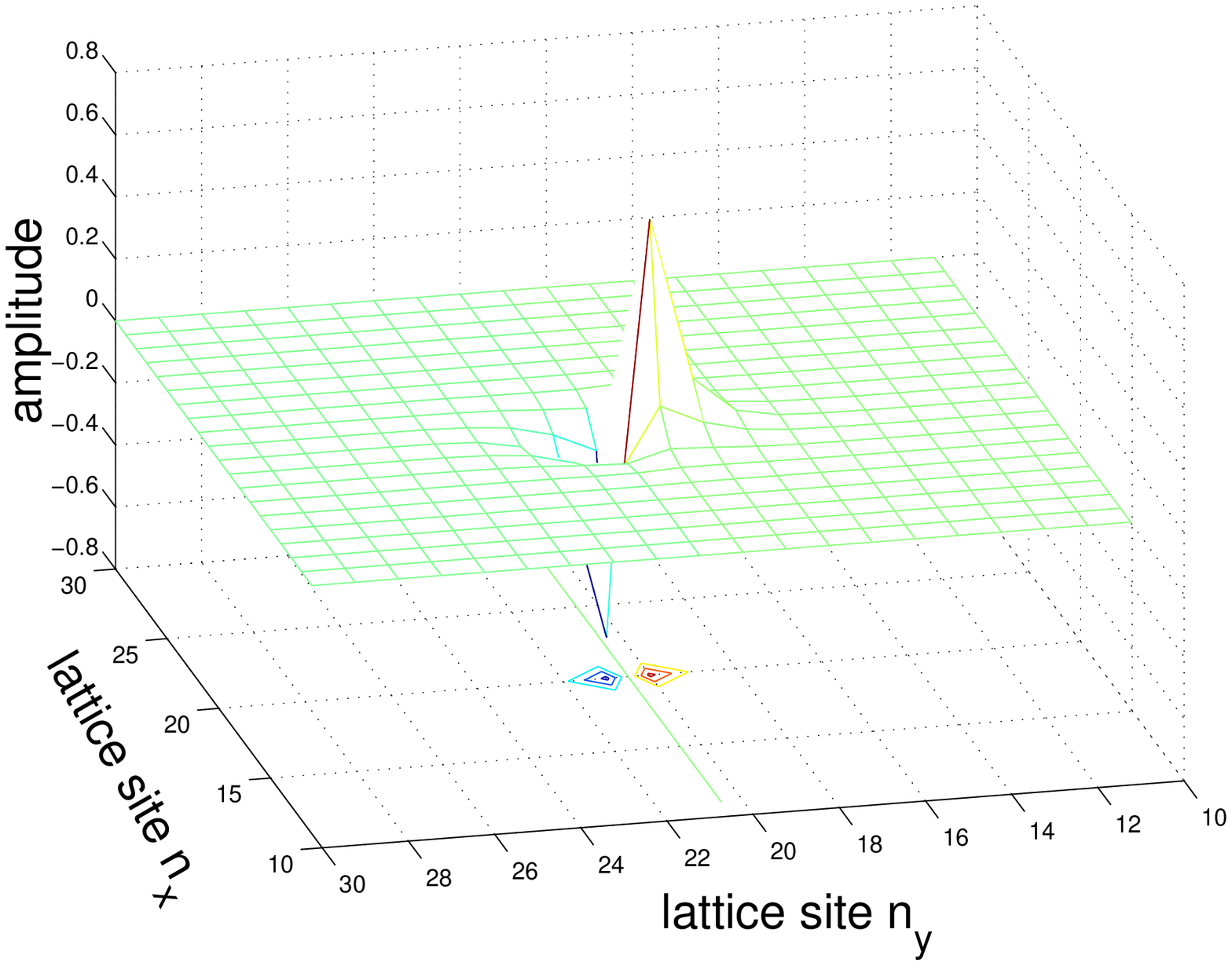,height=4.4cm,width=4.4cm} & \hspace{-0.2cm} 
\epsfig{file=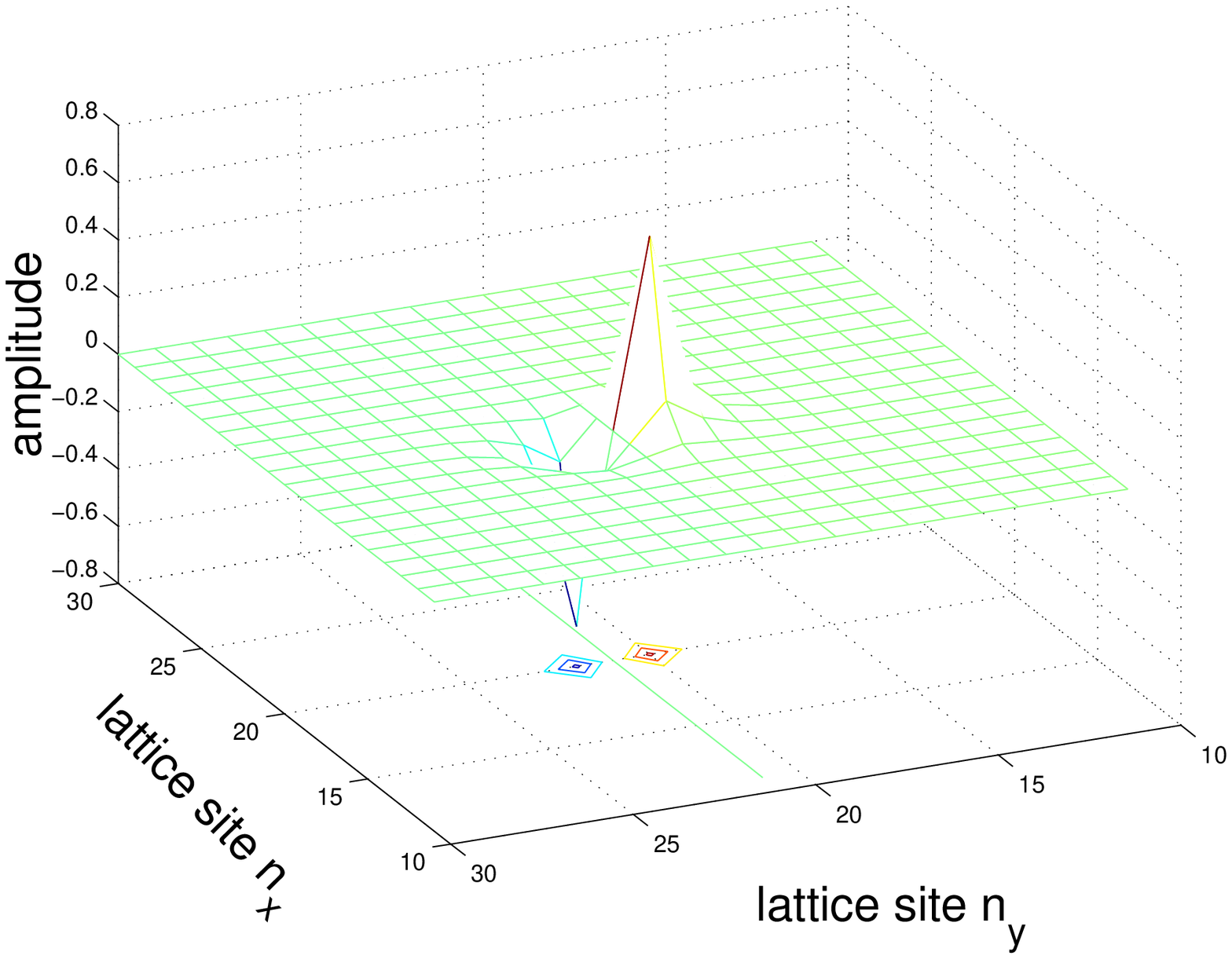,height=4.4cm,width=4.4cm} & \hspace{-0.2cm}
\epsfig{file=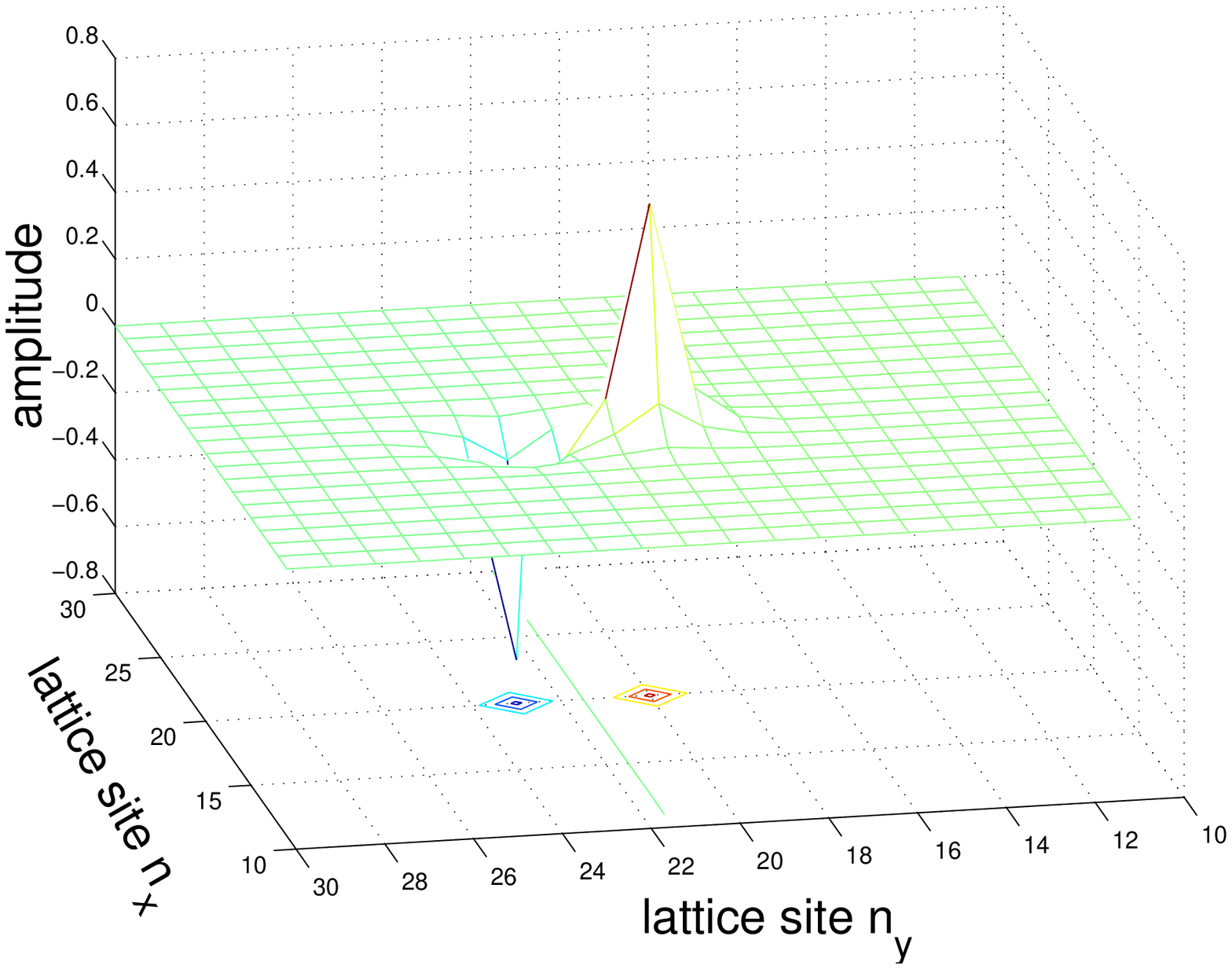,height=4.4cm,width=4.4cm}  \\
\epsfig{file=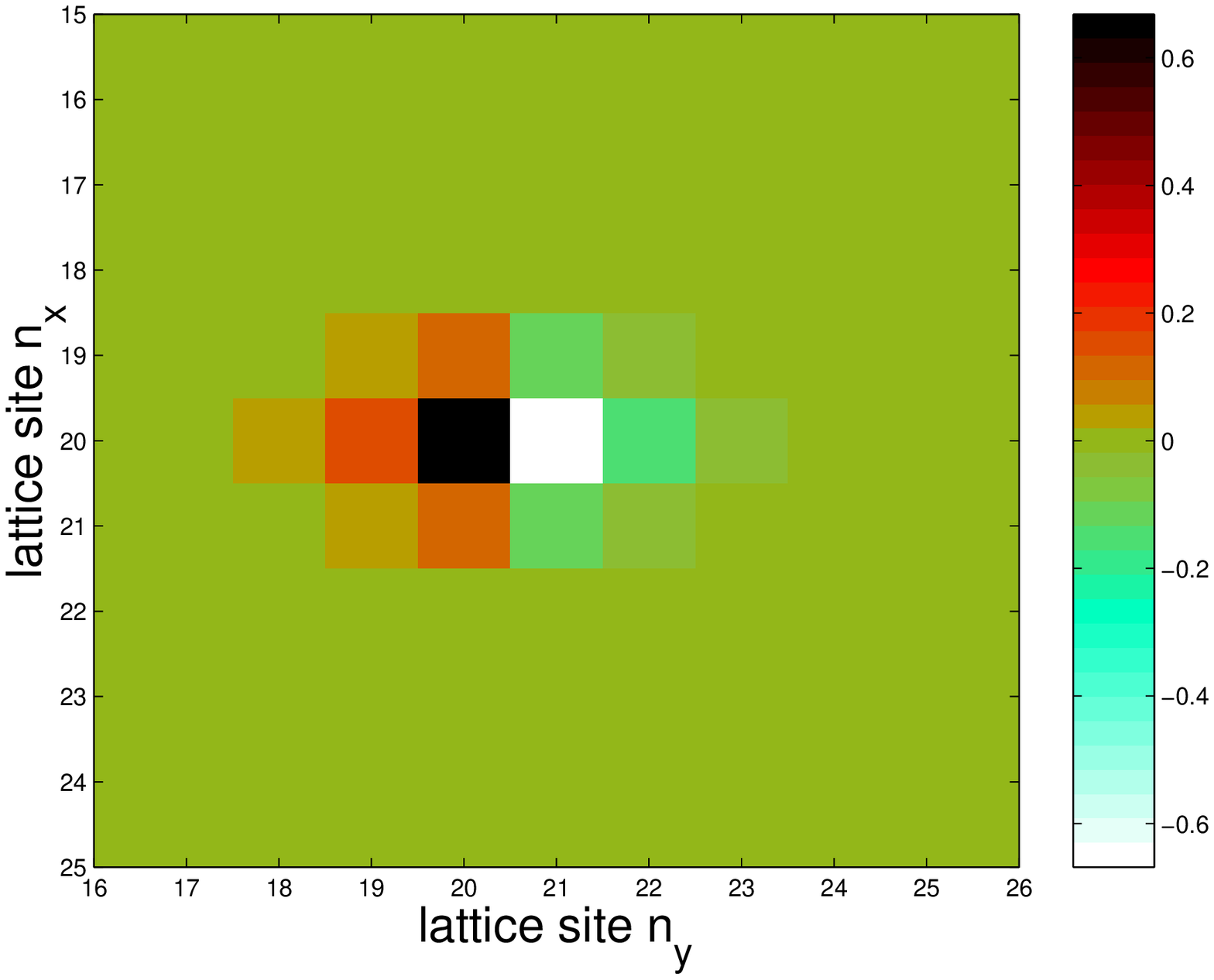,height=4.4cm,width=4.4cm} & \hspace{-0.2cm} 
\epsfig{file=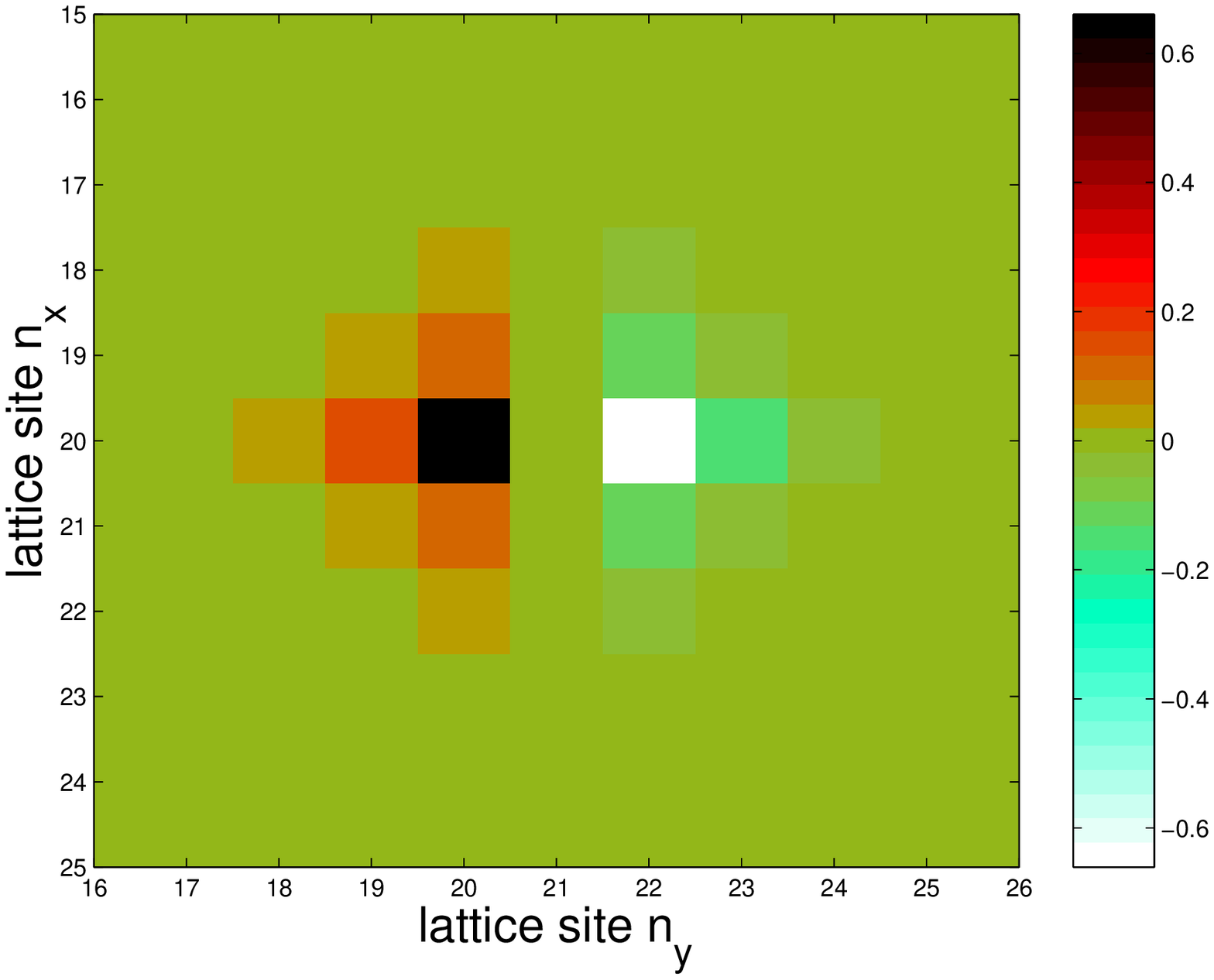,height=4.4cm,width=4.4cm} & \hspace{-0.2cm}
\epsfig{file=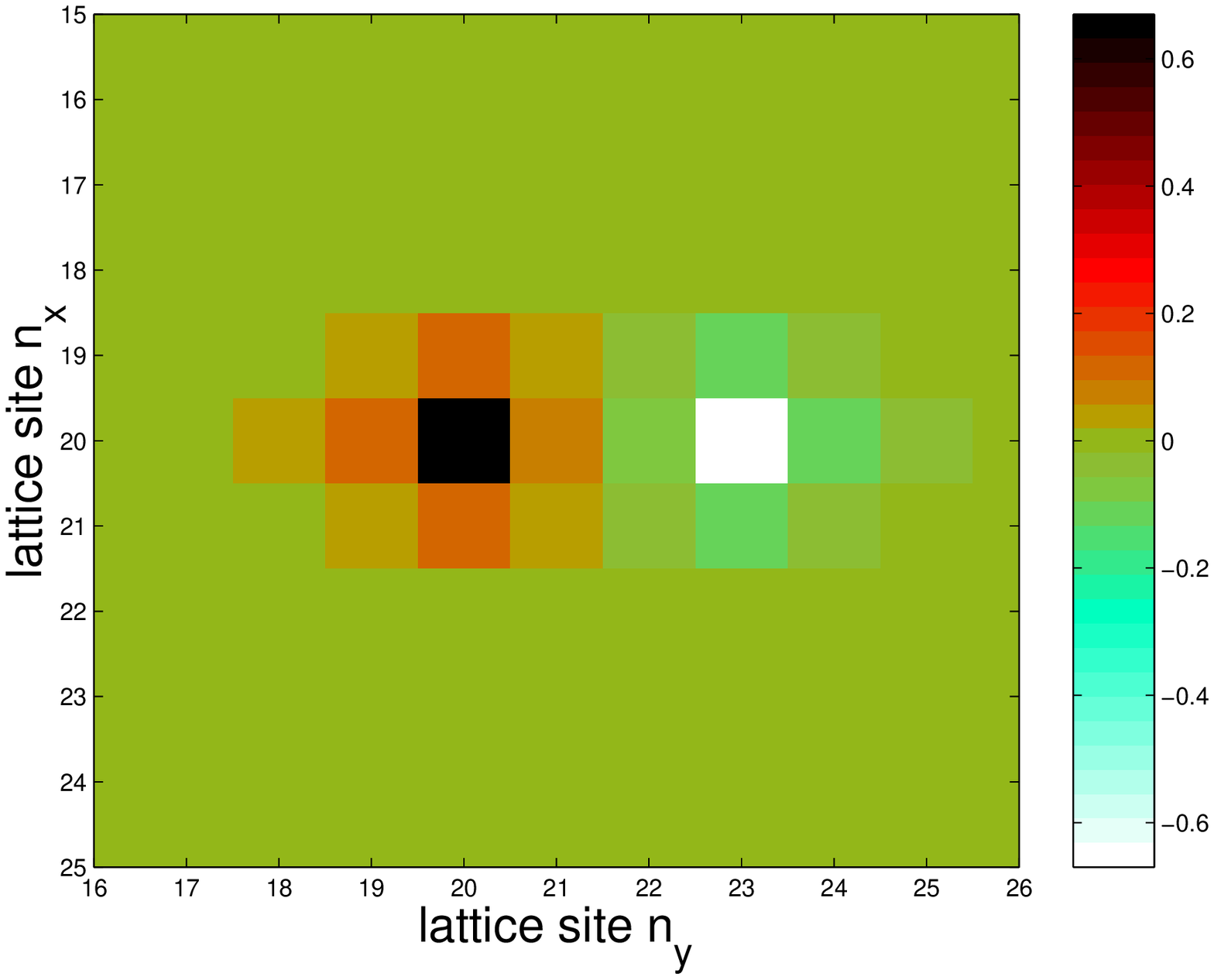,height=4.4cm,width=4.4cm}  \\
\epsfig{file=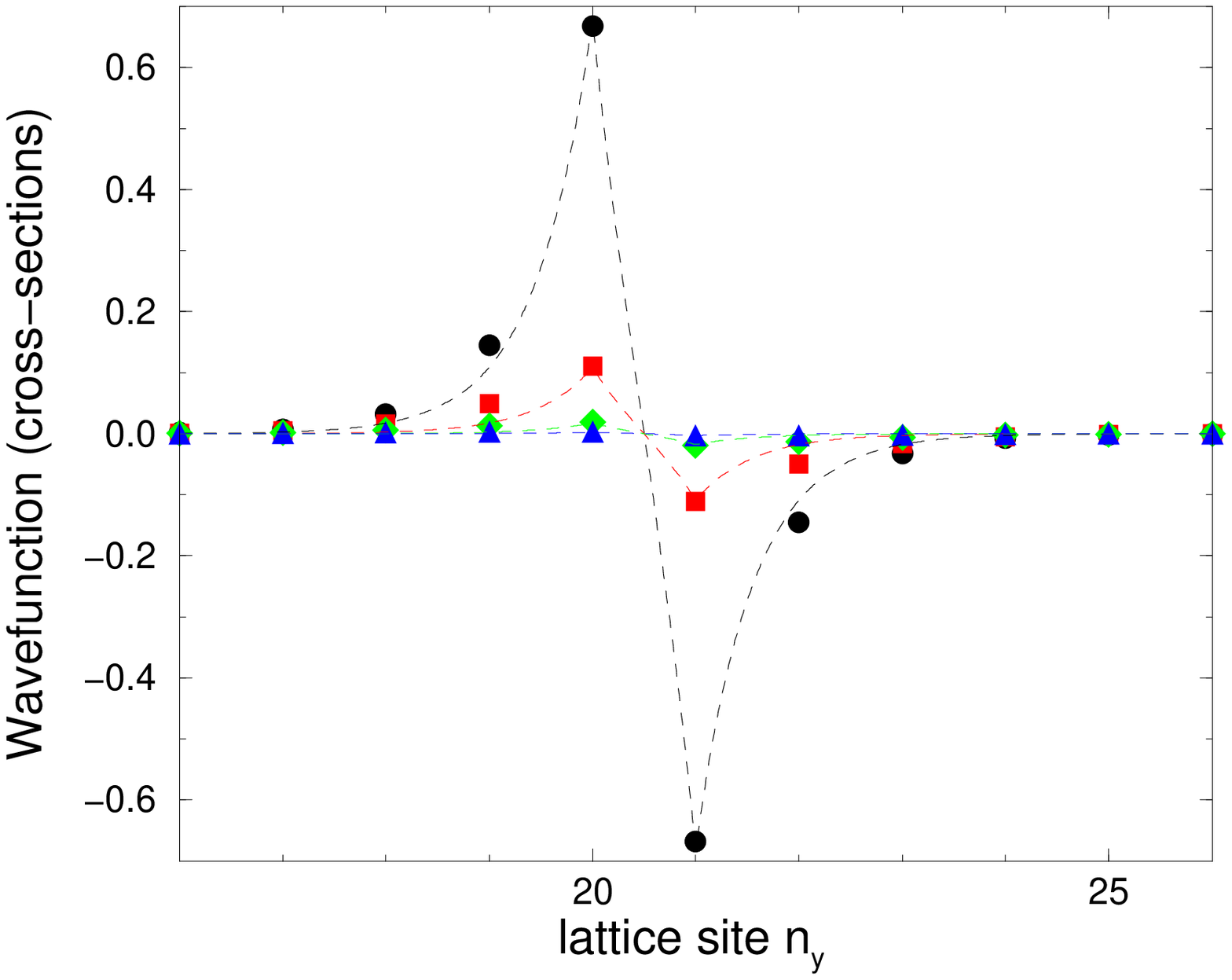,height=4.4cm,width=4.4cm} & \hspace{-0.2cm} 
\epsfig{file=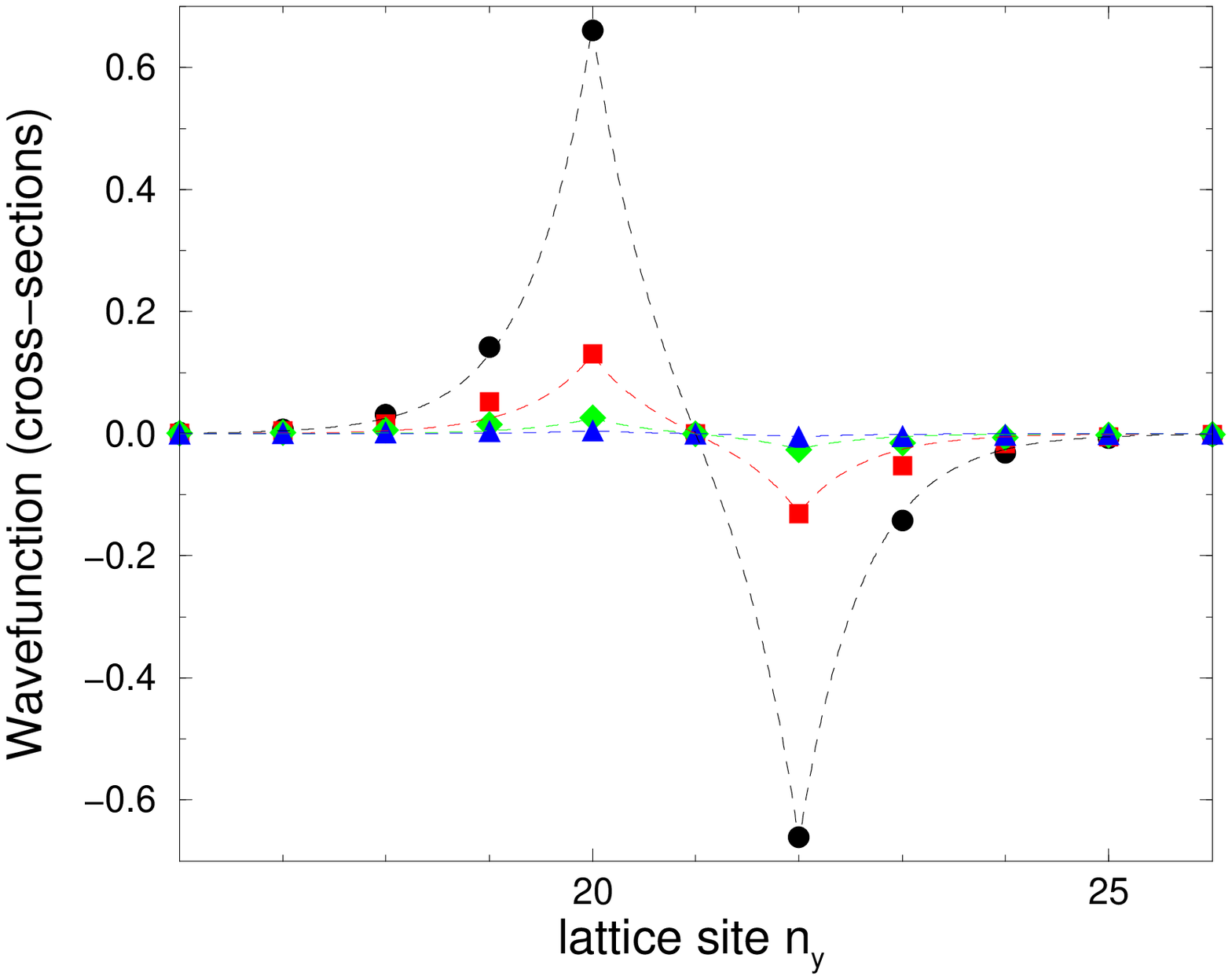,height=4.4cm,width=4.4cm} & \hspace{-0.2cm}
\epsfig{file=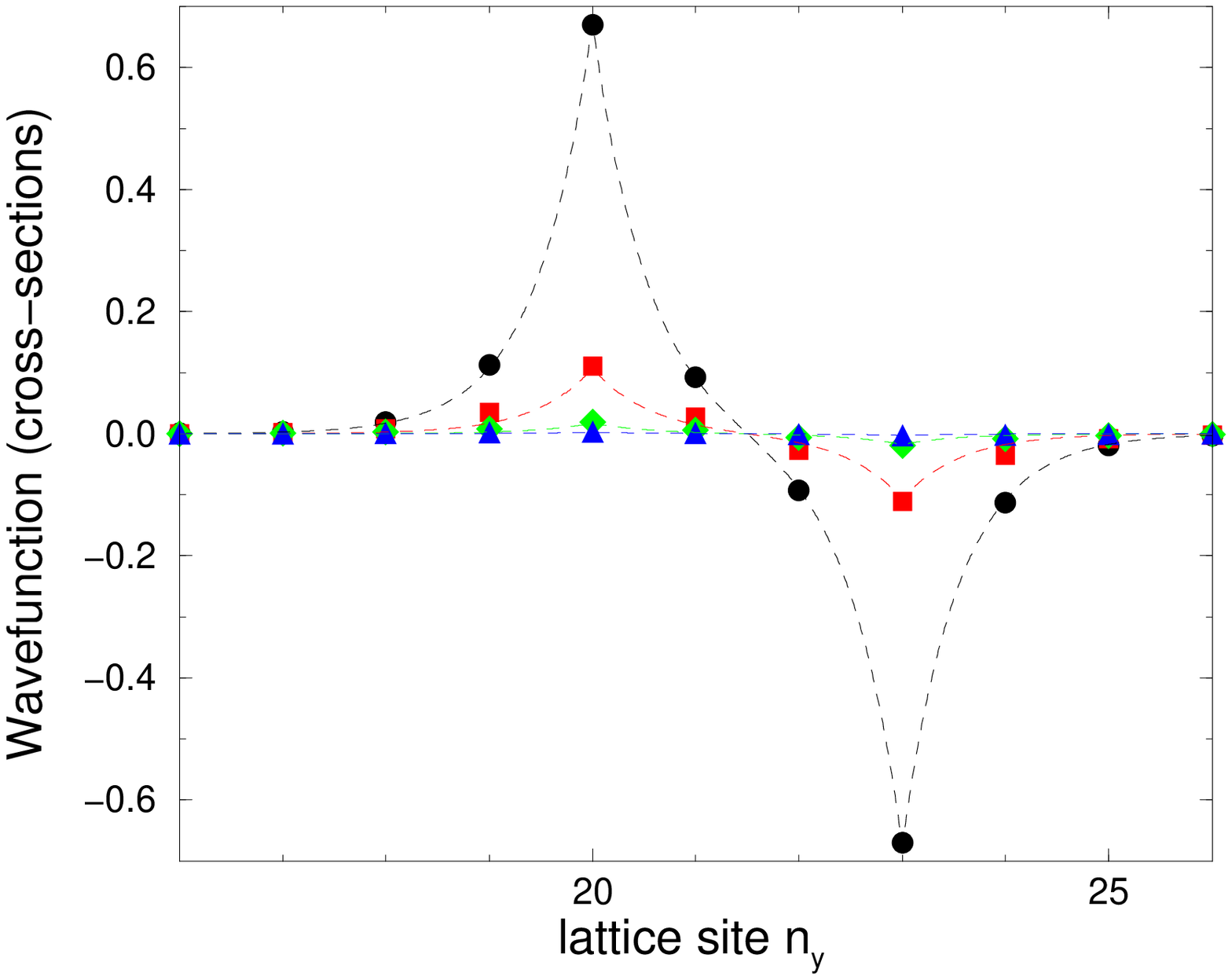,height=4.4cm,width=4.4cm}  \\
\end{tabular}
\end{center}
\caption{3D plots (first row) and density plots (second row) of
double-peaked antisymmetric solutions (with their peaks along a lattice axis)
of DNLS in 2D. {\it Left column:} interpeak separation $S=1$ lattice site,
$\chi=-14$. {\it Middle column:} interpeak separation $S=2$ sites,
$\chi=-12$. {\it Right column:} interpeak separation $S=3$ sites, $\chi=-14$.
Cross-sections of the wavefunctions are shown in the third row with points:
$\psi_{n_x=n_1,n_y}$ (filled circles), $\psi_{n_x=n_1+1,n_y}$ (filled
squares), $\psi_{n_x=n_1+2,n_y}$ (filled diamonds), $\psi_{n_x=n_1+3,n_y}$
(filled triangles), where $n_1=20$ is the x$-$coordinate of the two peaks.
Dashed lines show analytical approximations of the solutions using
Eq.~(\ref{dp2d}). }
\label{fDA2}
\end{figure}

DP solutions with the two peaks along a lattice axis
(let say the y-axis) are presented first. The corresponding branches
are calculated starting from the initial state
\be  \label{2d2as}
\psi_{n_x,n_y}^{(r=0)}= \frac{1}{\sqrt{2}}(\delta_{n_x,n_1} \delta_{n_y,n_2}
+ \delta_{n_x,n_1} \delta_{n_y,n_2+S})
\ee
for the symmetric and
\be
\psi_{n_x,n_y}^{(r=0)}= \frac{1}{\sqrt{2}}(\delta_{n_x,n_1} \delta_{n_y,n_2}
- \delta_{n_x,n_1} \delta_{n_y,n_2+S})
\ee
for the antisymmetric stationary states, respectively, where
$S= 1, 2, \ldots$ determines the interpeak separation along the
lattice axis. Figs. \ref{fDS2} and \ref{fDA2} show examples
for $S=1$ (left columns), $S=2$ (middle columns), and $S=3$
(right columns). Fig.~\ref{fFS2} contains branches of such solutions for
$S=1$, 2, 3, 4, 5, and 10. For $S=1$ the upper and lower single branches
of the DP states of Fig.~\ref{fFS2} are obtained.

\begin{figure}
\begin{center}
\begin{tabular}{ccc}
\epsfig{file=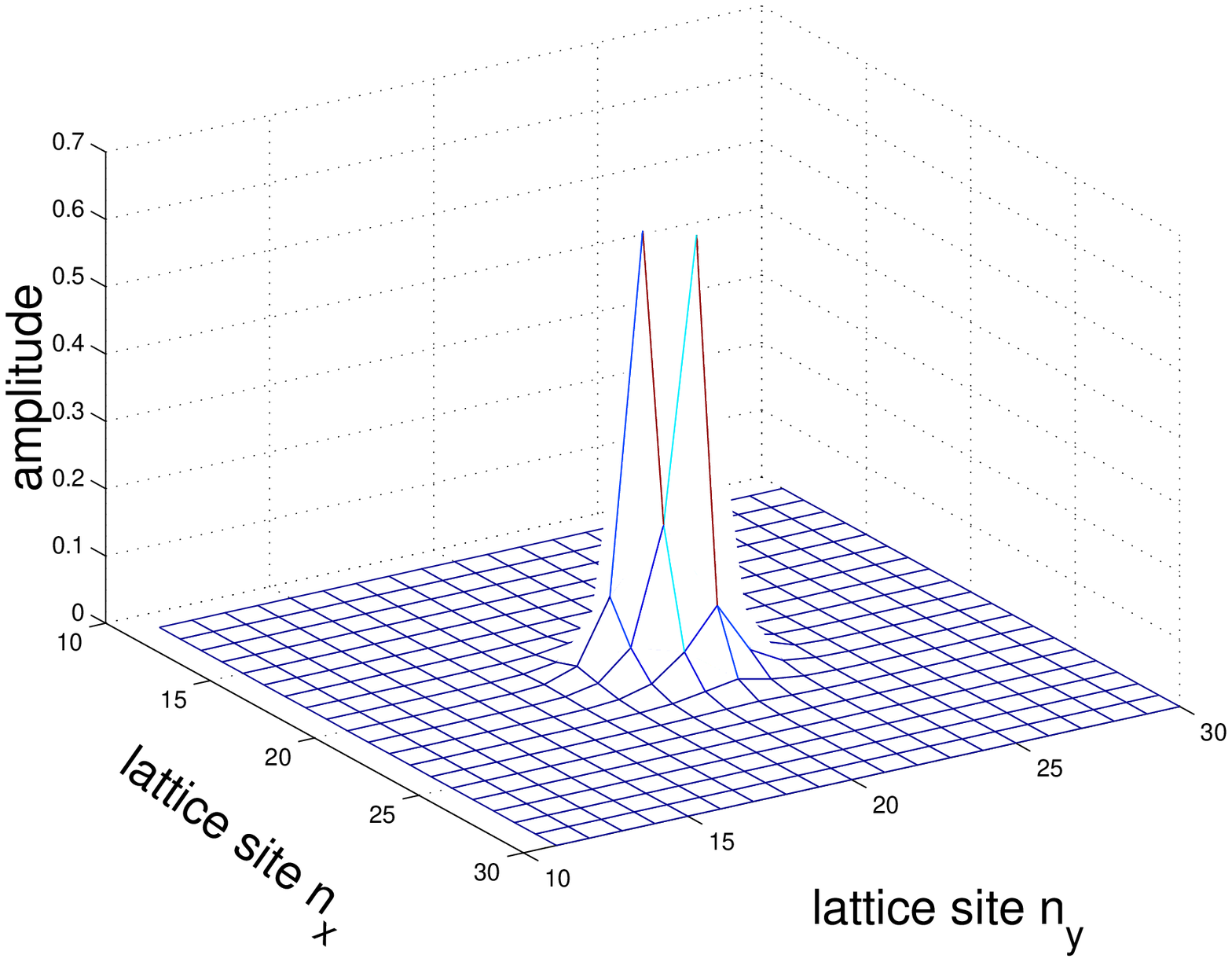,height=4.4cm,width=4.4cm} & \hspace{-0.2cm} 
\epsfig{file=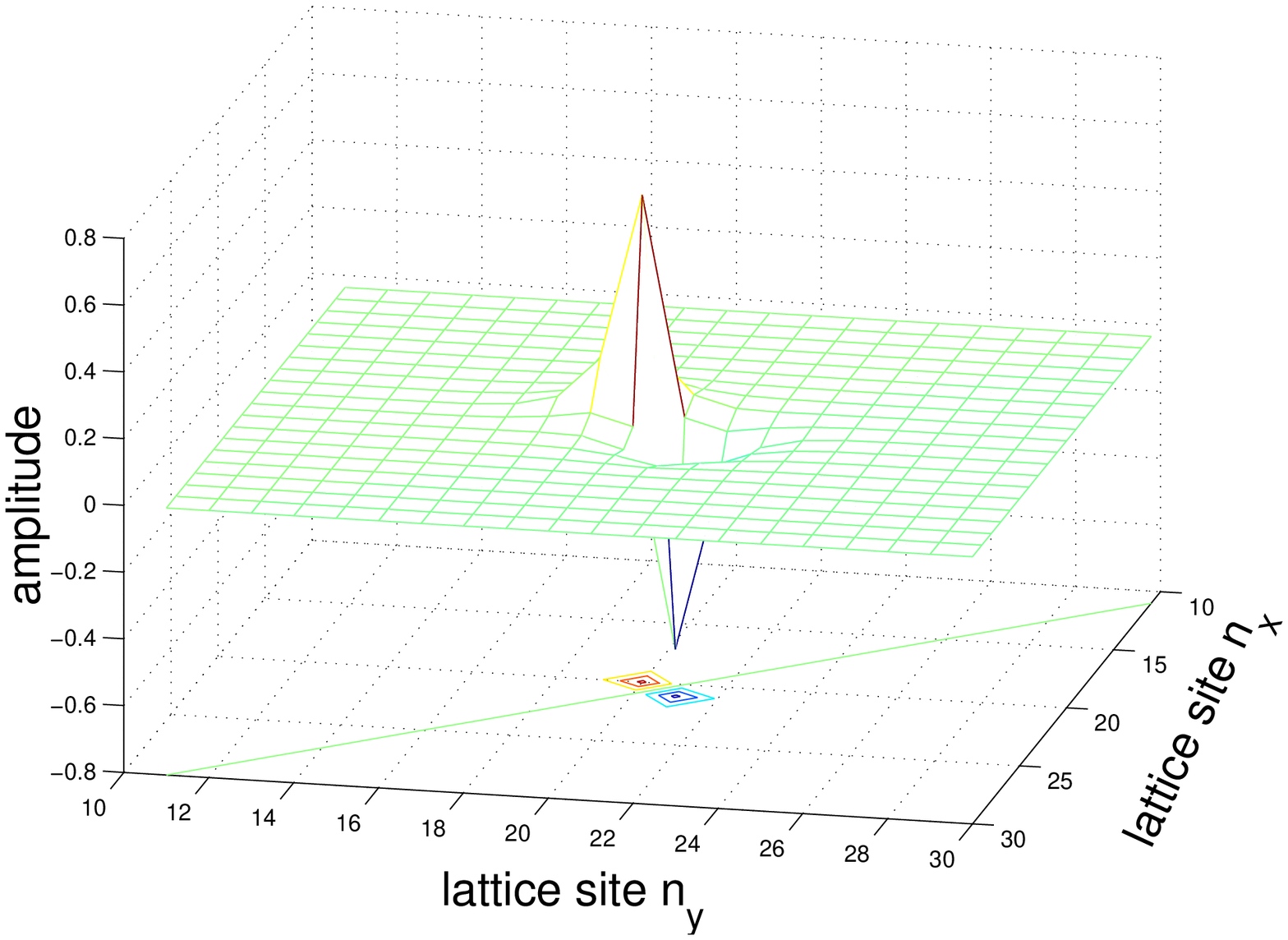,height=4.4cm,width=4.4cm} & \hspace{-0.2cm}
\epsfig{file=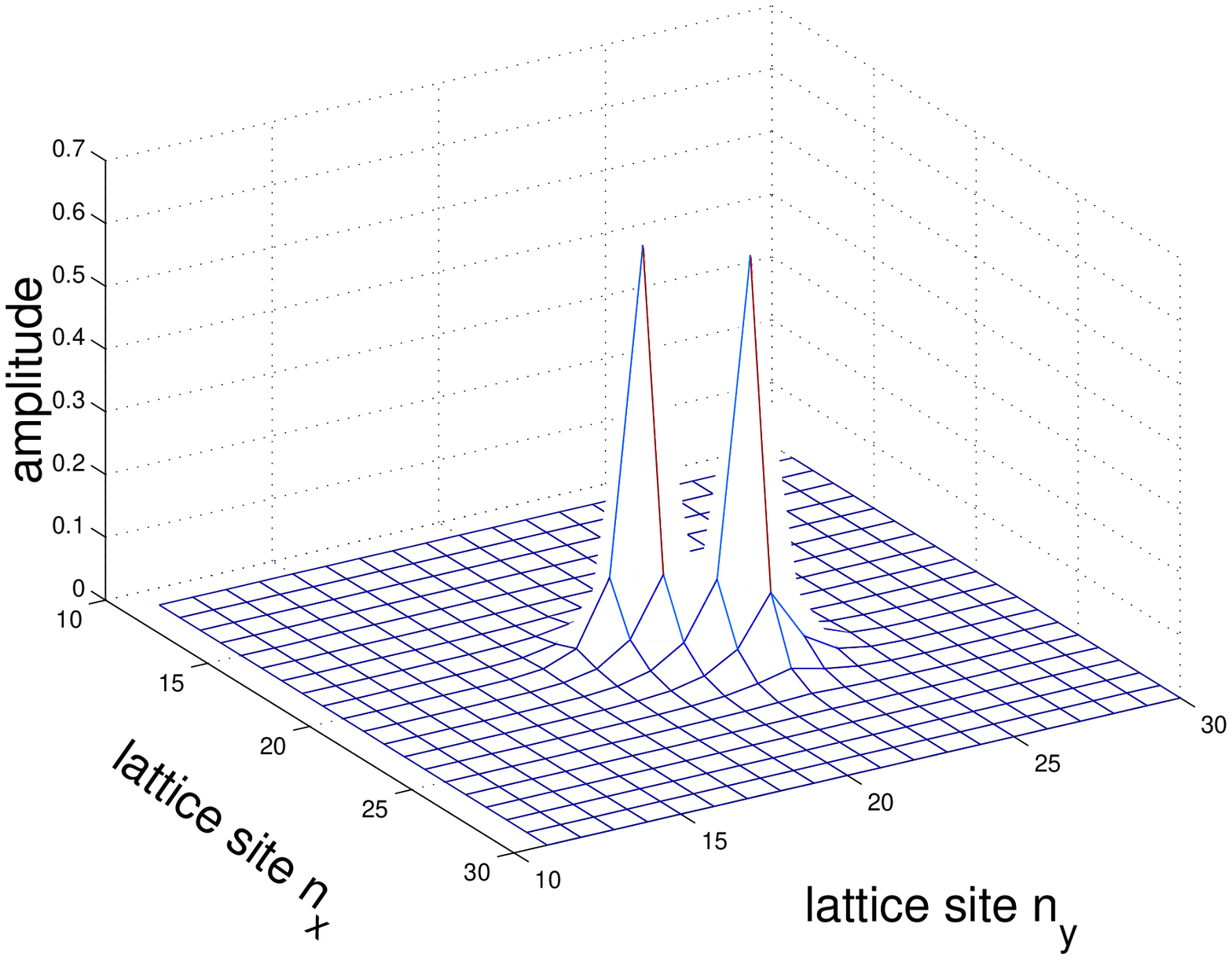,height=4.4cm,width=4.4cm}  \\
\epsfig{file=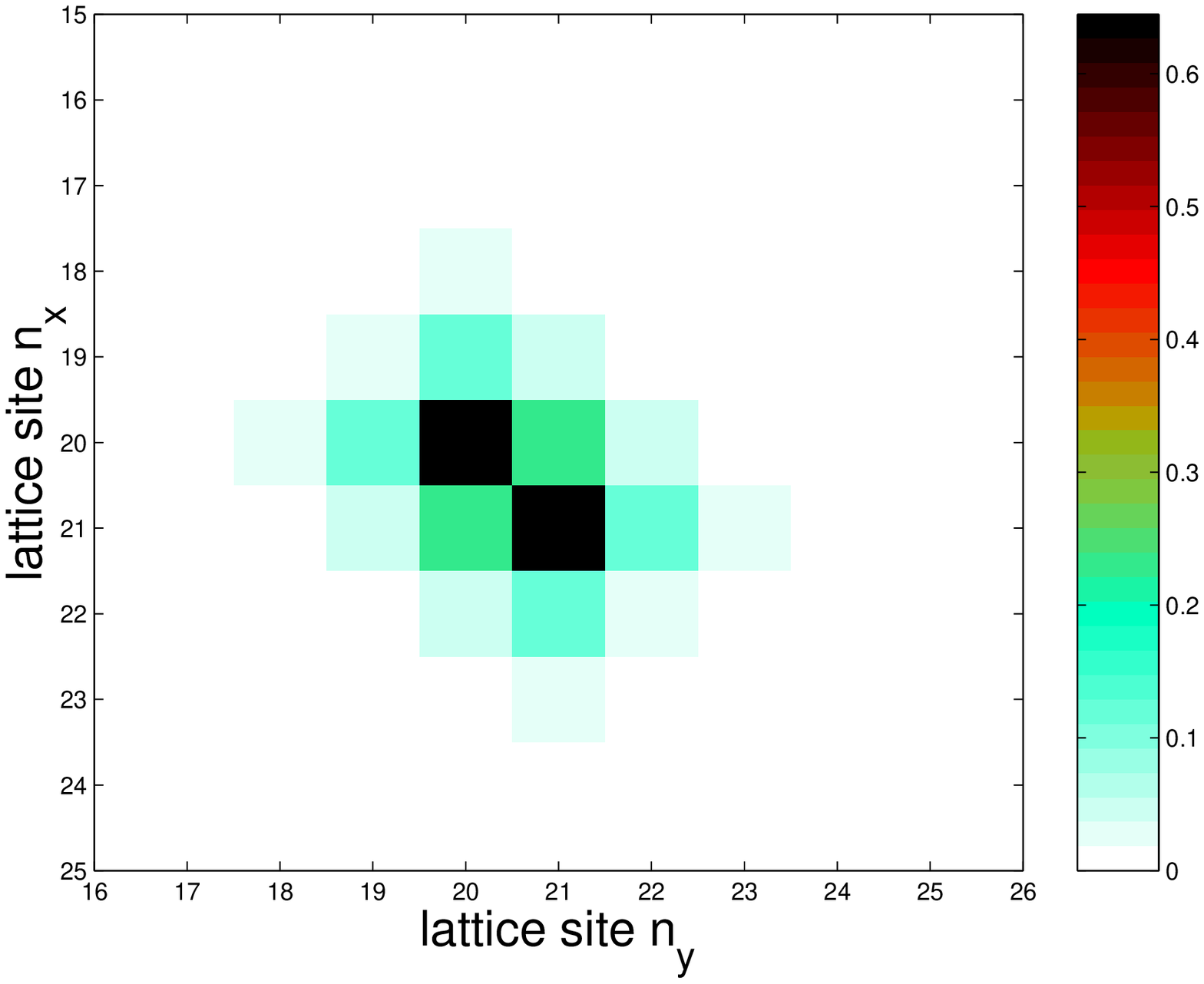,height=4.4cm,width=4.4cm} & \hspace{-0.2cm} 
\epsfig{file=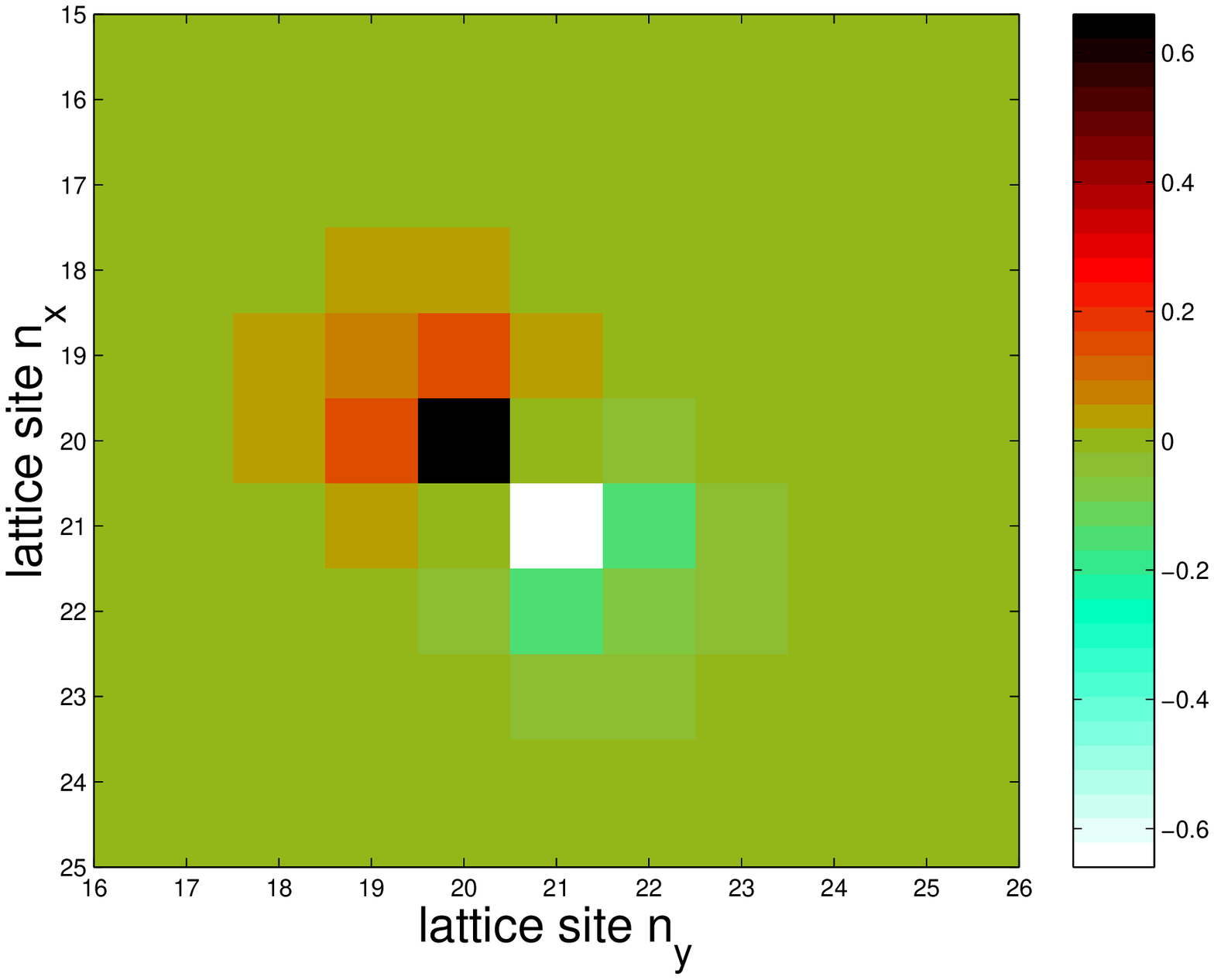,height=4.4cm,width=4.4cm} & \hspace{-0.2cm}
\epsfig{file=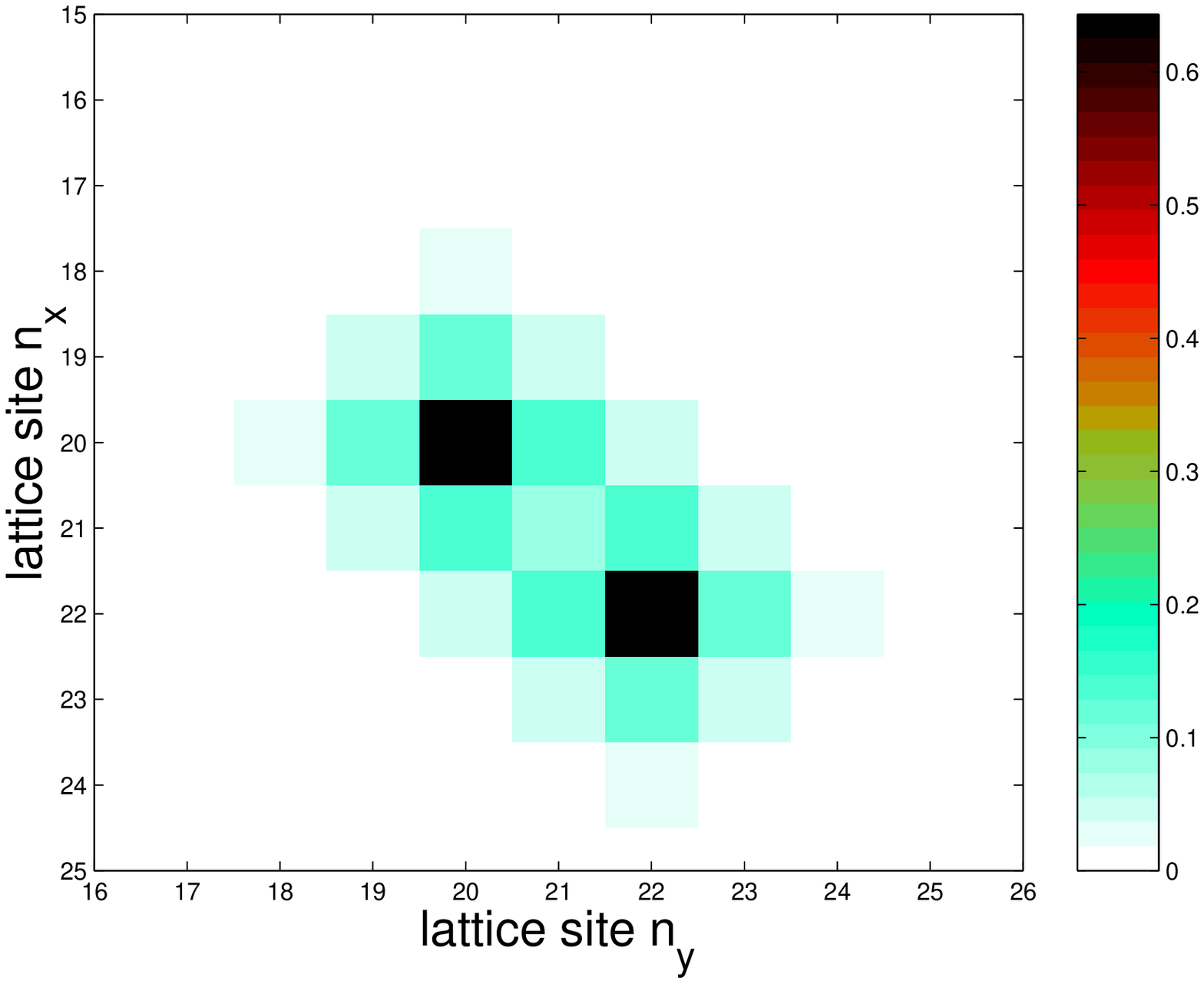,height=4.4cm,width=4.4cm}  \\
\epsfig{file=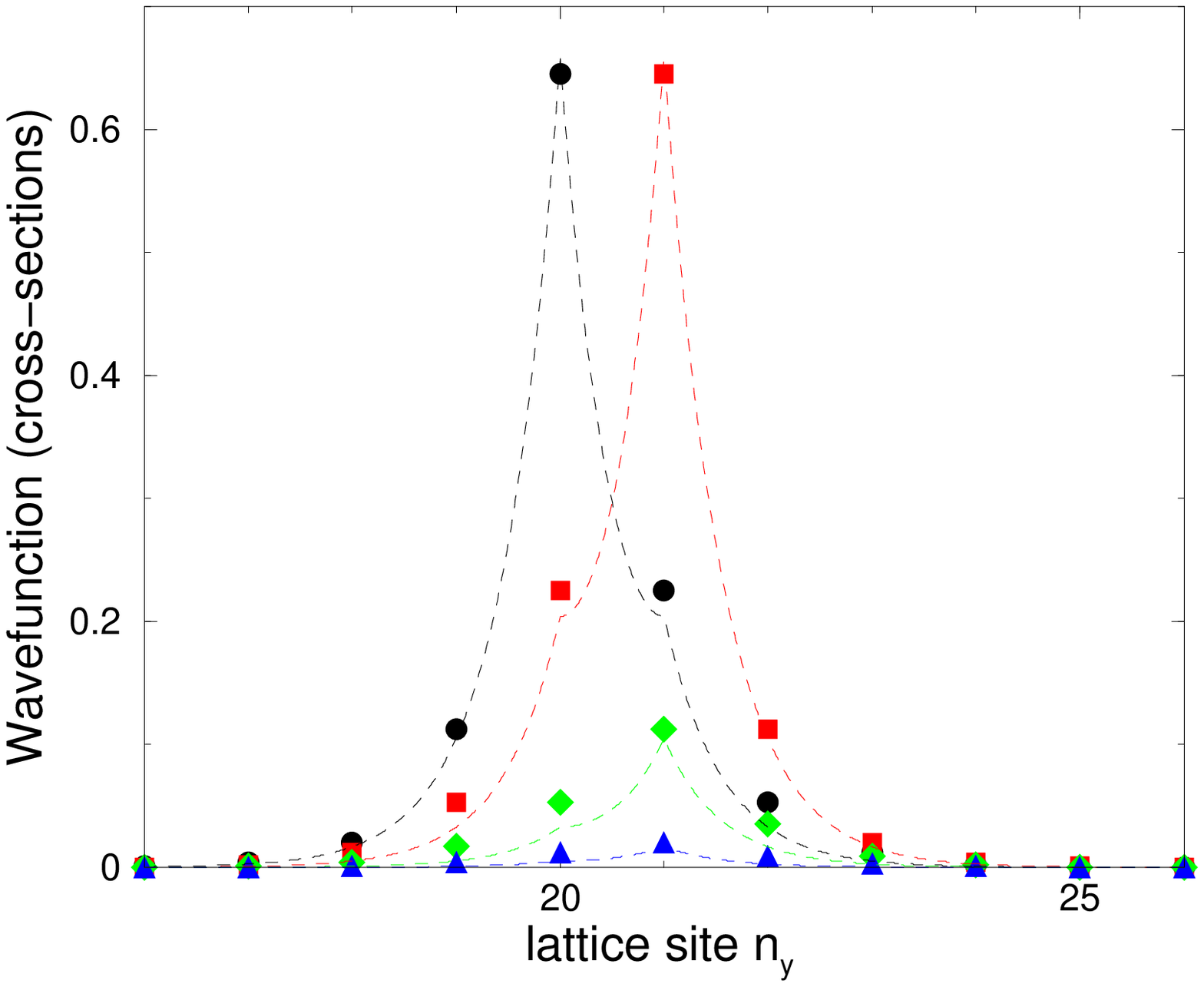,height=4.4cm,width=4.4cm} & \hspace{-0.2cm} 
\epsfig{file=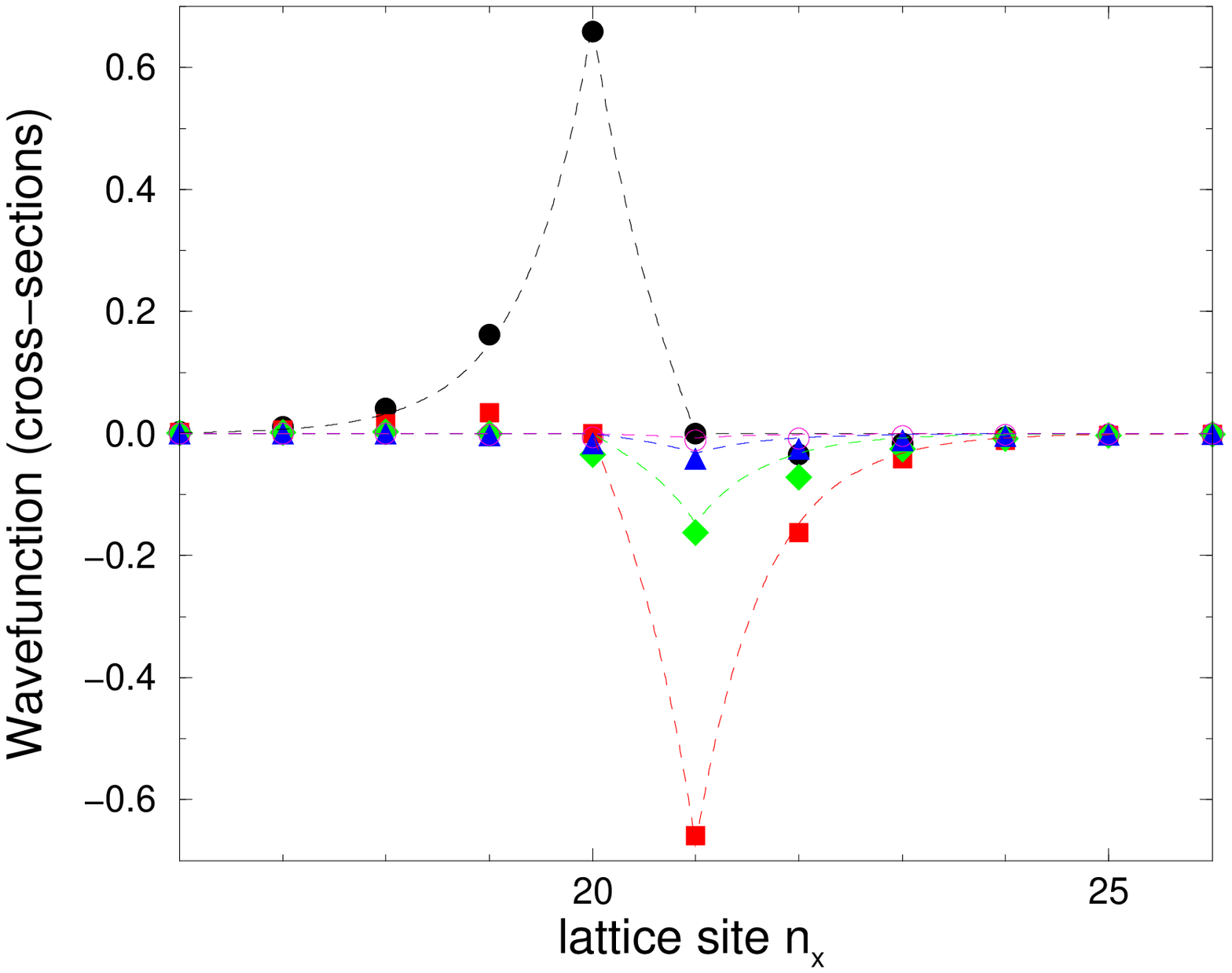,height=4.4cm,width=4.4cm} & \hspace{-0.2cm}
\epsfig{file=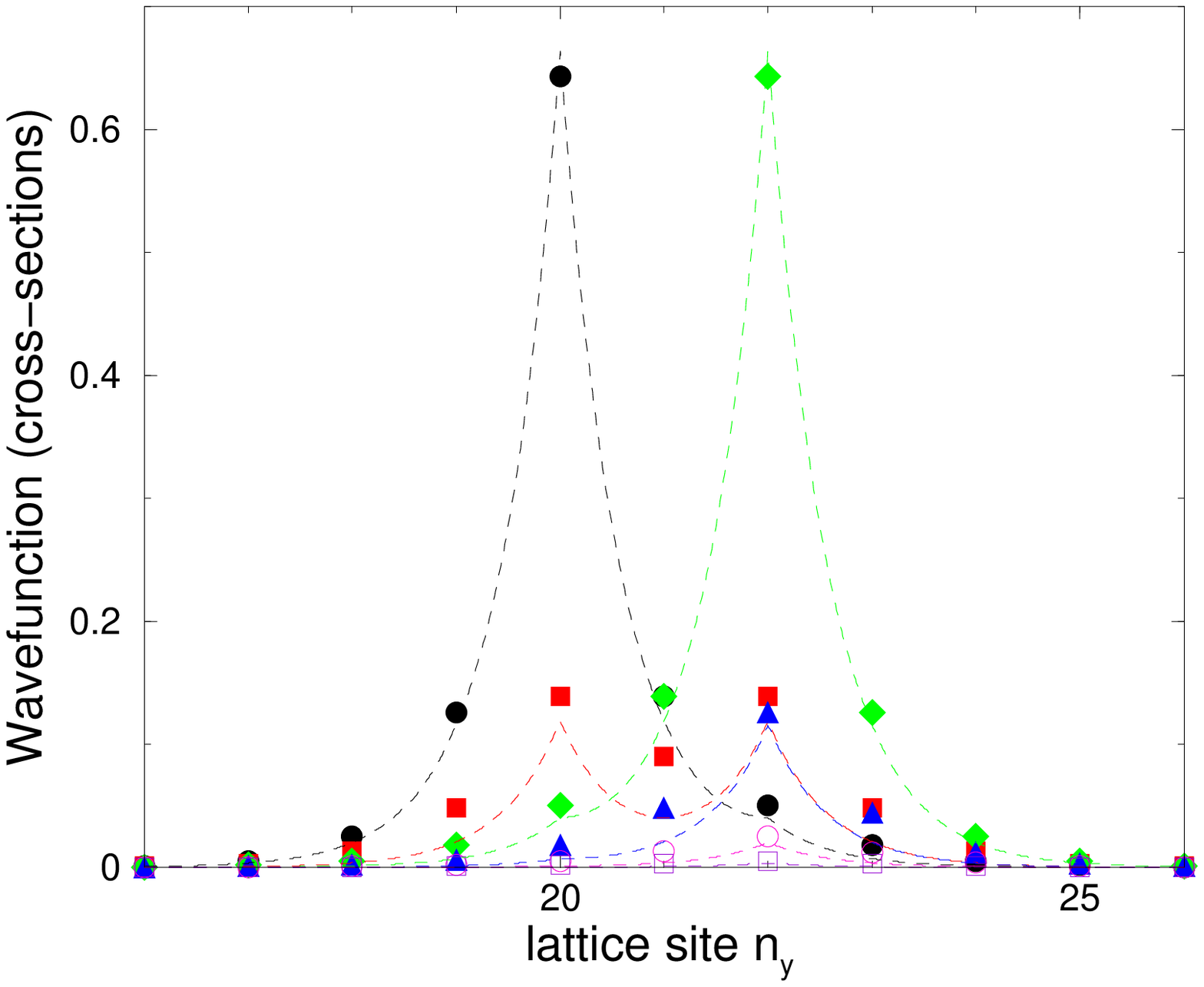,height=4.4cm,width=4.4cm}  \\
\end{tabular}
\end{center}
\caption{3D plots (first row) and density plots (second row) of
double-peaked symmetric and antisymmetric solutions (with their peaks along
the diagonal of the lattice axes) of DNLS in 2D.
{\it Left column:} symmetric state with $l=1$ and $\chi=-14.1$.
{\it Middle column:} antisymmetric state with $l=1$ and $\chi=-11$.
{\it Right column:} symmetric state with $l=2$ and $\chi=-13.1$.
Cross-sections of the wavefunctions are shown in the third row with points:
$\psi_{n_x=n_1,n_y}$ (filled circles), $\psi_{n_x=n_1+1,n_y}$ (filled squares),
$\psi_{n_x=n_1+2,n_y}$ (filled diamonds), $\psi_{n_x=n_1+3,n_y}$ (filled
triangles), $\psi_{n_x=n_1+4,n_y}$ (open circles, in the second and third
column), $\psi_{n_x=n_1+5,n_y}$ (open squares, in the third column), where
$n_1=20$ is the x$-$coordinate of the first peak. Dashed lines show
analytical approximations of the solutions using Eq.~(\ref{dp2d}). }
\label{fDd2}
\end{figure}

Branches of DP solutions with their two peaks along the
diagonal of the lattice axes have been also shown in Fig.~\ref{fFS2}.
These branches are obtained starting from the initial state
\be  \label{2d2dsa}
\psi_{n_x,n_y}^{(r=0)}= \frac{1}{\sqrt{2}}(\delta_{n_x,n_1} \delta_{n_y,n_2}
\pm \delta_{n_x,n_1+l} \delta_{n_y,n_2+l}),
\ee
where the plus sign gives the symmetric and the minus the antisymmetric,
respectively, stationary states. Fig.~\ref{fFS2} contains branches of these
solutions for $l=1$,
2, 3, and 10 and their frequencies are around $\omega = \frac{\chi}{2}$.
Some examples of such states are shown in Fig.~\ref{fDd2}.

Finally, many other DP solutions exist, with their peaks, of the
same or opposite sign, at random lattice sites (not aligned along a
lattice axis, or the diagonal). These can be obtained starting from
the initial state
\be
\psi_{n_x,n_y}^{(r=0)}= \frac{1}{\sqrt{2}}(\delta_{n_x,n_1} \delta_{n_y,n_2}
\pm \delta_{n_x,n_1+S_x} \delta_{n_y,n_2+S_y}),
\ee
with any combination of non-zero integers $S_x$, $S_y$ and $S_x \neq S_y$.
The corresponding branches are close to the middle branch of the 
DP solutions of Fig.~\ref{fFS2}.

As in the 1D case, one can obtain approximate analytical expressions for
the DP solutions of DNLS in 2D, by appropriate superpositions
of the single-peaked solutions (\ref{vap}). If $(n_1,n_2)$ is the lattice
site of the first peak and $(n_1+S_x,n_2+S_y)$ of the second one ($S_x$ and
$S_y$ are assumed to be positive), then
the corresponding approximate solution is
\be  \label{dp2d}
\psi_{n_x,n_y}^{DP} = \frac{1}{\sqrt{2(1\pm P)}} \frac{1-\zeta^2}{1+\zeta^2}
\left( \zeta^{|n_x-n_1|+|n_y-n_2|} \pm (-sgn\chi)^{S_x+S_y}
\zeta^{|n_x-n_1-S_x|+|n_y-n_2-S_y|} \right),
\ee
\be  \label{ov2}
\mbox{where} \hspace{0.3cm} P=(-sgn \chi)^{S_x+S_y} \; \frac{[(1+S_x)-(S_x-1) 
\zeta^2] [(1+S_y) - (S_y-1) \zeta^2] \zeta^{S_x+S_y}} {(1+\zeta^2)^2}
\ee
\be \label{dp2dz}
\mbox{and} \hspace{0.5cm} \zeta= -\frac{1}{\chi/2} - \frac{6}{(\chi/2)^3}
= -\frac{2}{\chi} - \frac{48}{\chi^3}.
\ee
Here also, the individual wavefunctions superimposed in this solution
should correspond to $\frac{\chi}{2}$, which has been taken into account
in the relation (\ref{dp2dz}) providing $\zeta$. The plus and minus signs
in Eq.~(\ref{dp2d}) correspond to symmetric and antisymmetric states,
respectively, apart from the case of positive $\chi$ and odd $S$ where it
is the other way around. The overlap $P$ of the two superimposed single-peaked
solutions is $P=\sum_{n_x,n_y} \psi_{n_x,n_y}^{SP[n_1,n_2]}(\frac{\chi}{2})
\psi_{n_x,n_y}^{SP[n_1+S_x,n_2+S_y]}(\frac{\chi}{2})$, where
$\psi_{n_x,n_y}^{SP[n_1,n_2]}(\frac{\chi}{2})$ is the SP solution
in 2D, centered at $(n_1,n_2)$  and corresponding to $\frac{\chi}{2}$.
Cross-sections of the approximate solution (\ref{dp2d}) are shown with
dashed lines in the third rows of Figs.~\ref{fDS2}-\ref{fDd2}.

\begin{figure}
\centerline{\hbox{\psfig{figure=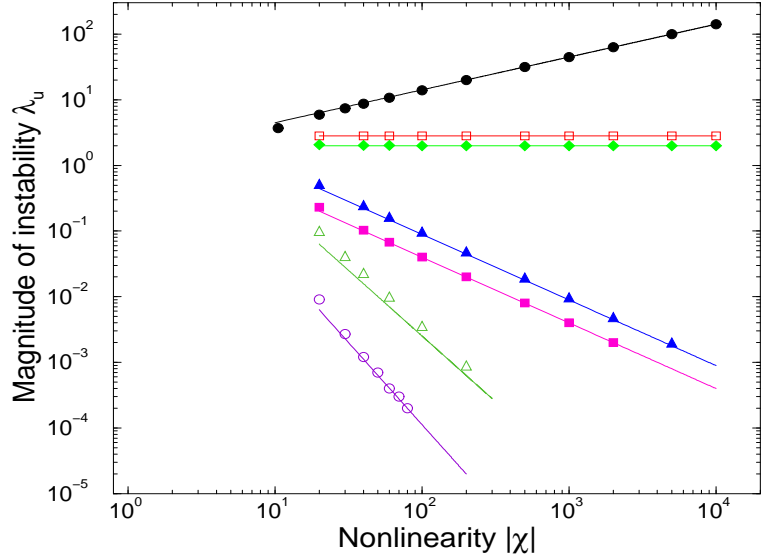,width=10cm,height=7.5cm}}}
\caption{Magnitude of instability $\lambda_u$ of symmetric double-peaked
solutions of DNLS for $\chi<0$ in 2D (points) as a function of the strength
of nonlinearity $|\chi|$, for various interpeak separations. Results
for solutions along a
lattice axis with  $S_x=1$, $S_y=0$ (filled circles), $S_x=2$, $S_y=0$ (filled
diamonds), $S_x=4$, $S_y=0$ (filled squares), $S_x=7$, $S_y=0$ (open circles),
and along the diagonal with $S_x=S_y=1$ (open squares), $S_x=S_y=2$ (filled
triangles), and $S_x=S_y=3$ (open triangles), are presented.
Lines show the power-law relation, Eq.(\ref{inst2}), derived for
large values of $|\chi|$.}
\label{fstab2}
\end{figure}

Concerning the stability of the DP stationary states, the picture
is similar like in 1D. For $\chi<0$, or $\chi>0$ and even $S$, all the
symmetric solutions are unstable, while the antisymmetric ones are in general
(at least for relatively large values of $|\chi|$) linearly stable. For
$\chi>0$ and odd $S$ the reverse is true. General arguments are
presented in the Appendix showing that close to the anti-continuous limit
any symmetric (antisymmetric) DP solution should be unstable (linearly
stable), except when $\chi>0$ and $S$ is odd, where it is linearly stable
(unstable). Further, this calculation allows to determine the
variation of the magnitude of instability $\lambda_u$ of the unstable
solution with the nonlinearity strength (for relatively large
values of $|\chi|$), depending on the interpeak distance. As it is shown in
the Appendix, if the separation of the two peaks in the 2D lattice is given
by $S_x$ and $S_y$, then
\be   \label{inst2}
\lambda_u = \sqrt{1+S_x S_y} \; \; 2^{(S_x+S_y)/2} \;
|\chi|^{1-\frac{S_x+S_y}{2}}, \hspace{1.0cm} \mbox{for} \; \; |\chi| \gg 1.
\ee
Fig.~\ref{fstab2} presents numerical results in the case of $\chi<0$
regarding the magnitude of instability
of symmetric DP states with different interpeak separations $S_x$,
$S_y$ (points), as well as a comparison with the power-paw (\ref{inst2}).

Besides the pair of stable or unstable discrete eigenvalues, there is also
the band of eigenvalues extending
approximately from $\frac{|\chi|}{2}-4$ to $\frac{|\chi|}{2}+4$, except for
the cases where the two peaks are located in first neighboring sites. Then,
for the symmetric states the band extends from $\frac{|\chi|}{2}-3$ to
$\frac{|\chi|}{2}+5$, while for the antisymmetric states extends from
$\frac{|\chi|}{2}-5$ to $\frac{|\chi|}{2}+3$. As $|\chi|$ decreases, the
band moves towards zero and the states which are stable at large $|\chi|$
become unstable when
the band collides with the discrete eigenvalues lying on the real axis.

\subsection{ Quadruple-peaked solutions}

In this subsection QP solutions of high symmetry are presented, where
their four peaks are on lattice sites forming a square of edge equal
to $l$ lattice constants. Such solutions may be symmetric, antisymmetric
along both lattice axes, or symmetric along the one axis and antisymmetric
along the other one, and at various interpeak distances $l$.

\begin{figure}
\begin{center}
\begin{tabular}{ccc}
\epsfig{file=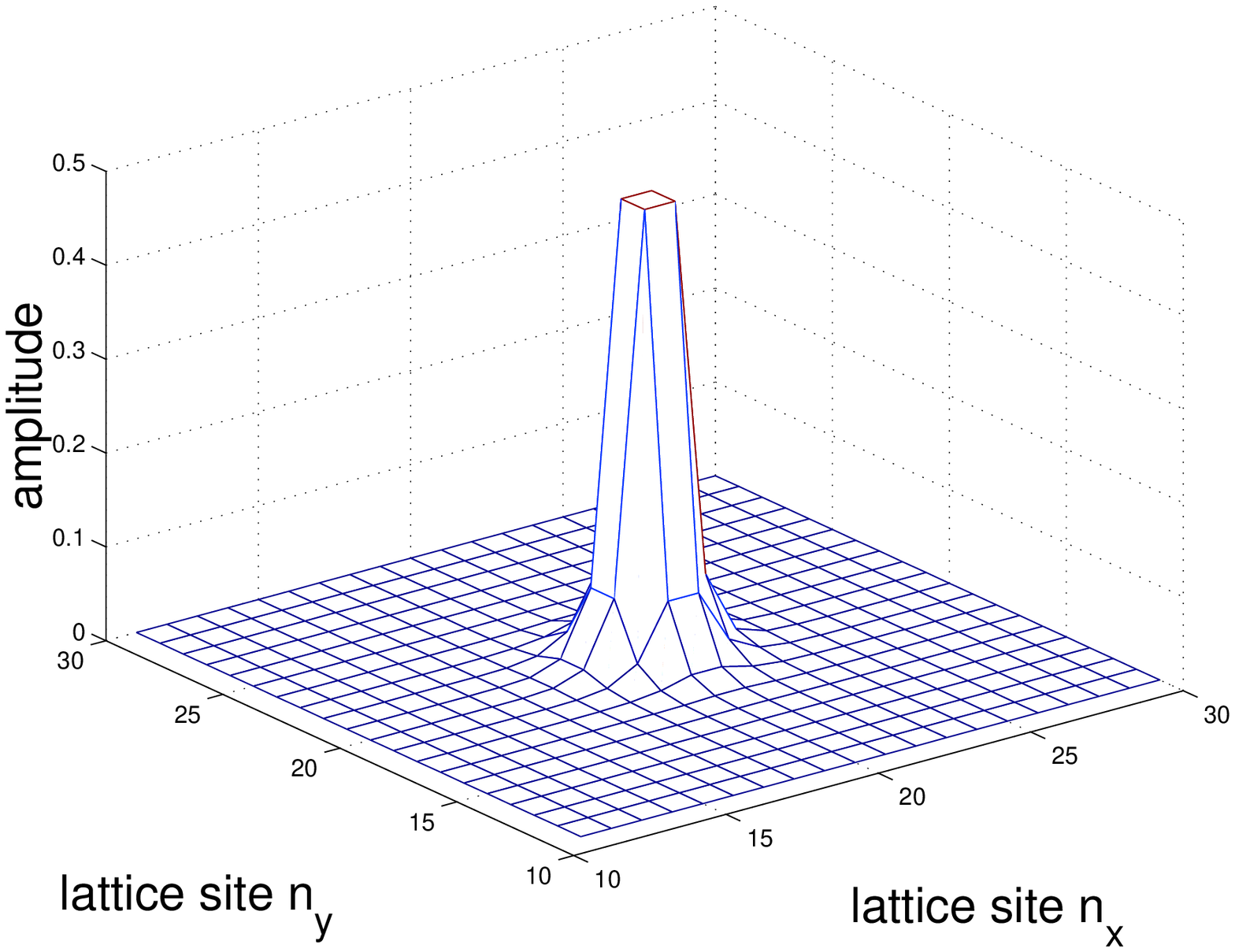,height=4.4cm,width=4.4cm} & \hspace{-0.2cm} 
\epsfig{file=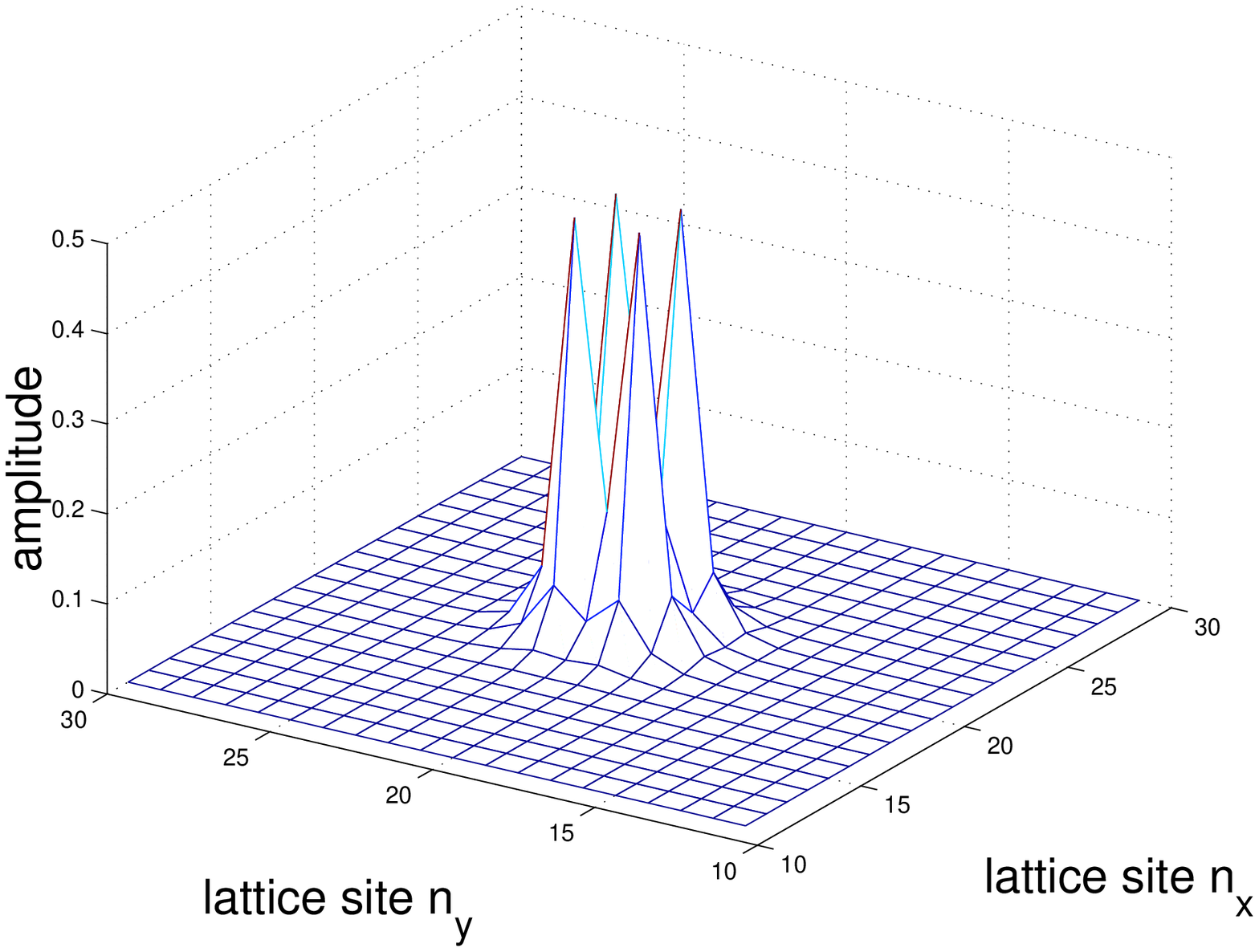,height=4.4cm,width=4.4cm} & \hspace{-0.2cm}
\epsfig{file=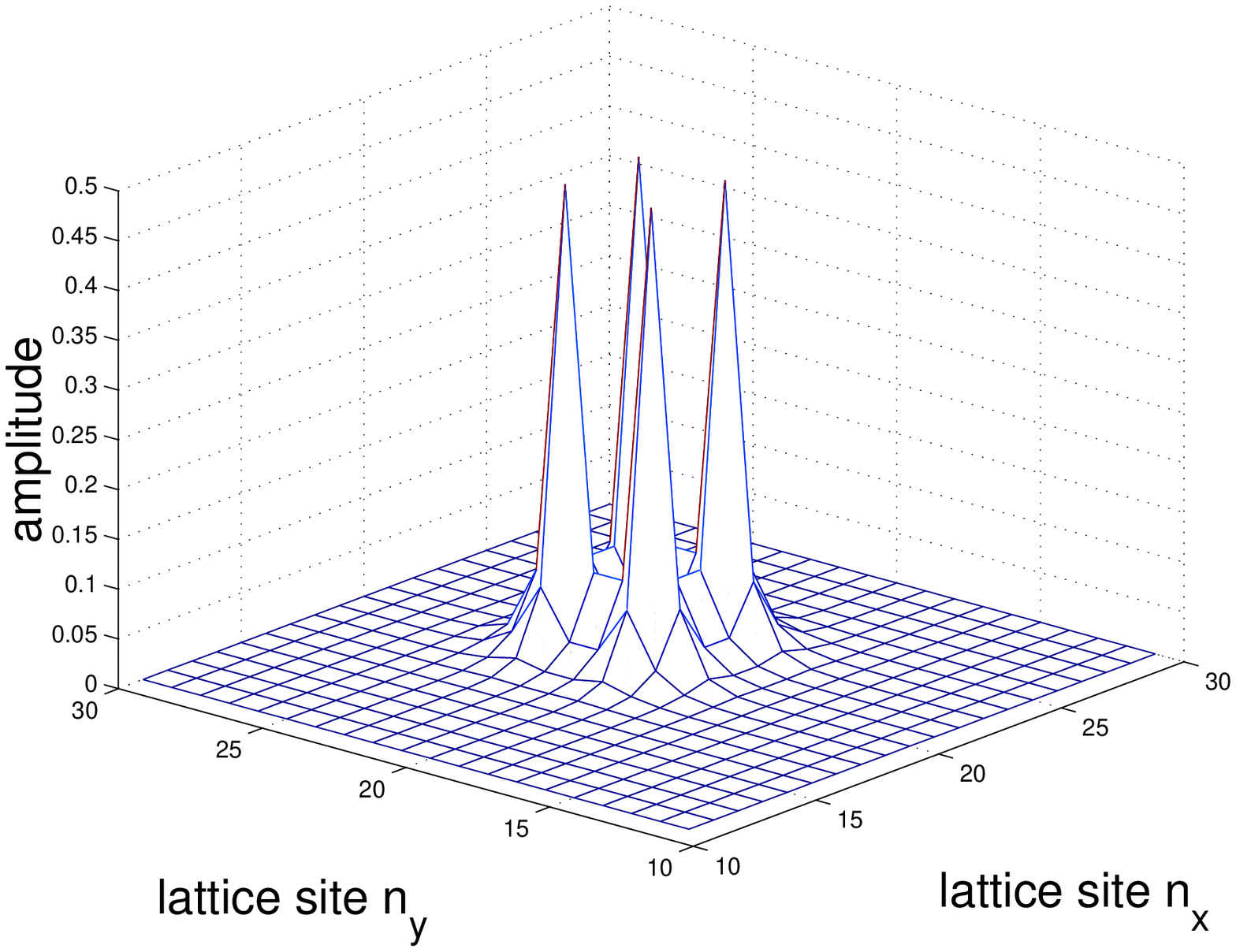,height=4.4cm,width=4.4cm}  \\
\epsfig{file=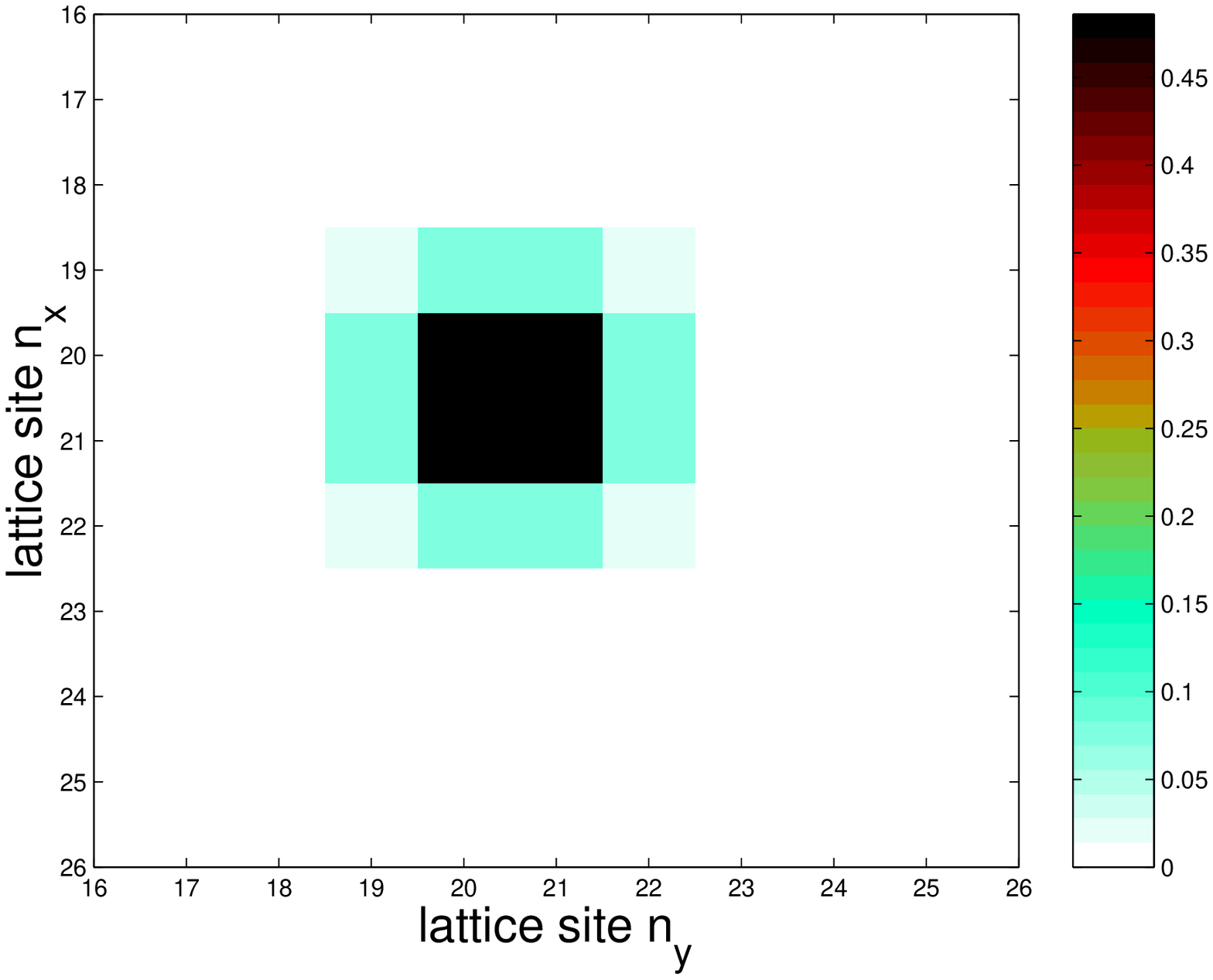,height=4.4cm,width=4.4cm} & \hspace{-0.2cm} 
\epsfig{file=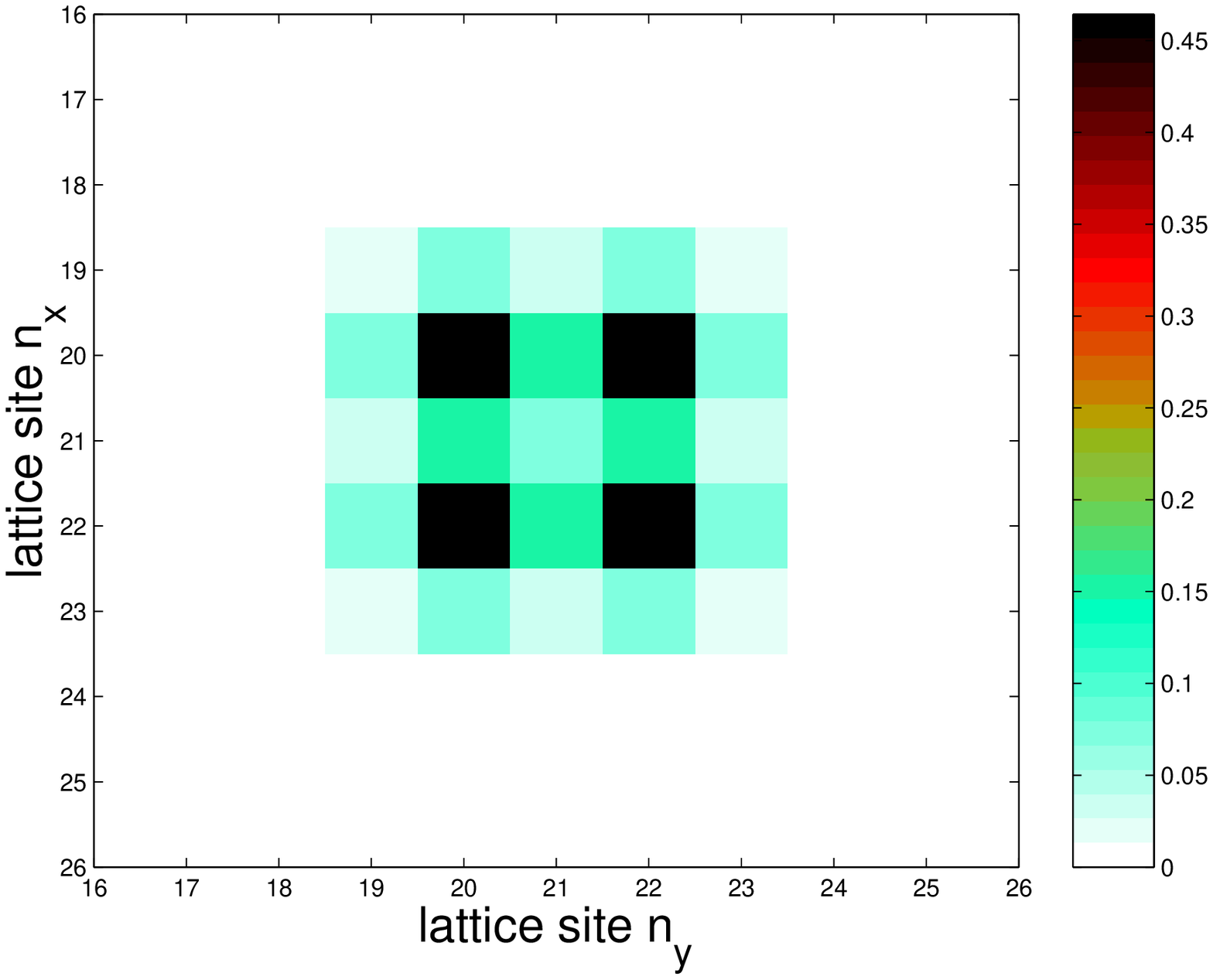,height=4.4cm,width=4.4cm} & \hspace{-0.2cm}
\epsfig{file=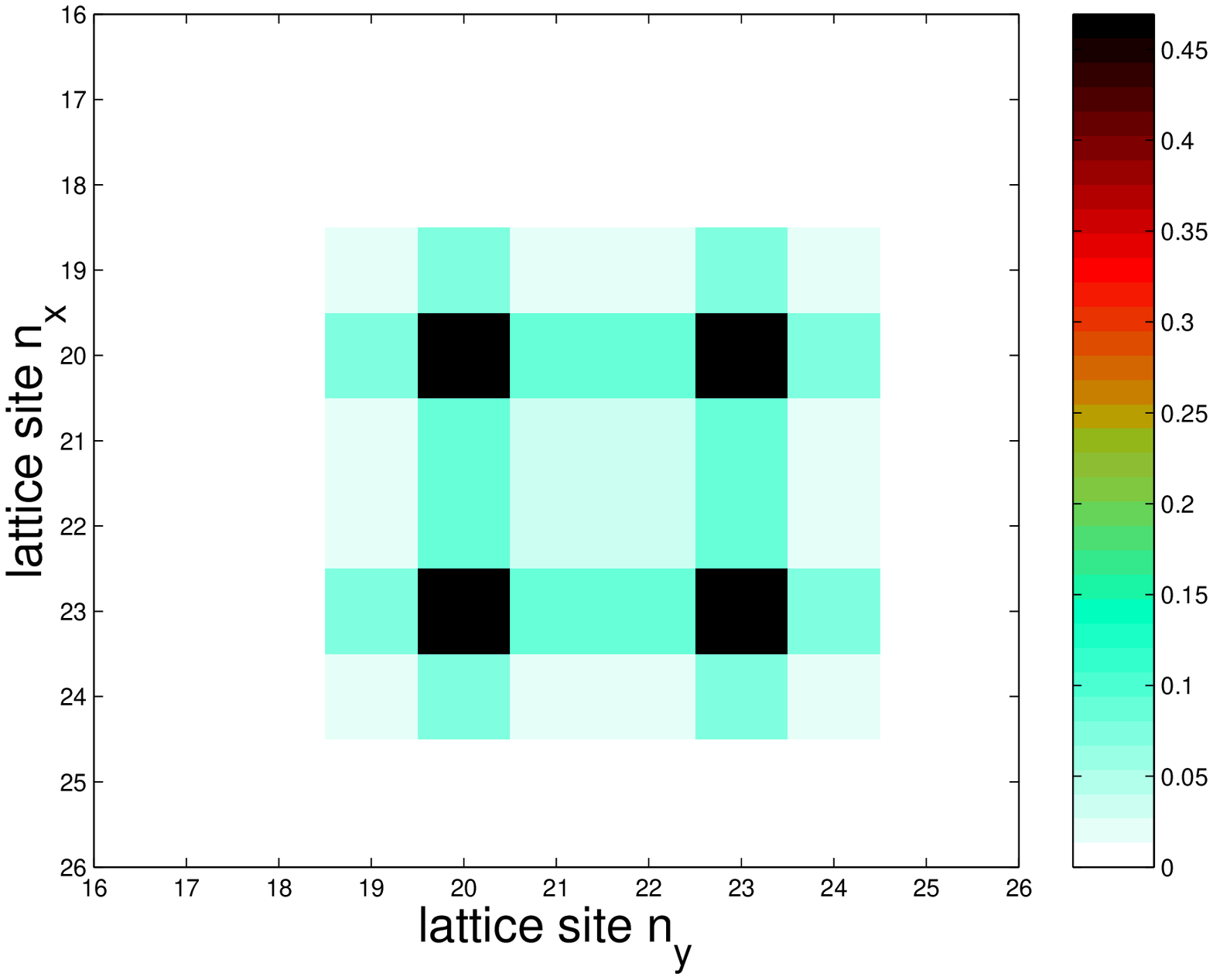,height=4.4cm,width=4.4cm}  \\
\epsfig{file=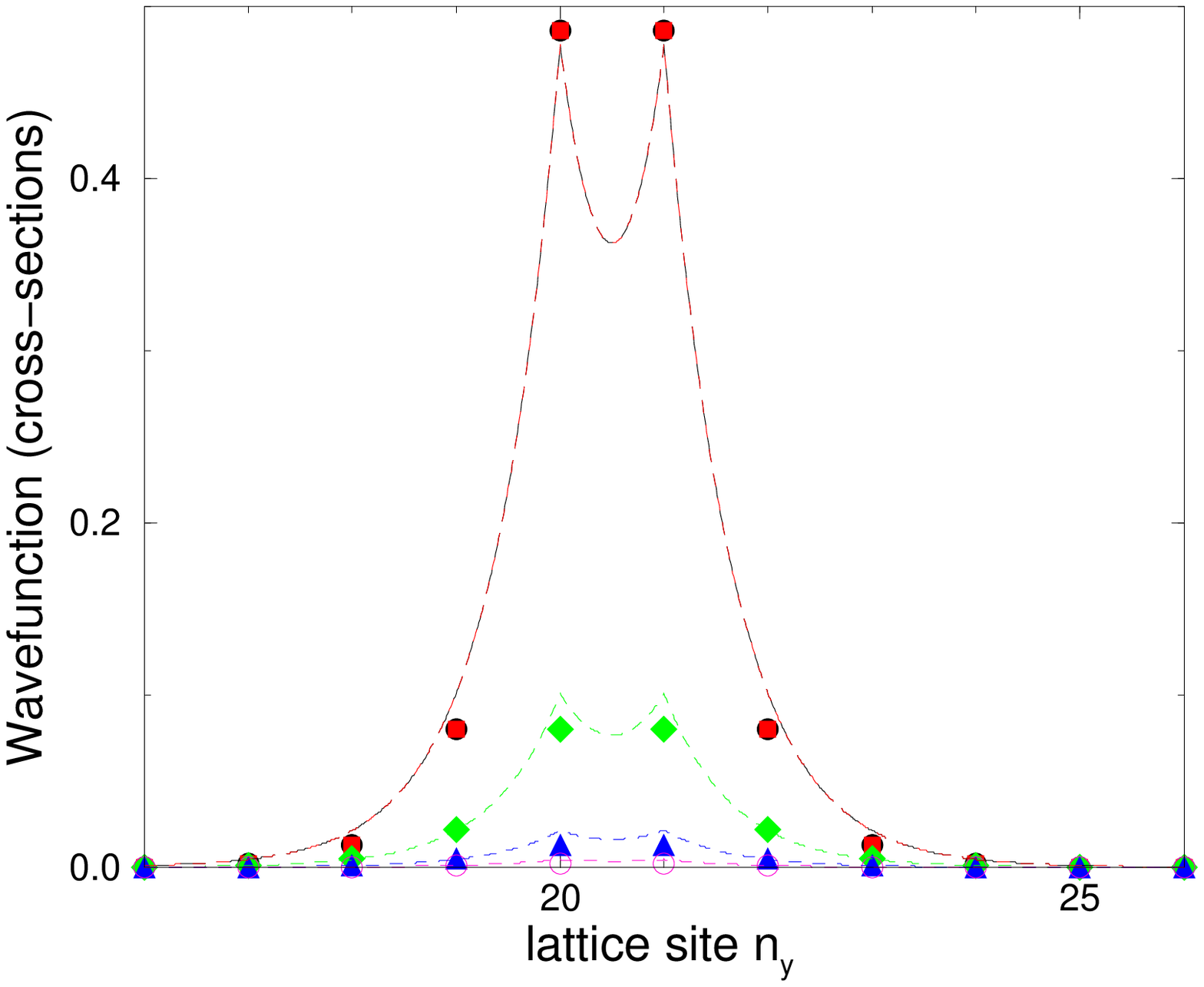,height=4.4cm,width=4.4cm} & \hspace{-0.2cm} 
\epsfig{file=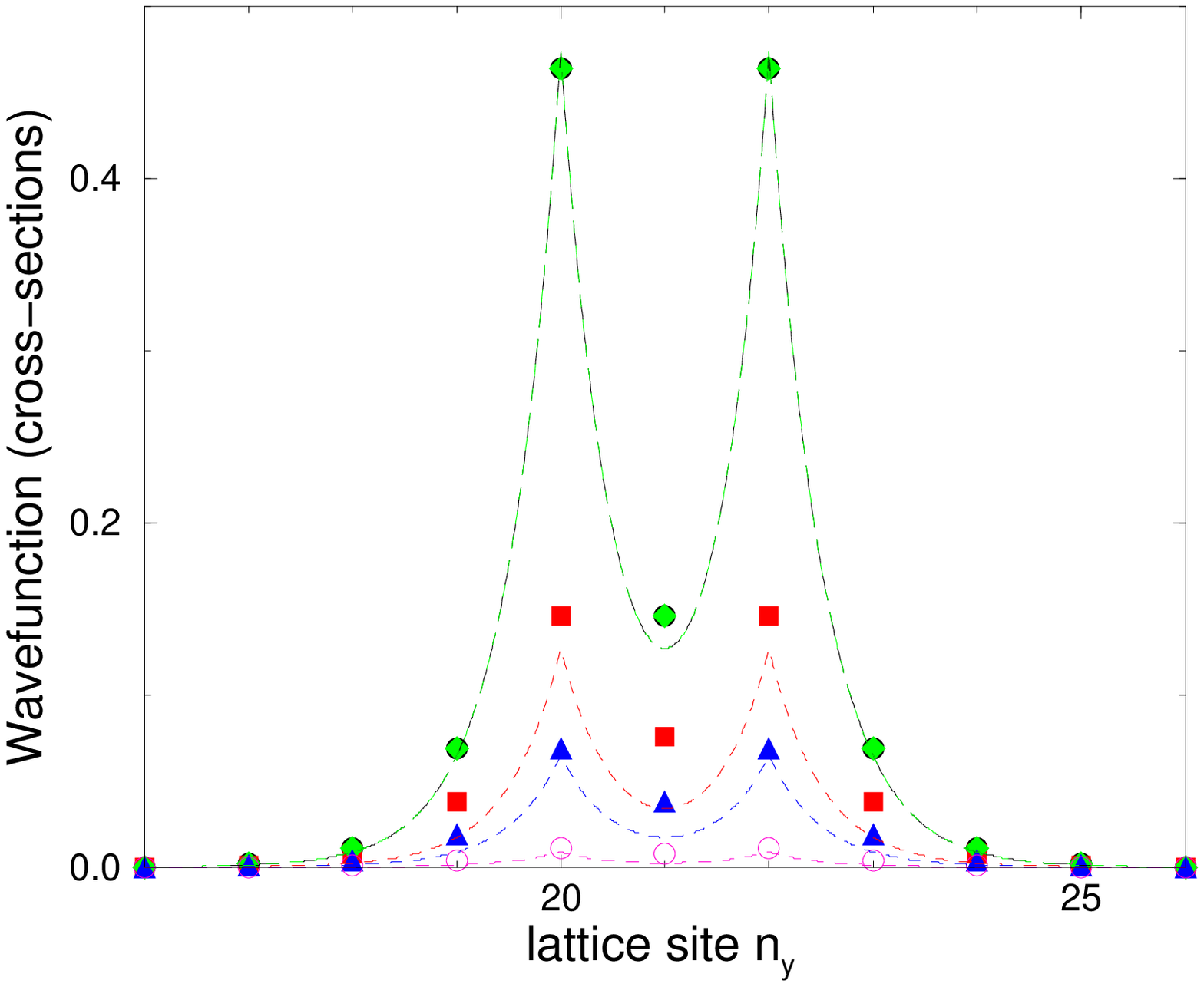,height=4.4cm,width=4.4cm} & \hspace{-0.2cm}
\epsfig{file=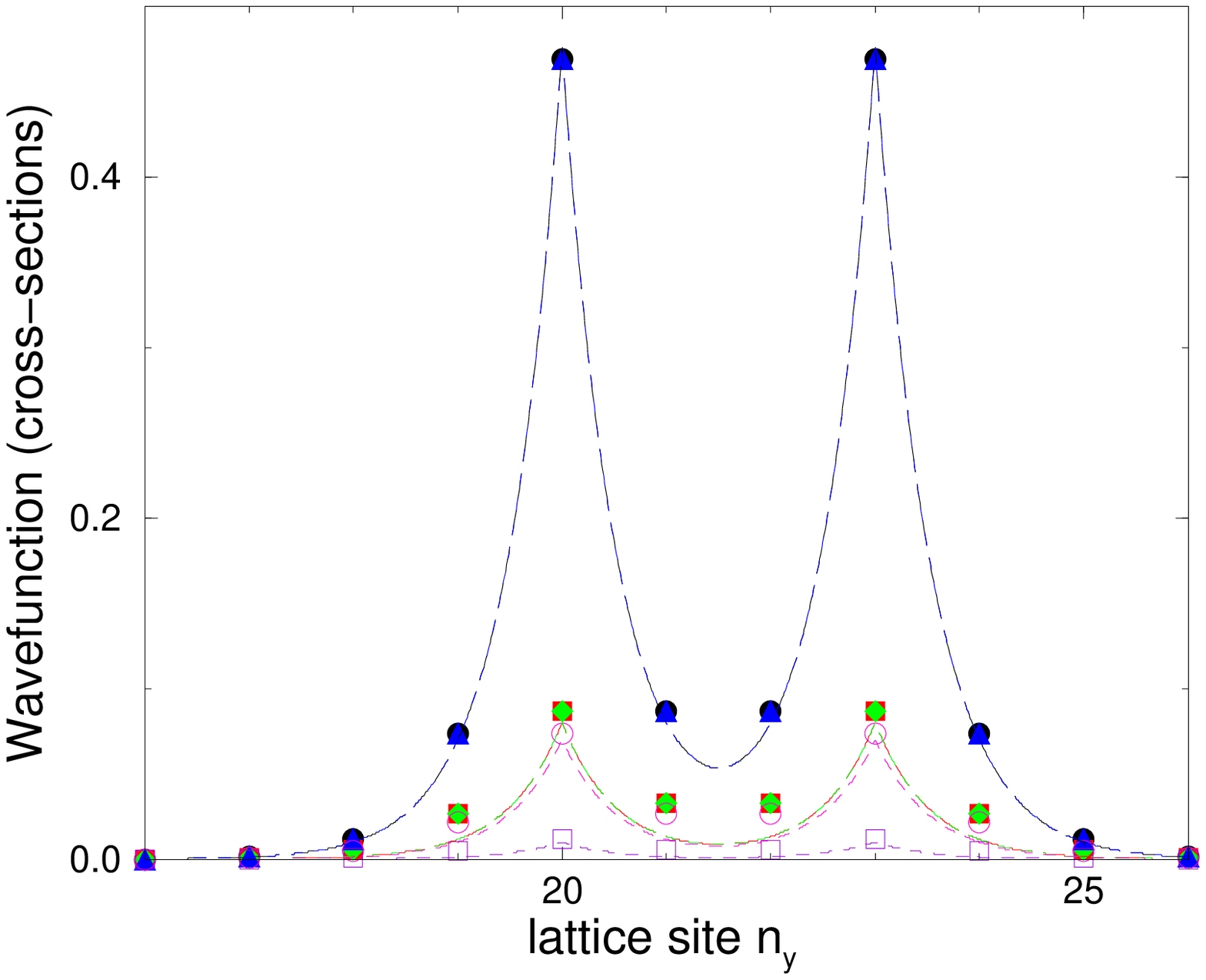,height=4.4cm,width=4.4cm}  \\
\end{tabular}
\end{center}
\caption{3D plots (first row) and density plots (second row)
of quadruple-peaked symmetric solutions of DNLS in 2D.
{\it Left column:} interpeak separation $l=1$ lattice site, $\chi=-22.5$.
{\it Middle column:} interpeak separation $l=2$ sites, $\chi=-32$.
{\it Right column:} interpeak separation $l=3$ sites, $\chi=-30$.
Cross-sections of the wavefunction are shown in the third row with points:
$\psi_{n_x=n_1,n_y}$ (filled circles), $\psi_{n_x=n_1+1,n_y}$ (filled squares),
$\psi_{n_x=n_1+2,n_y}$ (filled diamonds), $\psi_{n_x=n_1+3,n_y}$ (filled
triangles), $\psi_{n_x=n_1+4,n_y}$ (open circles), $\psi_{n_x=n_1+5,n_y}$
(open squares, in the third column), where $n_1=20$ is the x$-$coordinate of
the first peak. Dashed lines show analytical approximations of the solutions
using Eq.~(\ref{qp2d}). }
\label{fQS2}
\end{figure}

The former are calculated from the initial state
\be   \label{qs2}
\psi_{n_x,n_y}^{(r=0)}= \frac{1}{\sqrt{4}}(\delta_{n_x,n_1} \delta_{n_y,n_2}
+ \delta_{n_x,n_1} \delta_{n_y,n_2+l} + \delta_{n_x,n_1+l} \delta_{n_y,n_2+l}
+ \delta_{n_x,n_1+l} \delta_{n_y,n_2}).
\ee
Fig.~\ref{fQS2} shows such symmetric stationary states for $l=1$, 2, and 3.
The solutions with $l=1$ give the single branch with frequencies around
$\omega = \frac{\chi}{4}-2$ in Fig.~\ref{fFS2}. These solutions are unstable
for negative $\chi$.
Linear stability analysis shows three purely imaginary pairs of opposite
eigenvalues (two of these pairs are degenerate). For the case $l=1$ ($l>2$),
as $|\chi|$ increases the magnitude of the unstable eigenvalues increases
(decreases). For $l=2$ the unstable eigenvalues approach constant values
(around $\pm 2.83$ the most unstable one and around $\pm 2$ the doubly
degenerate eigenvalues) for $|\chi| \gg 1$. However, for relatively large
$\chi>0$ the symmetric solutions corresponding to odd values of $l$ are
linearly stable with three real discrete pairs of eigenvalues. The band of
eigenvalues extends approximately from $\frac{|\chi|}{4}-4$ to
$\frac{|\chi|}{4}+4$, apart from the case of $l=1$, where it extends from
$\frac{|\chi|}{4}-2$ to $\frac{|\chi|}{4}+6$.

\begin{figure}
\begin{center}
\begin{tabular}{ccc}
\epsfig{file=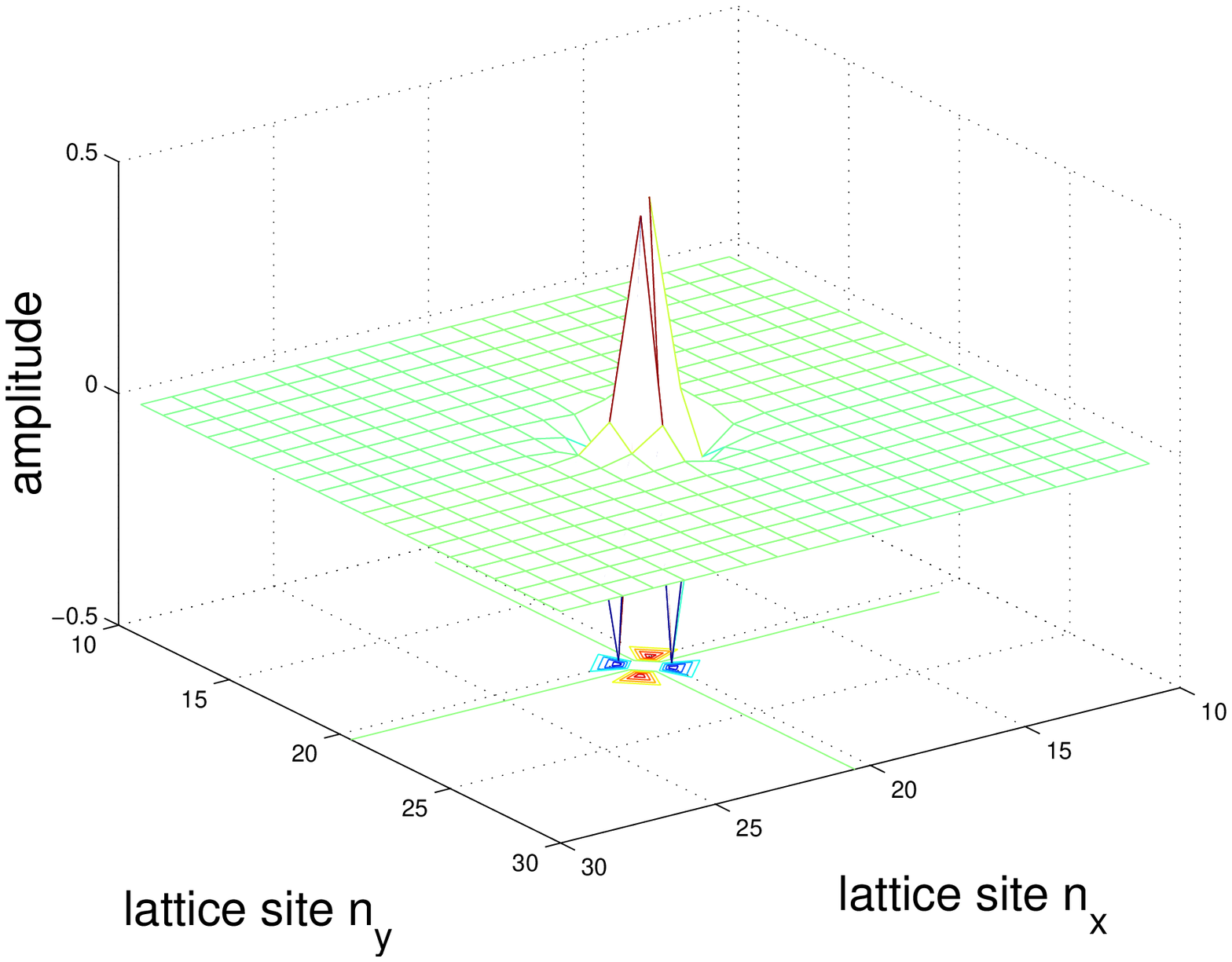,height=4.4cm,width=4.4cm} & \hspace{-0.2cm} 
\epsfig{file=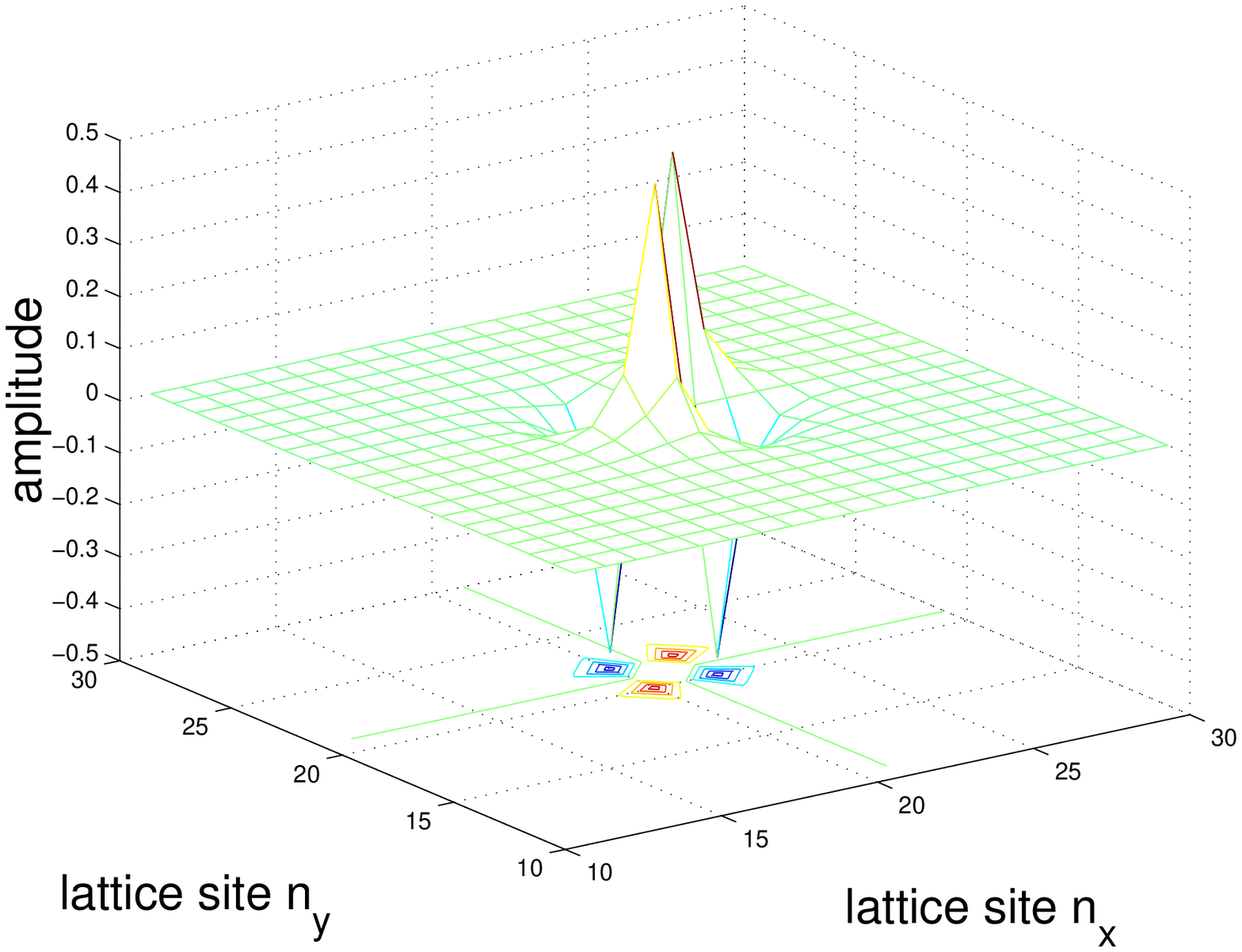,height=4.4cm,width=4.4cm} & \hspace{-0.2cm}
\epsfig{file=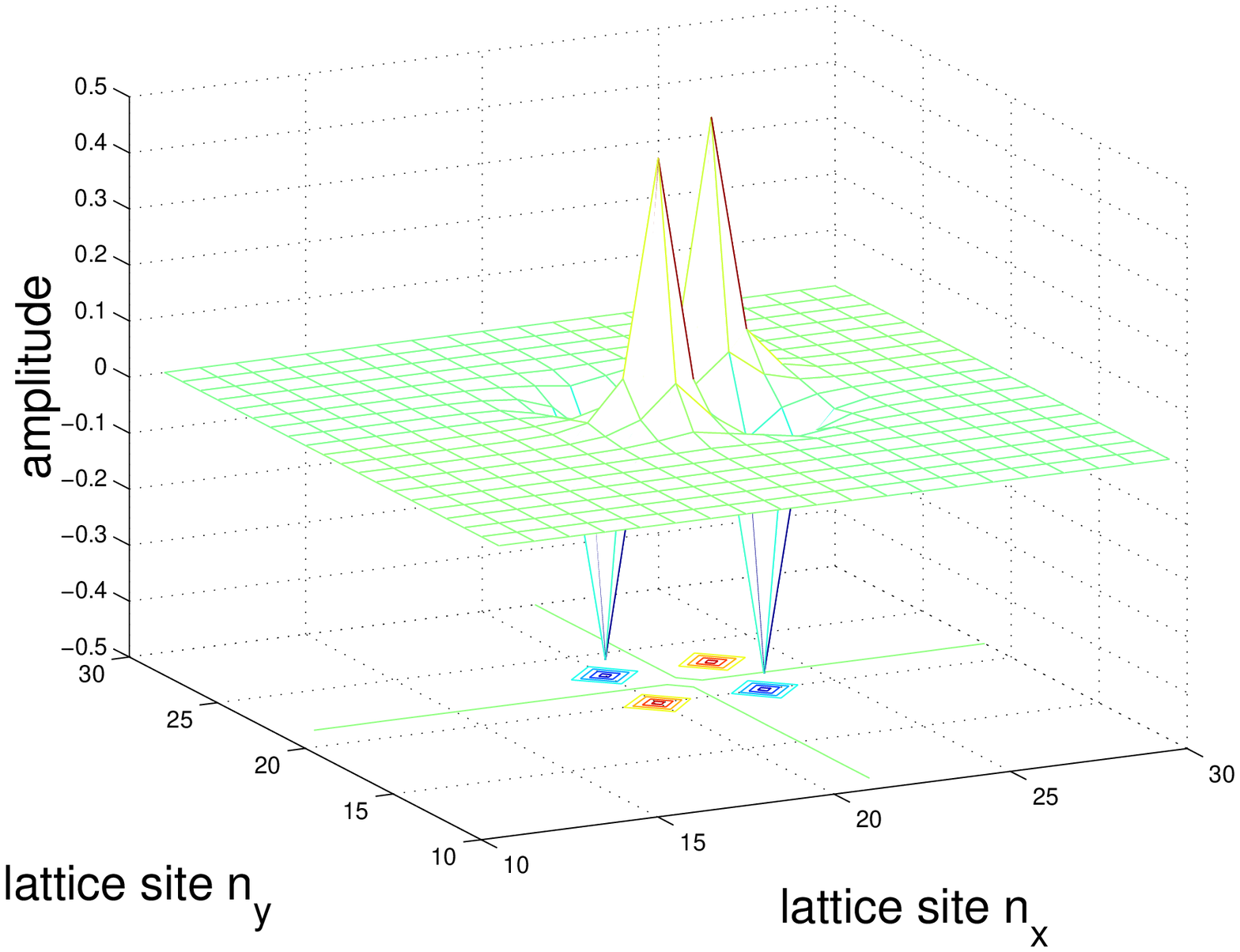,height=4.4cm,width=4.4cm}  \\
\epsfig{file=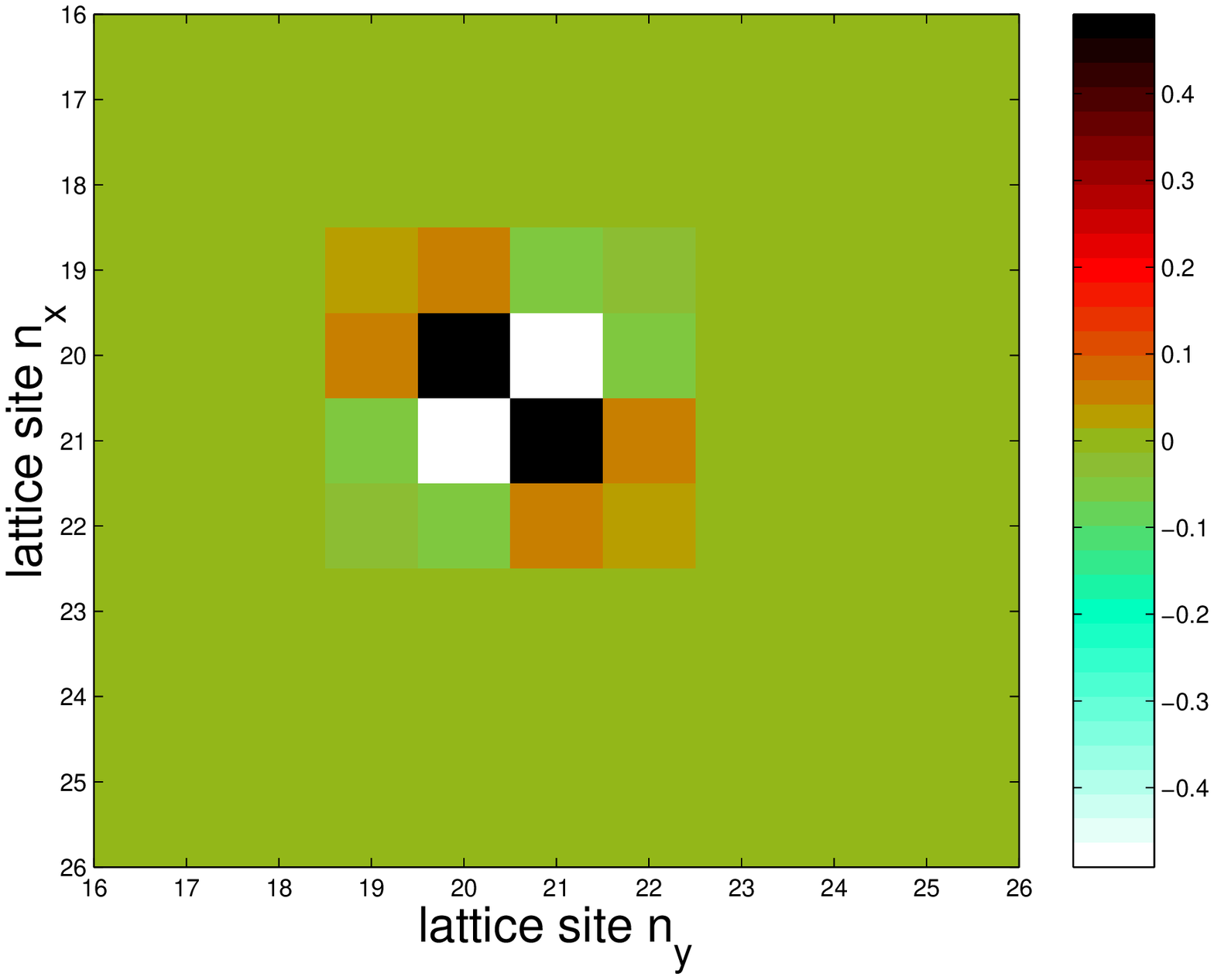,height=4.4cm,width=4.4cm} & \hspace{-0.2cm} 
\epsfig{file=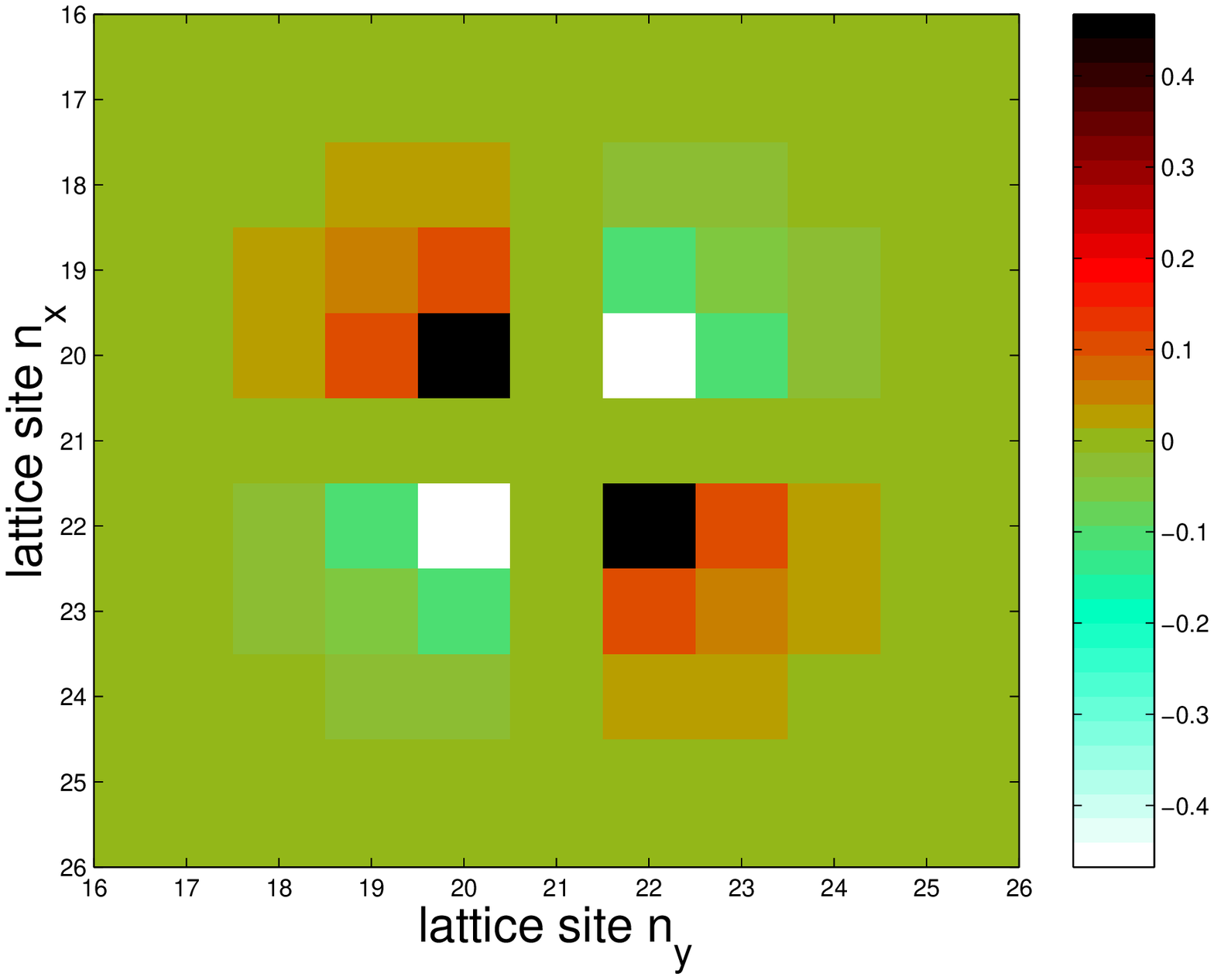,height=4.4cm,width=4.4cm} & \hspace{-0.2cm}
\epsfig{file=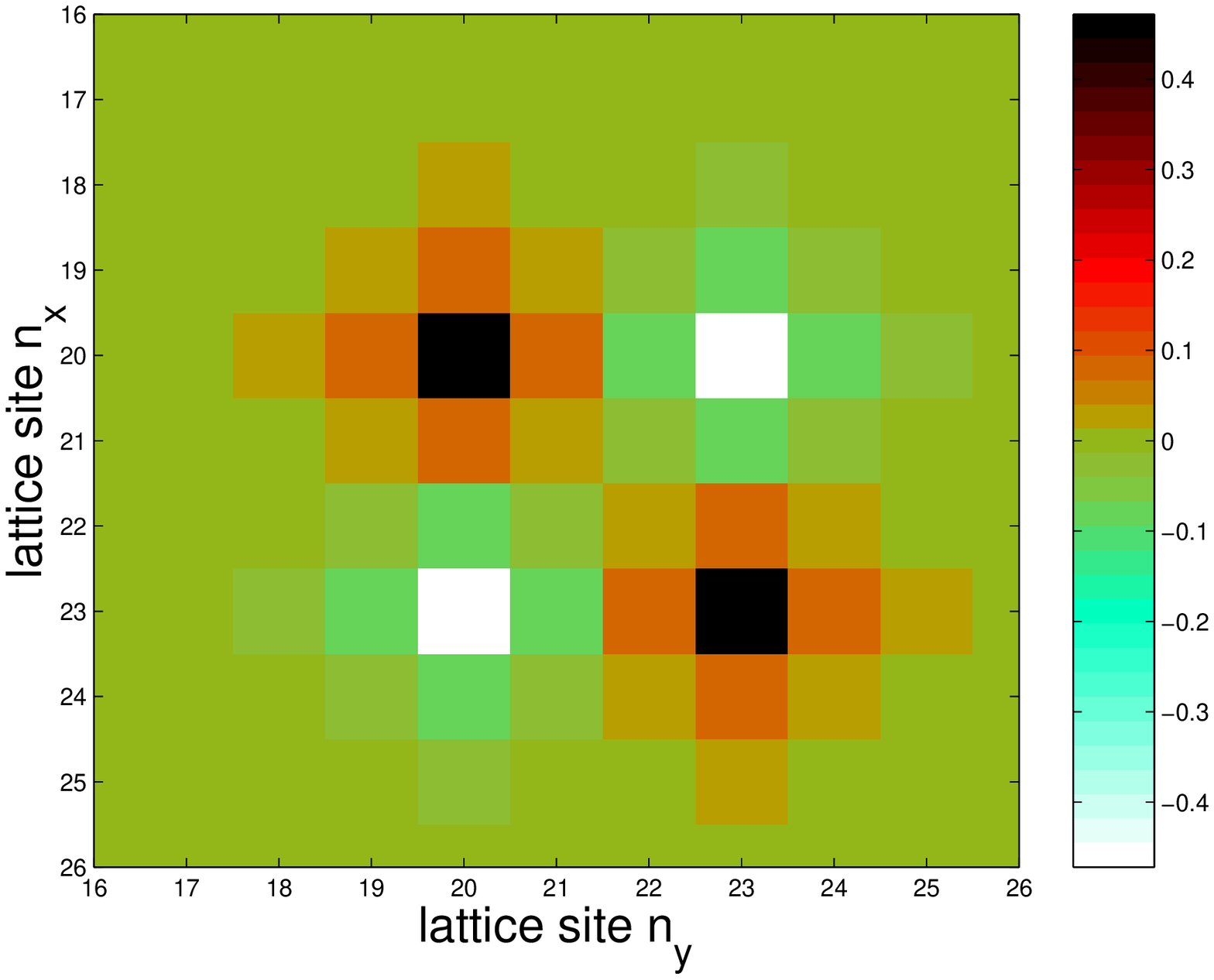,height=4.4cm,width=4.4cm}  \\
\epsfig{file=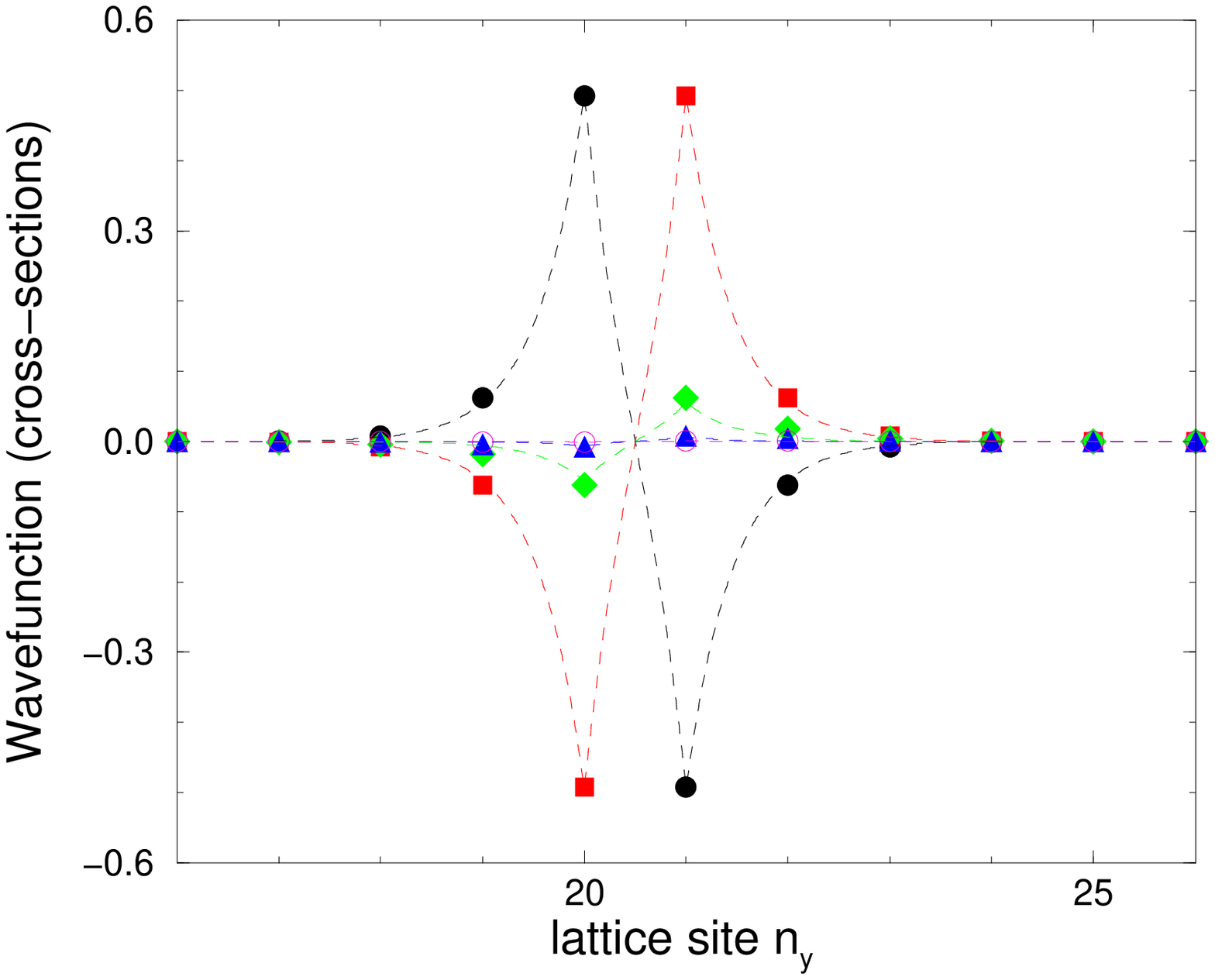,height=4.4cm,width=4.4cm} & \hspace{-0.2cm} 
\epsfig{file=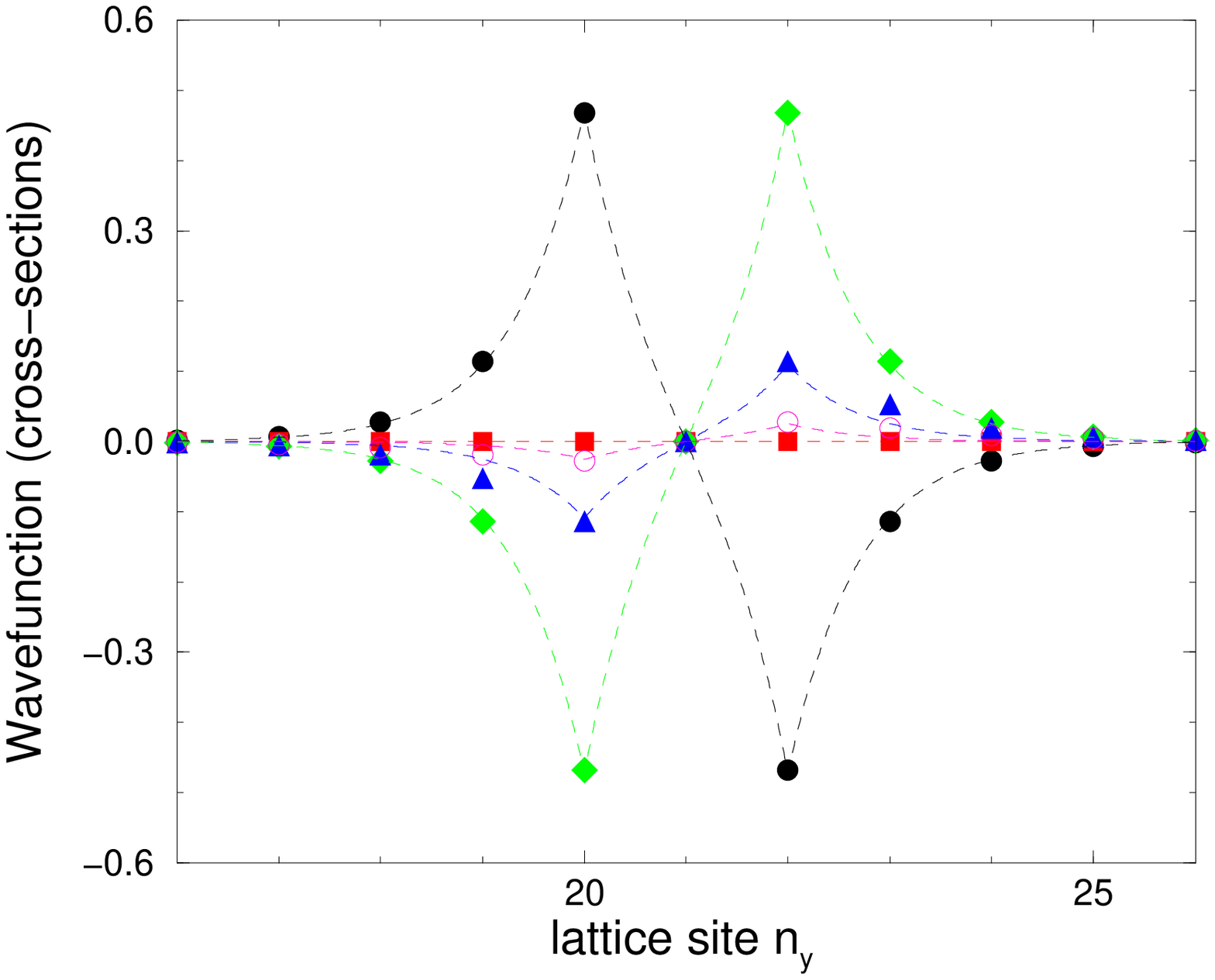,height=4.4cm,width=4.4cm} & \hspace{-0.2cm}
\epsfig{file=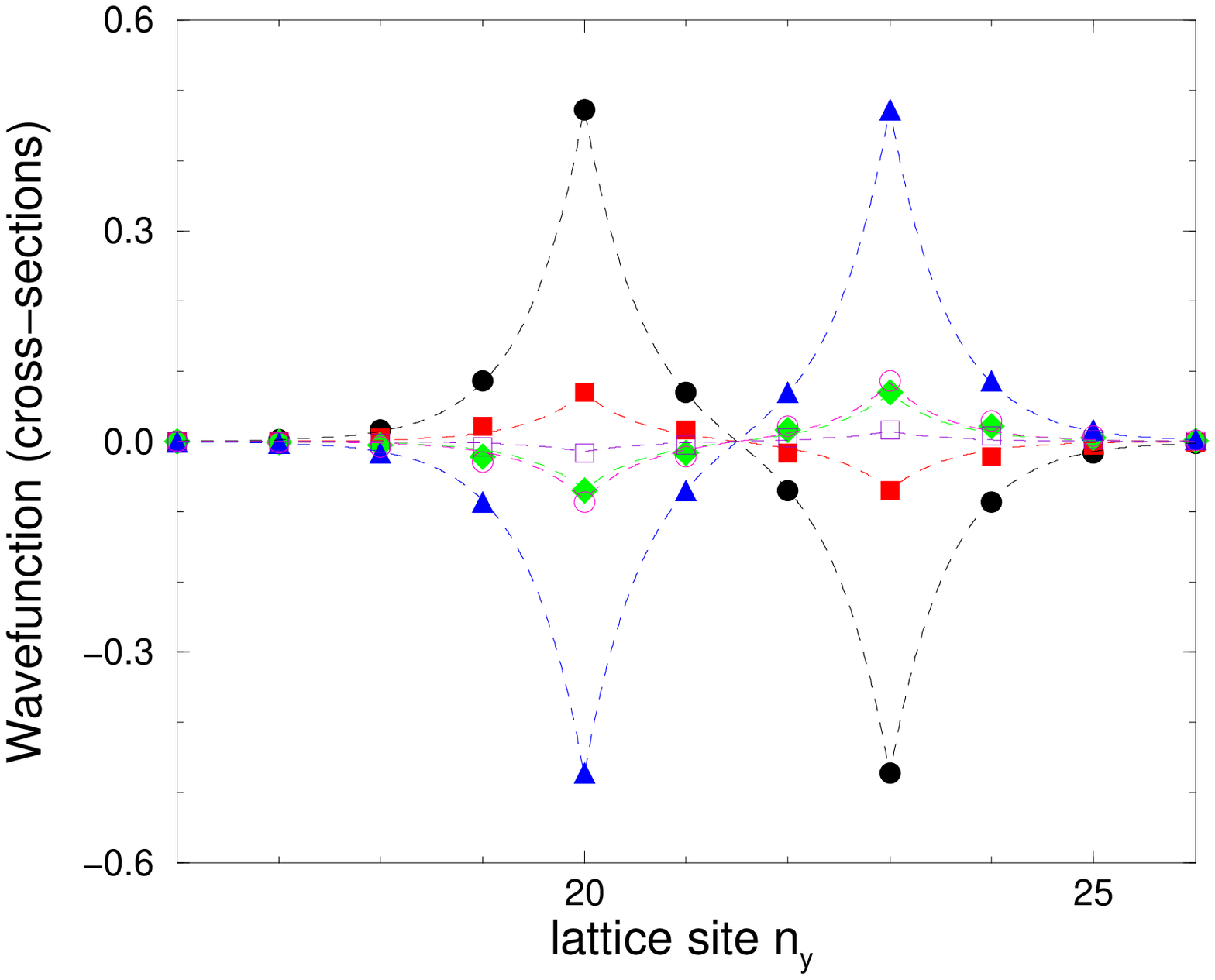,height=4.4cm,width=4.4cm}  \\
\end{tabular}
\end{center}
\caption{ 3D plots (first row) and density plots (second row)
of quadruple-peaked antisymmetric solutions of DNLS in 2D.
{\it Left column:} interpeak separation $l=1$ lattice site, $\chi=-38$.
{\it Middle column:} interpeak separation $l=2$ sites, $\chi=-21$.
{\it Right column:} interpeak separation $l=3$ sites, $\chi=-26$.
Cross-sections of the wavefunction are shown in the third row with points:
$\psi_{n_x=n_1,n_y}$ (filled circles), $\psi_{n_x=n_1+1,n_y}$ (filled squares),
$\psi_{n_x=n_1+2,n_y}$ (filled diamonds), $\psi_{n_x=n_1+3,n_y}$ (filled
triangles), $\psi_{n_x=n_1+4,n_y}$ (open circles), $\psi_{n_x=n_1+5,n_y}$
(open squares, in the third column), where $n_1=20$ is the x$-$coordinate of
the first peak. Dashed lines show analytical approximations of the solutions
using Eq.~(\ref{qp2d}).}
\label{fQA2}
\end{figure}

The antisymmetric along both axes QP solutions are obtained
from alternating signs on neighboring peaks:
\be   \label{qa2}
\psi_{n_x,n_y}^{(r=0)}= \frac{1}{\sqrt{4}}(\delta_{n_x,n_1} \delta_{n_y,n_2}
- \delta_{n_x,n_1} \delta_{n_y,n_2+l} + \delta_{n_x,n_1+l} \delta_{n_y,n_2+l}
- \delta_{n_x,n_1+l} \delta_{n_y,n_2}).
\ee
Some examples are shown in Fig.~\ref{fQA2}. The branch of these solutions
with $l=1$ gives the single branch with frequencies around
$\omega = \frac{\chi}{4}+2$ in Fig.~\ref{fFS2}. For $\chi<0$ these
solutions are linearly
stable for large values of $|\chi|$. In this case there exist three discrete
real eigenvalues (two of them are degenerate) and the band extends from
$\frac{|\chi|}{4}-4$ to $\frac{|\chi|}{4}+4$, apart from the case of $l=1$,
where it extends from $\frac{|\chi|}{4}-6$ to $\frac{|\chi|}{4}+2$. As long as
the discrete eigenvalues are outside of the band the solution is linearly
stable. When they collide instabilities develop. The dependence of the
discrete eigenvalues on $|\chi|$ is the usual: for $l=1$ ($l>2$) they
increase (decrease) with $|\chi|$, while for $l=2$ they slightly vary,
approaching constant values at $|\chi| \gg 1$. Regarding the solutions shown
in Fig.~\ref{fQA2} the last one (for $l=3$, in the right column) is linearly
stable and the other two are unstable. As previously, the opposite sign of
$\chi$ ($\chi>0$) does not change anything regarding the stability eigenvalues
of symmetric and fully antisymmetric QP states, when $l$ is even. On the
contrary, for odd $l$ these families of solutions interchange stability
eigenvalues when $\chi \rightarrow -\chi$.

\begin{figure}
\begin{center}
\begin{tabular}{ccc}
\epsfig{file=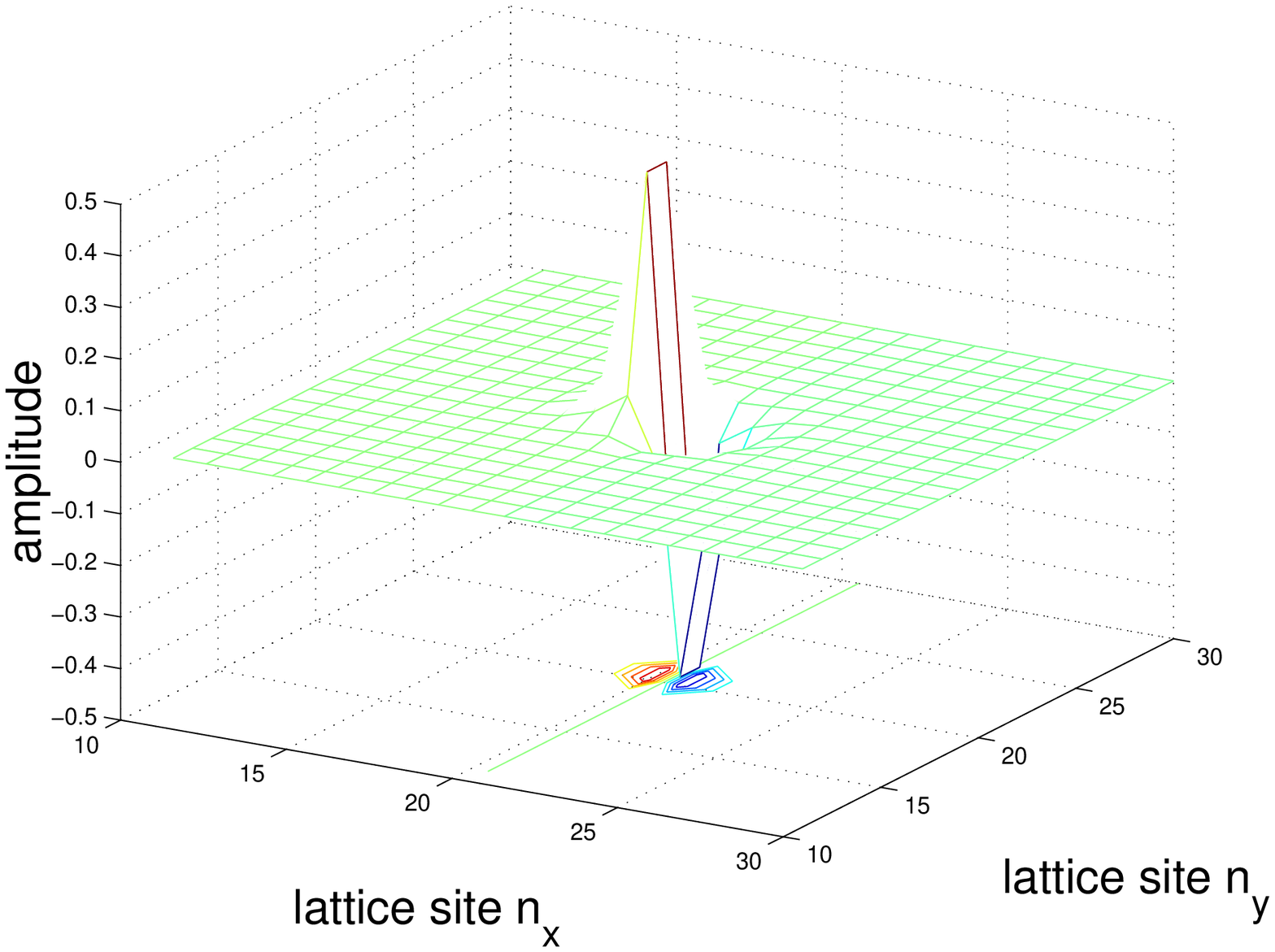,height=4.4cm,width=4.4cm} & \hspace{-0.2cm} 
\epsfig{file=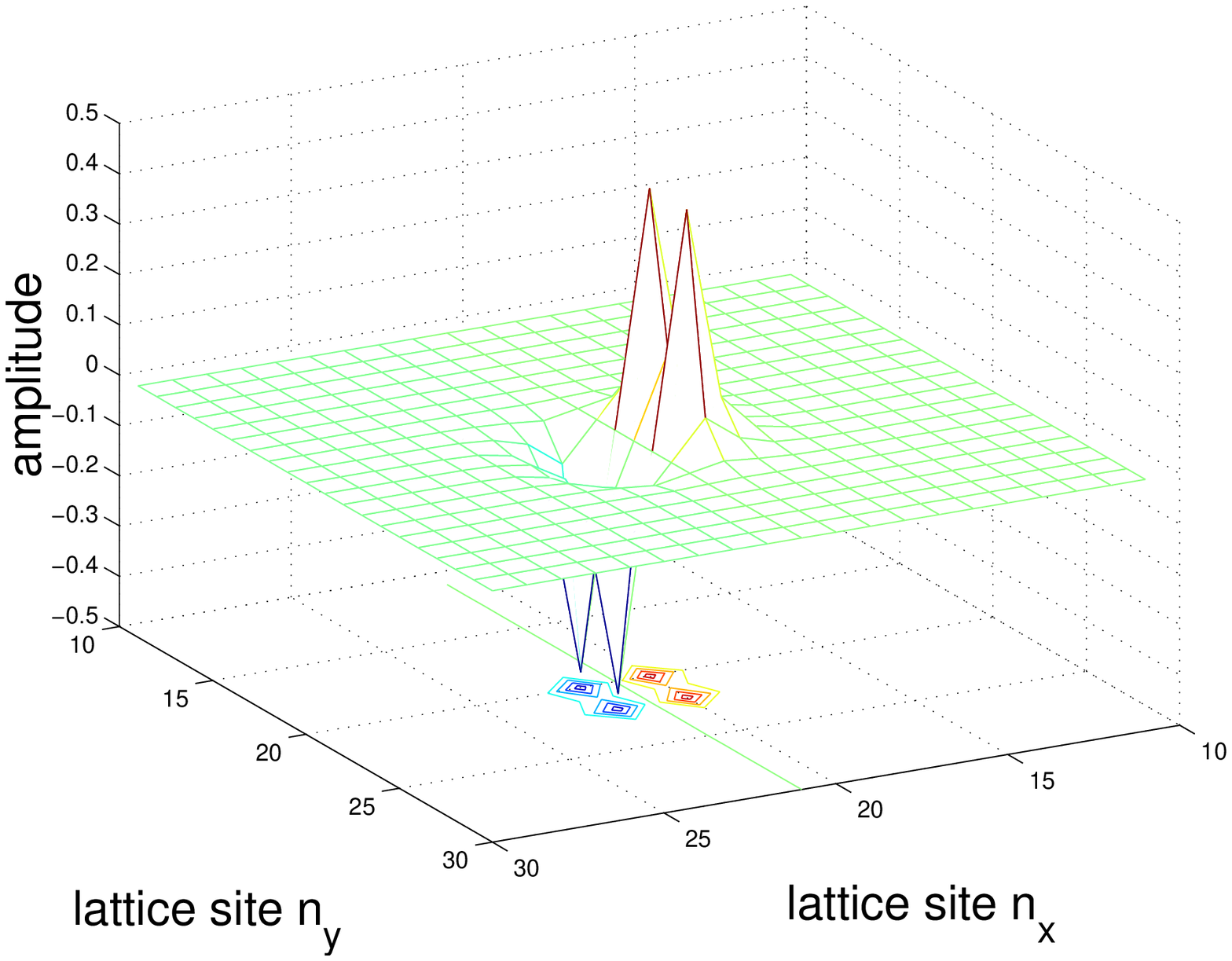,height=4.4cm,width=4.4cm} & \hspace{-0.2cm}
\epsfig{file=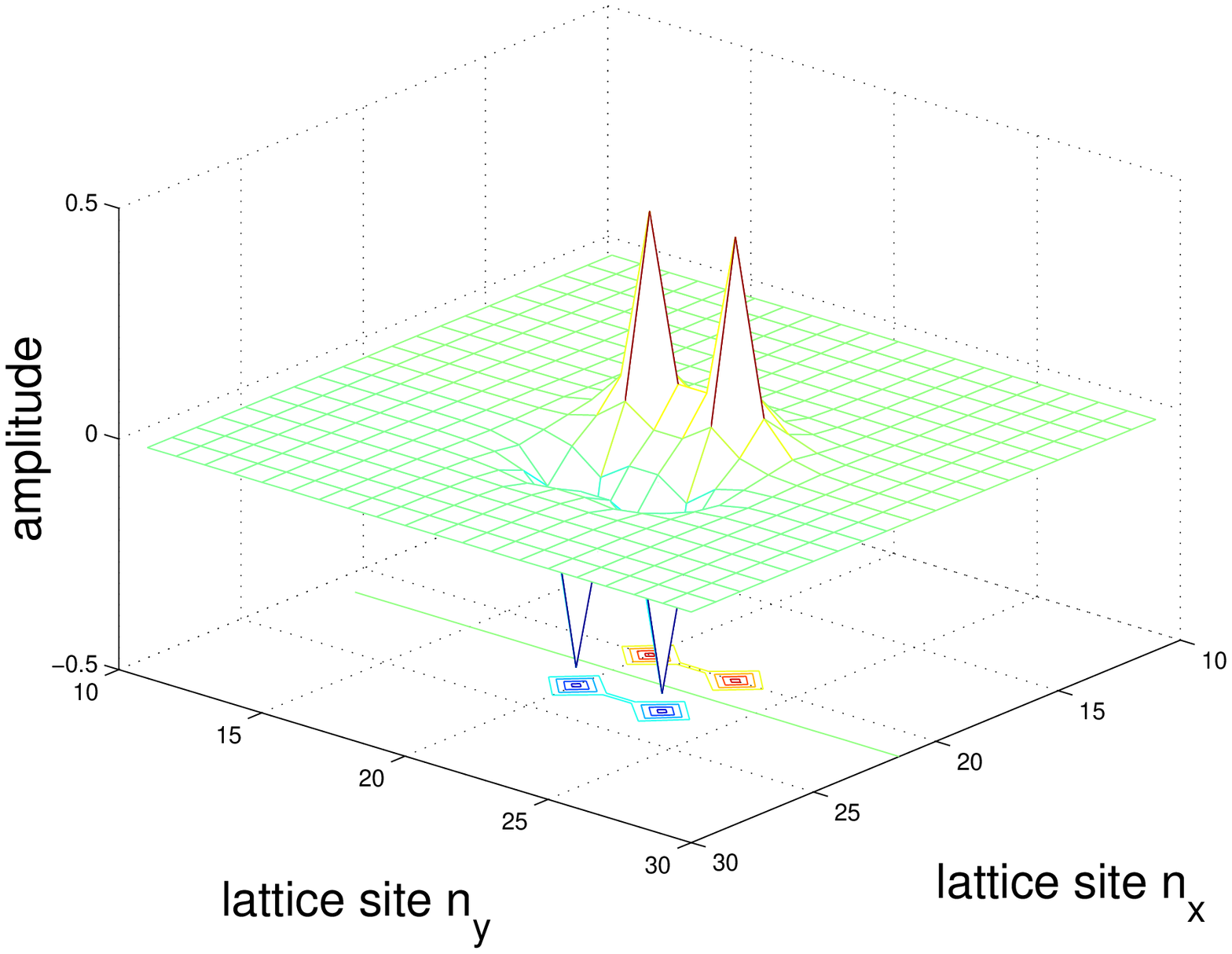,height=4.4cm,width=4.4cm}  \\
\epsfig{file=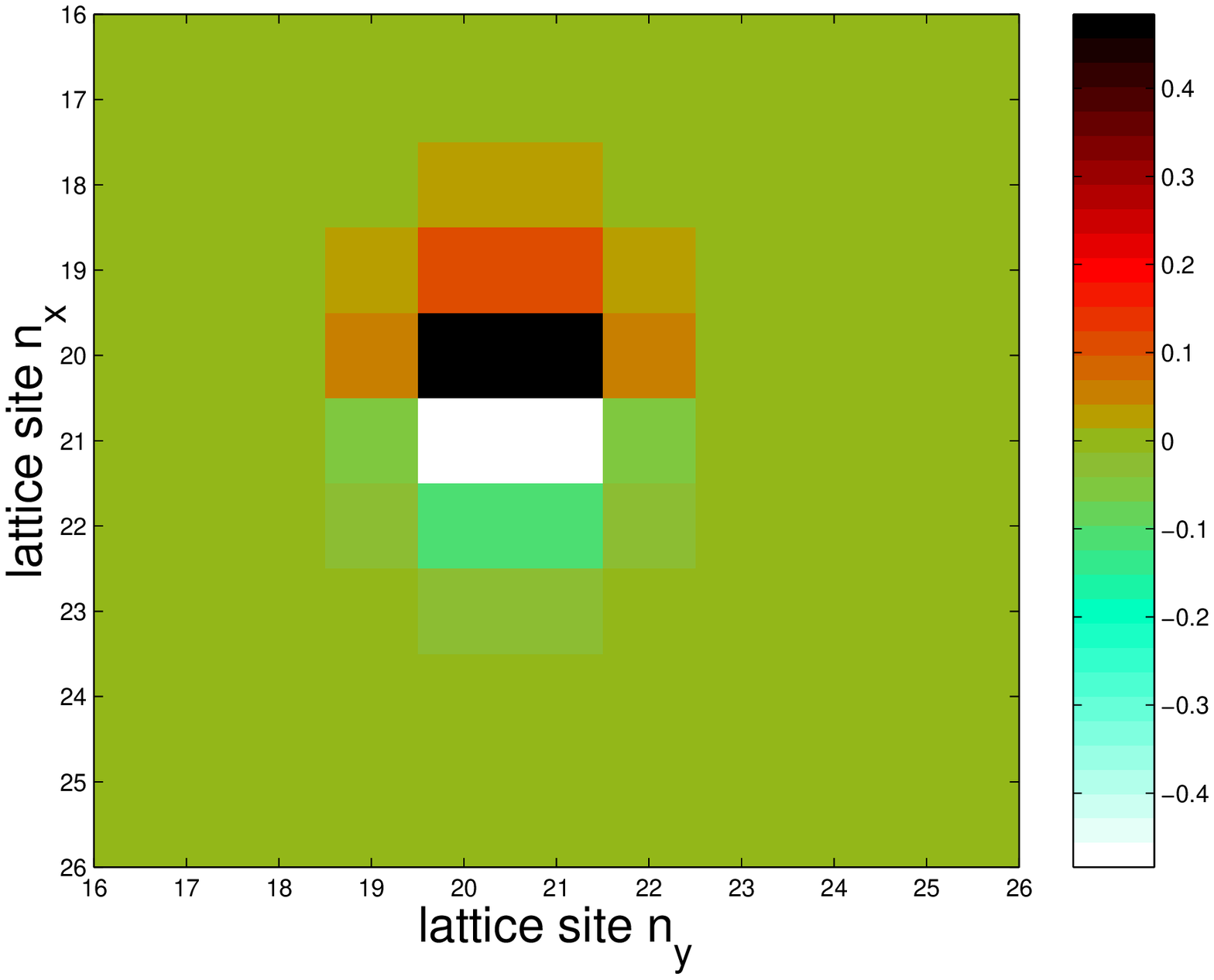,height=4.4cm,width=4.4cm} & \hspace{-0.2cm} 
\epsfig{file=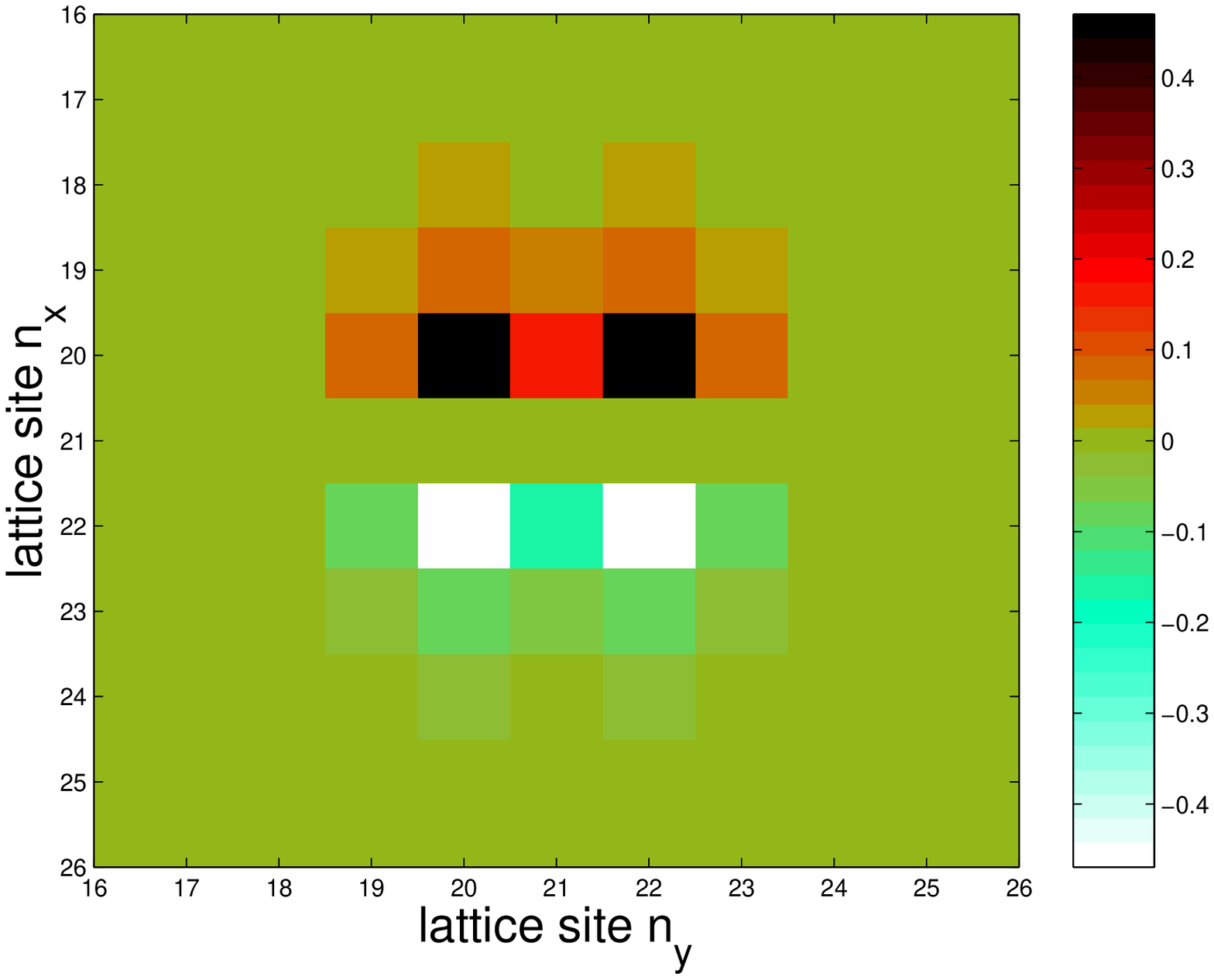,height=4.4cm,width=4.4cm} & \hspace{-0.2cm}
\epsfig{file=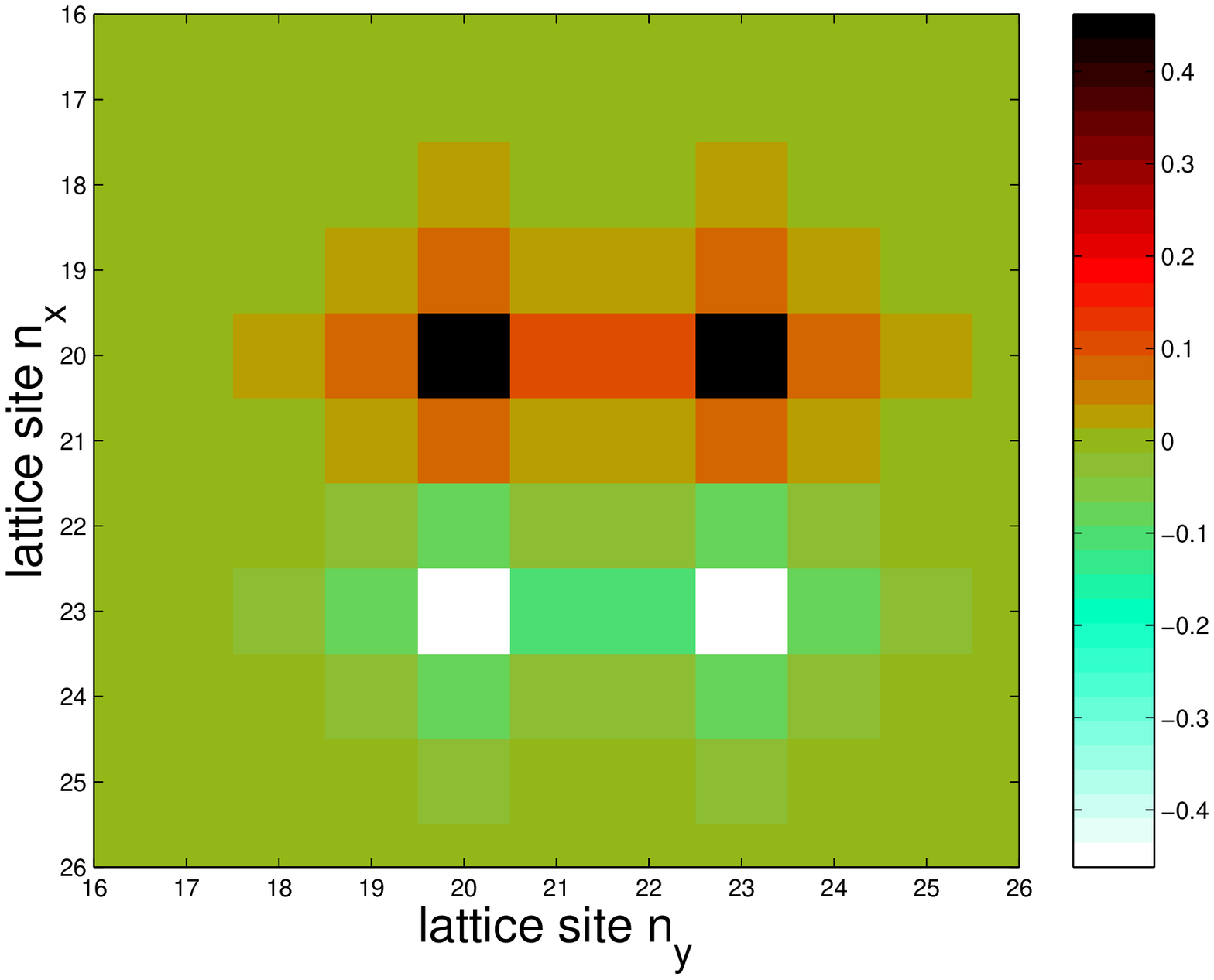,height=4.4cm,width=4.4cm}  \\
\epsfig{file=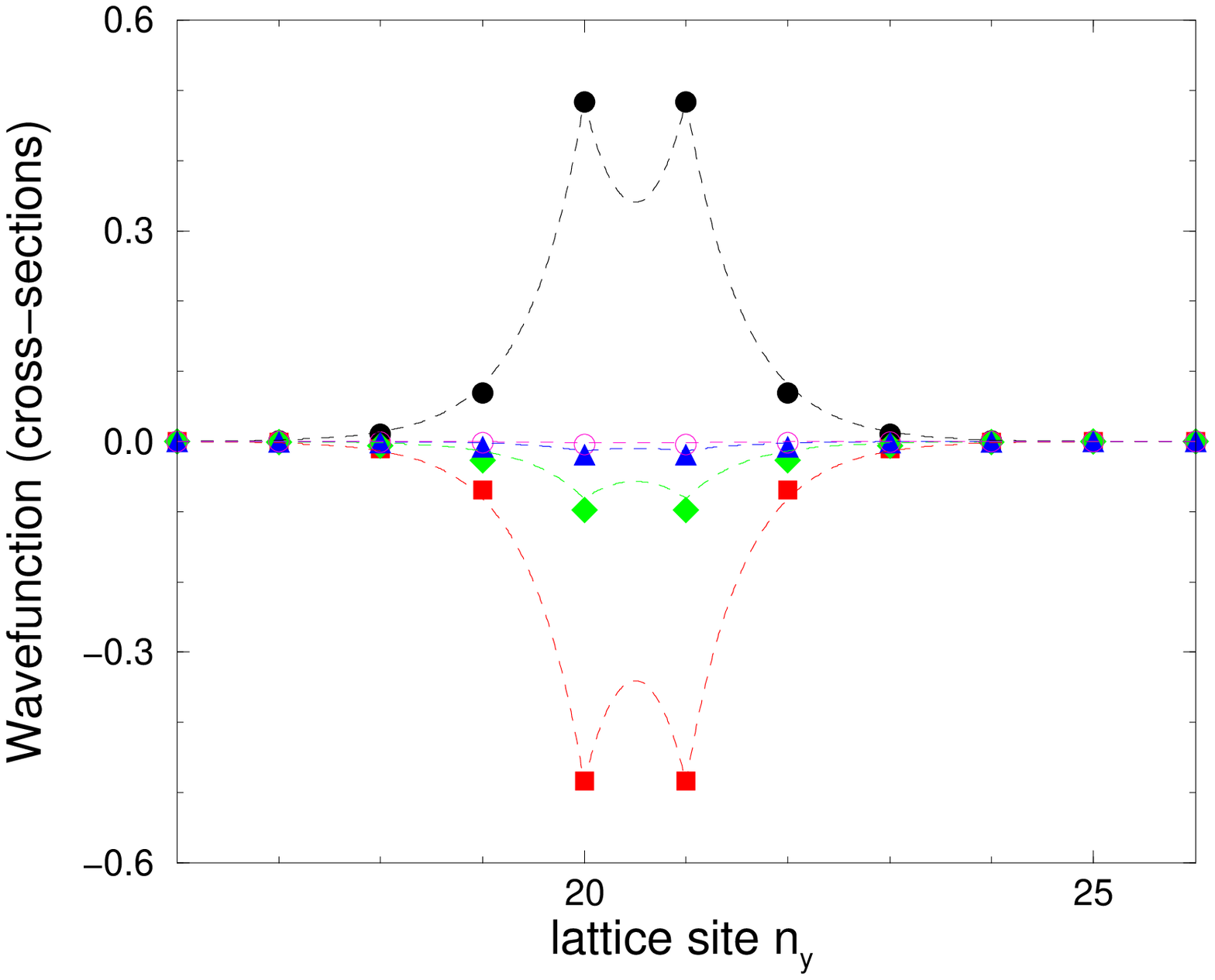,height=4.4cm,width=4.4cm} & \hspace{-0.2cm} 
\epsfig{file=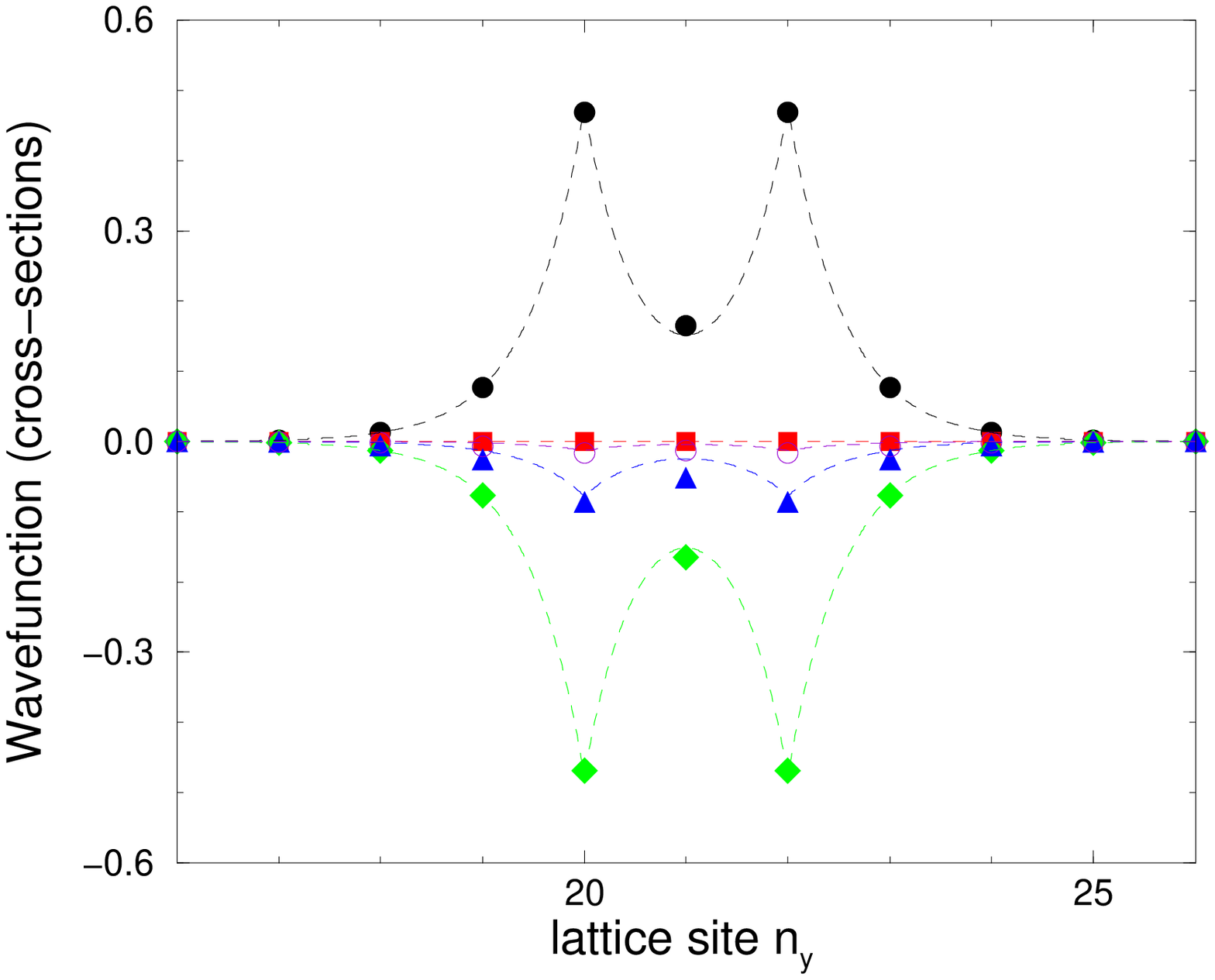,height=4.4cm,width=4.4cm} & \hspace{-0.2cm}
\epsfig{file=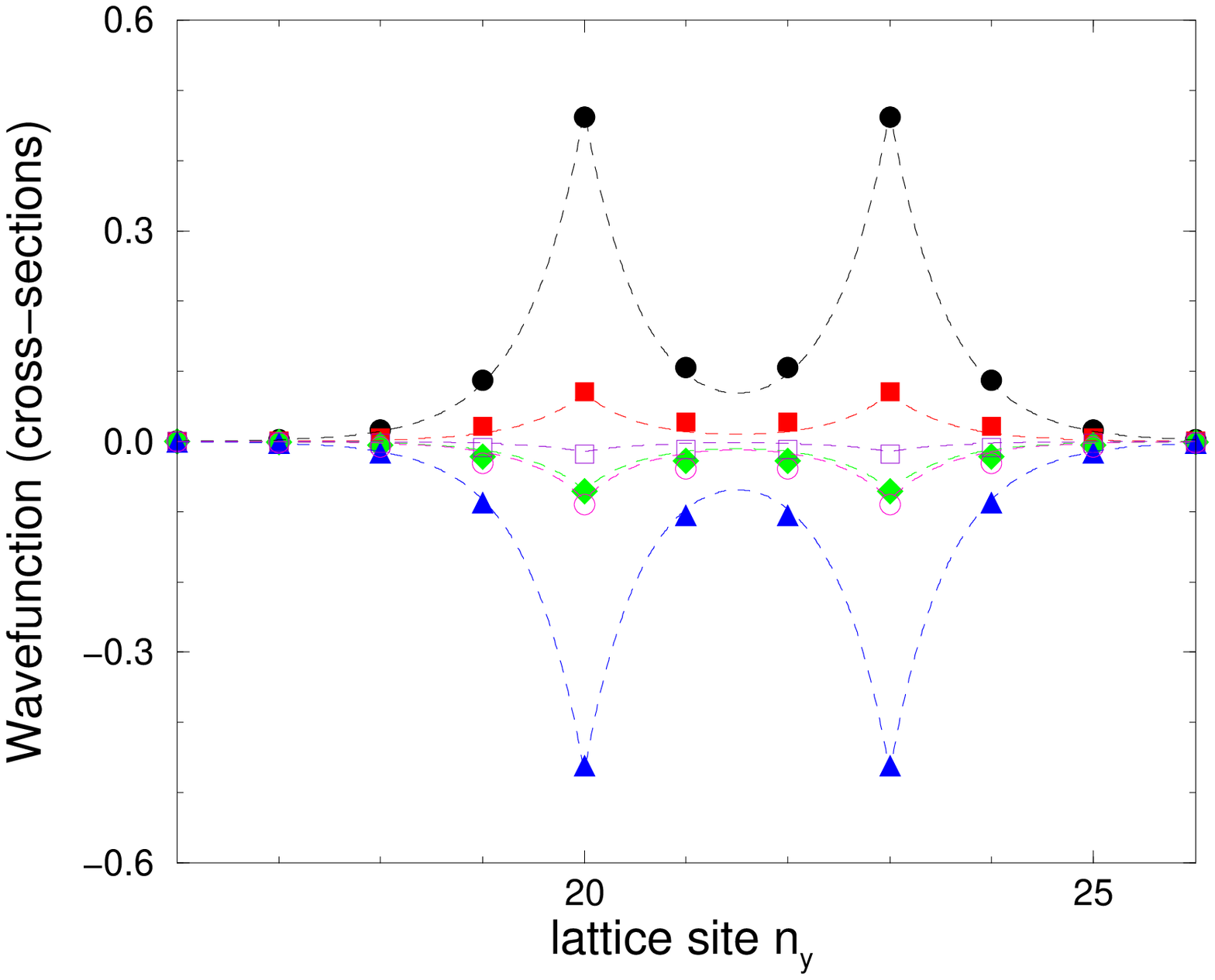,height=4.4cm,width=4.4cm}  \\
\end{tabular}
\end{center}
\caption{3D plots (first row) and density plots (second row)
of quadruple-peaked symmetric/antisymmetric solutions of DNLS in 2D.
{\it Left column:} interpeak separation $l=1$ lattice site, $\chi=-27$.
{\it Middle column:} interpeak separation $l=2$ sites, $\chi=-27.5$.
{\it Right column:} interpeak separation $l=3$ sites, $\chi=-26$.
Cross-sections of the wavefunction are shown in the third row with points:
$\psi_{n_x=n_1,n_y}$ (filled circles), $\psi_{n_x=n_1+1,n_y}$ (filled squares),
$\psi_{n_x=n_1+2,n_y}$ (filled diamonds), $\psi_{n_x=n_1+3,n_y}$ (filled
triangles), $\psi_{n_x=n_1+4,n_y}$ (open circles), $\psi_{n_x=n_1+5,n_y}$
(open squares, in the third column), where $n_1=20$ is the x$-$coordinate of
the first peak. Dashed lines show analytical approximations of the solutions
using Eq.~(\ref{qp2db}).}
\label{fQSA2}
\end{figure}

The last example of QP states, which are symmetric along one
lattice axis and antisymmetric along the other one, are obtained from the
initial state
\be  \label{qsa2}
\psi_{n_x,n_y}^{(r=0)}= \frac{1}{\sqrt{4}}(\delta_{n_x,n_1} \delta_{n_y,n_2}
+ \delta_{n_x,n_1} \delta_{n_y,n_2+l} - \delta_{n_x,n_1+l} \delta_{n_y,n_2+l}
- \delta_{n_x,n_1+l} \delta_{n_y,n_2}).
\ee
Three cases (for $l=1$, 2, and 3) are presented in Fig.~\ref{fQSA2}.
These solutions are always unstable since there are two pairs of purely
imaginary eigenvalues $\pm i \beta$ and $\pm i \gamma$. There is also a
discrete eigenvalue $\delta$ which is real for large values of $|\chi|$,
but when it collides with the band (which extends approximately from
$\frac{|\chi|}{4}-4$ to $\frac{|\chi|}{4}+4$) one more instability is
developed. For $|\chi| \gg 1$, $\beta$ and $\delta$ tend to the same value
and they show the typical dependence on $|\chi|$: they increase (decrease)
for $l=1$ ($l>2$) and they tend to 2 for $l=2$. This picture does not
change for positive $\chi$, regardless whether $l$ is even or odd.

In Fig.~\ref{fFS2} branches of solutions of the symmetry obtained from
(\ref{qs2}) and
(\ref{qa2}) are shown for $l=1$, 2, 3, 4, 5, 6, 7, 10, and 11 and from
(\ref{qsa2}) for $l=1$, 2, 3, 4, 5, 10, and 11. Of course there are many
more quadruplets having their four peaks in a rectangle or in random lattice
sites, with any combination of signs, which are congested around the middle
quasi-degenerate branch of the QP solutions of Fig.~\ref{fFS2}.
An example of antisymmetric along {\it both diagonals} QP
solution, forming a square with its edges (of length equal to
$\sqrt{2}$) not along the axes, as the states presented above, but
along the diagonals, is reported as a quasivortex in Ref.~\cite{KMCF}

Once more, approximate analytical expressions can be derived for the
QP solutions in 2D, by superimposing four single-peaked solutions
of Eq. (\ref{vap}), each one corresponding to $\frac{\chi}{4}$. Regarding
the high symmetry solutions presented above, if $(n_1,n_2)$ is the position
of the first peak and $l$ the interpeak distance along the lattice axes
($l>0$), then these stationary states can be approximated by
\begin{eqnarray} 
\psi_{n_x,n_y}^{QP}=\frac{1}{2\sqrt{1 \pm 2P+P^2}} \frac{1-\zeta^2}{1+\zeta^2}
 & \left( \zeta^{|n_x-n_1|+|n_y-n_2|} \pm (-sgn\chi)^l
\zeta^{|n_x-n_1|+|n_y-n_2-l|} \right. \nonumber \\  &  \left. +
\zeta^{|n_x-n_1-l|+|n_y-n_2-l|} \pm (-sgn\chi)^l \zeta^{|n_x-n_1-l|+|n_y-n_2|}
 \right)  \hspace{1.0cm}  \label{qp2d}   \end{eqnarray}
and
\begin{eqnarray}
\psi_{n_x,n_y}^{QP} = \frac{1}{2\sqrt{1-P^2}} \frac{1-\zeta^2}{1+\zeta^2}
& \left( \zeta^{|n_x-n_1|+|n_y-n_2|} + \zeta^{|n_x-n_1|+|n_y-n_2-l|}
\right. \nonumber \\  &  \left.
- \zeta^{|n_x-n_1-l|+|n_y-n_2-l|} - \zeta^{|n_x-n_1-l|+|n_y-n_2|} \right),
\hspace{1.0cm}  \label{qp2db}   \end{eqnarray}
\be
\mbox{where} \hspace{0.3cm}
P=(-sgn\chi)^l \; \frac{(1+l) \zeta^l - (l-1) \zeta^{l+2}}{1+\zeta^2}
\ee
\be
\mbox{and} \hspace{0.3cm} \zeta = -\frac{1}{\chi/4} - \frac{6}{(\chi/4)^3}
= -\frac{4}{\chi} - \frac{384}{\chi^3}.
\ee
Plus (minus) signs in Eq.~(\ref{qp2d}) provide the symmetric (fully
antisymmetric along both axes) QP solutions, except when $\chi$ is positive
and $l$ odd, where it is the reverse. Eq.~(\ref{qp2db})
gives the solutions which are symmetric along one axis and antisymmetric
along the other one, like those of Fig.~\ref{fQSA2}.
$P=\sum_{n_x,n_y} \psi_{n_x,n_y}^{SP[n_1,n_2]}(\frac{\chi}{4})
\psi_{n_x,n_y}^{SP[n_1,n_2+l]}(\frac{\chi}{4})$ is the overlap of two
neighboring single-peaked states at distance $l$. The overlap of two
single-peaked states across the diagonal of the square,
$\sum_{n_x,n_y} \psi_{n_x,n_y}^{SP[n_1,n_2]}(\frac{\chi}{4})
\psi_{n_x,n_y}^{SP[n_1+l,n_2+l]}(\frac{\chi}{4})$, is equal to $P^2$.
Cross-sections of Eqs. (\ref{qp2d}) and (\ref{qp2db}) are shown in the
third rows of Figs. \ref{fQS2}-\ref{fQSA2} with dashed lines.

\section{Conclusions}

Multi-peaked localized excited states are discussed for the one- and
two-dimensional discrete nonlinear Schr\"odinger equation. Their numerical
calculation is achieved by iterations of a simple map, where trivial
initial states rapidly converge to the desired solution.
Examples have been presented of symmetric and antisymmetric states
with different interpeak separations for double-peaked solutions in 1D,
as well as for double-peaked states along a lattice axis or along the
diagonal and quadruple-peaked states on a square in 2D. Analytical
approximations and the linear stability of the solutions have been discussed.
For strong nonlinearities, the symmetric double-peaked states are unstable
and the antisymmetric linearly stable, except for the case of positive
nonlinearity and odd interpeak separation, where the situation is reversed.
An interesting application that such multi-peaked solutions may have,
concerns their potential use for information encoding and transfer
in optical lattices \cite{KETC}. Multi-peaked solutions have been also
discussed in different contexts \cite{BE,fuentes,pan}.

The classification of these stationary states is based
on their origin at the anti-continuous limit, even though sometimes
their name may be misleading. For example, the symmetric double-peaked
states in 1D and 2D and the symmetric quadruplet in 2D with interpeak
separations equal to one lattice constant, can be viewed as solutions
having one peak. However, this classification is very useful for
organizing the discrete frequency and energy spectrum of DNLS.
The concept of anti-continuous limit \cite{sergeAC} facilitates the
interpretation of the structure of this spectrum.

Some of the stationary states presented in this work have been also
discussed in other studies and different names are attributed to them.
The double-peaked solution in 1D with interpeak separation $S=1$
has been named Page mode, even mode, or centered-between-sites \cite{KC}.
Antisymmetric double-peaked states in 1D with $S=1$ or $S=2$ are
also known as twisted modes \cite{DKL}, while their analogues in 2D
with their peaks along the diagonal have been discussed in \cite{KMB}.
In Ref. \cite{kevr} the symmetric double-peaked state along a lattice
axis with $S=1$ in 2D and the symmetric quadruple-peaked state
with interpeak separation $l=1$ have been named hybrid mode and
Page-like mode, respectively.
The double-peaked symmetric stationary solution along the diagonal with
$l=1$ in 2D has been also known \cite{rodrigo}.

{\bf Acknowledgements}

The author would like to thank S. Aubry for useful and stimulating
suggestions, R. Vicencio and M. Johansson for interesting discussions
and for bringing to his attention relevant references, and the referees
for useful comments.
This work is dedicated to Serge on the occasion of his 60th birthday,
with a strong appreciation for the valuable knowledge and fruitful ideas
that he has tried to communicate to us and many wishes for even more
breakthroughs on nonlinear physics in the future.

\section*{ Appendix A}

\subsection*{A.1 Linear stability of a stationary state}

Considering small deviations $\delta \psi_n(t)$ from a stationary solution
of Eq.~(\ref{statsol}), i.e.
$\Psi_n(t) = \left( \psi_n +\delta \psi_n(t) \right) \cdot e^{-i\omega_0 t}$
(where $\psi_n$ is time-independent and real and $\omega_0$ is the
corresponding frequency), substituting in DNLS Eq.~(\ref{dnls}), and
linearizing in respect to the complex small perturbations
$\delta \psi_n(t)$, one obtains
\be  \label{ldnls}
 i \frac{d \delta \psi_n}{dt} = (-\omega_0 + \chi \psi_n^2) \delta \psi_n
- \sum_{\delta} \delta \psi_{n+\delta} + 2 \chi \psi_n^2 Re(\delta \psi_n),
\ee
where $Re(\delta \psi_n)$ is the real part of $\delta \psi_n(t)$.
Substituting solutions of the form
\be  \label{lsol}
\delta \psi_n(t) = a_n \sin(\omega t) + i b_n \cos(\omega t)
\ee
in the linearized equation (\ref{ldnls}), yields the following coupled
system
\begin{eqnarray}  \label{ls1}
\omega a_n= & (-\omega_0 +\chi \psi_n^2) b_n & - \sum_\delta b_{n+\delta} \\
\omega b_n= & (-\omega_0 +3\chi \psi_n^2) a_n & - \sum_\delta a_{n+\delta}.
\label{ls2}    \end{eqnarray}
This implies that the stability eigenvalues $\omega$ can be obtained
as the eigenvalues of the $2L \times 2L$ (where $L$ is the total number of
lattice sites) matrix M:
\be
M \left(
\begin{array}{c}
A \\ B
\end{array} \right) = \left(
\begin{array}{cc}
{\it 0} & M_1 \\ M_2 & {\it 0}
\end{array}
\right)  \left(
\begin{array}{c}
A \\ B
\end{array} \right) = \omega \left(
\begin{array}{c}
A \\ B
\end{array} \right),
\ee
where the $2L \times 1$ column
$(A,B)^T \equiv (a_1, \ldots,a_L,b_1,\ldots,b_L)^T$ corresponds to the
eigenvectors, ${\it 0}$ denotes the $L \times L$ zero matrix and $M_1$,
$M_2$ are the tight-binding $L \times L$ matrices given through the
right-hand-sides
of Eqs. (\ref{ls1}) and (\ref{ls2}), respectively. $M_1$ and $M_2$ differ
only in their diagonal elements; $M_{1_{ii}}=-\omega_0 +\chi \psi_i^2$ and
$M_{2_{ii}}=-\omega_0 +3\chi \psi_i^2$, while the non-diagonal matrix
elements are zero, except when they correspond to first neighboring sites
where they are equal to $-1$.

From Eq.~(\ref{lsol}) we see that if there is an eigenvalue of $M$ with
non-zero imaginary part, then the stationary solution is unstable. It can be
easily verified that any stationary solution has a stability eigenvalue
$\omega=0$, with eigenvector $a_n=0$, $b_n=\psi_n$, since Eq.~(\ref{ls2})
is trivially satisfied and Eq.~(\ref{sdnls}) provides a solution making
zero the right-hand-side of (\ref{ls1}). The eigenvalue $\omega=0$ is
doubly degenerate possessing a second eigenvector known as growth mode
\cite{JA2}.

\subsection*{A.2 Obtaining the linear stability eigenvalues through a
tight-binding Hamiltonian}

Here it is shown that the
eigenvalues of the linear stability problem can be calculated through the
energy eigenvalues and eigenvectors of a tight-binding
Hamiltonian\footnote{ a similar procedure has been used in
Ref.~\cite{KAT} for the linear stability of a slightly more complicated
system, viz. the polarons of the adiabatic Holstein model} with an on-site
potential determined by the stationary state $\psi_n$. In particular,
consider the Hamiltonian eigenvalue problem
\be    \label{etb}
H_l \phi_n^\nu = -\sum_\delta \phi_{n+\delta}^\nu + U_n \phi_n^\nu
= E_\nu \phi_n^\nu, \hspace{0.5cm} \mbox{with potential} \; \;
U_n = \chi \psi_n^2,
\ee
where $E_\nu$ and $\phi_n^\nu$ are the eigenvalues and the corresponding
eigenvectors. The potential $U_n$
has the shape of the stationary state under discussion, multiplied by
the nonlinearity parameter. Comparing Eqs. (\ref{sdnls}) and (\ref{etb})
one obtains that $\psi_n$ is an eigenvector $\phi_n^{\nu=0}$ of $H_l$
with $E_{\nu=0}=\omega_0$.

Expressing the eigenvectors $a_n$ and $b_n$ of the linear stability
system (\ref{ls1}), (\ref{ls2}) in the complete basis $\phi_n^\nu$
of the Hamiltonian $H_l$, i.e. $a_n = \sum_\nu a_\nu \phi_n^\nu$ and
$b_n = \sum_\nu b_\nu \phi_n^\nu$,
substituting in the system (\ref{ls1}), (\ref{ls2}), and using
Eq.~(\ref{etb}) and the relation
$\sum_n \phi_n^\nu \phi_n^{\nu^\prime}=\delta_{\nu,\nu^\prime}$,
yields that for $\nu \neq 0$ is $b_\nu=\frac{\omega}{E_\nu-E_0}a_\nu$ and
\be \label{sptb}
\left[ (E_\nu-E_0)-\frac{\omega^2}{E_\nu-E_0}+2\chi
\sum_n(\phi_n^0 \phi_n^\nu)^2 \right] a_\nu + 2\chi
\sum_{\nu^\prime \neq \nu} a_{\nu^\prime} \left(
\sum_n (\phi_n^0)^2 \phi_n^\nu \phi_n^{\nu^\prime} \right) =0.
\ee
These $L-1$ equations (since $\nu \neq 0$) provide the $2L-2$ non-zero
stability eigenvalues $\omega=\pm \sqrt{\omega^2}$. We see that the
eigenvalues of the linear stability analysis appear as pairs of opposite
sign and the $\omega^2$ result from the diagonalization of the
$(L-1) \times (L-1)$ matrix
\begin{eqnarray}  \label{dtb}
K_{\nu,\nu^\prime}= & \left[  (E_\nu-E_0)^2 + 2 \chi (E_\nu-E_0)
\sum_n(\phi_n^0 \phi_n^\nu)^2  \right]  \delta_{\nu,\nu^\prime} & \\
\nonumber   & + (1-\delta_{\nu,\nu^\prime}) 2 \chi (E_\nu-E_0)
\sum_n (\phi_n^0)^2 \phi_n^\nu \phi_n^{\nu^\prime}, & \hspace{0.3cm}
\mbox{where} \; \; \nu,\nu^\prime \neq 0.
\end{eqnarray}
$K_{\nu,\nu^\prime}$ is constructed from the eigenspectrum of $H_l$.
Using this matrix, analytical results are obtained in the following
subsection for the linear stability
eigenvalues of double-peaked solutions of DNLS for large values of
$|\chi|$, and the Eqs. (\ref{inst1}) and (\ref{inst2}) are derived.

\subsection*{A.3  Application: double-peaked stationary states close to the
anti-continuous limit}

For a DP stationary solution the potential $U_n$ of
Eq.~(\ref{etb}) is a double well for negative $\chi$ and a double barrier
for positive $\chi$. If $|\chi| \gg 1$, then $\psi_n$ is localized at
almost two lattice sites, where the two peaks are located, and each well
(barrier) of $U_n$ is very deep (high) and narrow, with a strength about
$\frac{\chi}{2}$. In this case the Hamiltonian $H_l$, Eq.~(\ref{etb}), is
equivalent to a tight-binding problem with two equal impurities with
large on-site energies
$\frac{\chi}{2}$. Then the energy spectrum has two discrete eigenvalues
(the $E_0=\omega_0$ and a second one, denoted by $E_1$) and, for an
extended system, all the other eigenvalues belong to the continuous band
from $-2d$ to $2d$ \cite{econ}. The eigenvectors of the continuous spectrum
are proportional to $\frac{1}{\sqrt{L}}$ and therefore negligibly small
at any lattice site. Since the sums over the lattice sites $n$ that appear in
$K_{\nu,\nu^\prime}$ in Eq.~(\ref{dtb}) contain the tightly localized
around two sites wavefunction $\phi_n^0 \equiv \psi_n$, they can be
neglected when they involve an eigenstate of the continuous.
The $\phi_n^\nu, \phi_n^{\nu^\prime}$ of $K_{\nu,\nu^\prime}$
do not include the $\nu=0$ and thus there is only one localized eigenstate
among them; the $\phi_n^1$ corresponding to $E_1$. As a result the
non-diagonal matrix elements of $K_{\nu,\nu^\prime}$ can be neglected, since
they contain products of two eigenstates $\phi_n^\nu \phi_n^{\nu^\prime}$
and at least one of them it belongs to the continuum. The diagonal ones,
providing directly the stability eigenvalues $\omega^2$, are
\be  \label{fcon}
\omega^2 = K_{\nu,\nu} = (E_\nu-E_0)^2,
\hspace{0.8cm} \mbox{for} \hspace{0.2cm} \nu \neq 1 \hspace{1.2cm} \mbox{and}
\ee
\be \label{fdis}
\omega^2 = K_{1,1} = (E_1-E_0)^2 +2 \chi (E_1-E_0)
\sum_n(\phi_n^0 \phi_n^1)^2, \hspace{0.8cm} \mbox{for} \hspace{0.2cm} \nu=1
\ee
Eq. (\ref{fcon}) gives two bands of real eigenvalues $\omega=\pm |E_\nu-E_0|$,
symmetrically positioned around zero, and Eq.~(\ref{fdis}) a discrete
pair of eigenvalues $\omega=\pm \sqrt{K_{1,1}}$.

The discrete eigenspectrum of $H_l$ is given by
\be  \label{distb}
\phi_n^\pm = \frac{1}{\sqrt{2(1\pm P)}} \left( g_n^{[n_0]} \pm
(-sgn\chi)^S g_n^{[n_0+S]} \right) \longrightarrow  E_\pm =
\frac{\epsilon\pm\upsilon}{1\pm P} \approx\epsilon \pm\upsilon\mp\epsilon P,
\ee
where $g_n^{[n_0]}, g_n^{[n_0+S]}$ represent single-peaked wavefunctions,
centered at $n_0$ and $n_0+S$, respectively. Here, $n$, $n_0$, and $S$
($S$ or its components are assumed positive)
have to be understood as integers in 1D and pairs or sums of integers in 2D,
e.g. $n\rightarrow (n_x,n_y)$, $n_0 \rightarrow(n_1,n_2)$,
$S\rightarrow (S_x,S_y)$, $(-sgn\chi)^S \rightarrow (-sgn\chi)^{S_x+S_y}$,
etc. The other quantities in the energy eigenvalues $E_\pm$ of
Eq.~(\ref{distb}) are $\epsilon= \langle g_n^{[n_0]}
|H_l|g_n^{[n_0]} \rangle=\langle g_n^{[n_0+S]}|H_l|g_n^{[n_0+S]} \rangle$
and the small overlap integrals
$\upsilon= (-sgn\chi)^S \langle g_n^{[n_0]}|H_l|g_n^{[n_0+S]} \rangle$ and
$P= (-sgn\chi)^S \langle g_n^{[n_0]}|g_n^{[n_0+S]} \rangle$.
For $\chi$ negative, i.e. attractive potential $U_n$, the upper (lower) signs
in Eq.~(\ref{distb}) correspond to the ground (first excited) state of
$H_l$, while for positive $\chi$, i.e. repulsive $U_n$, they provide the
highest-energy (second highest-energy) state. Note that for positive $\chi$
and odd $S$ the highest-energy state is the antisymmetric one because of the
$(-sgn\chi)^S$ term in (\ref{distb}), in contrary to the case of even $S$.
It is convenient to distinguish three cases: \\
(i) negative $\chi$, \\ (ii) positive $\chi$ and even $S$ (or even $S_x+S_y$),
and \\ (iii) positive $\chi$ and odd $S$ (or odd $S_x+S_y$). \\
For symmetric DP stationary states $\psi_n$, the $\phi_n^+,E_+$ of
Eq.~(\ref{distb}) correspond to $\phi_n^0 \equiv \psi_n$ and $E_0$ and the
$\phi_n^-,E_-$ to $\phi_n^1$ and $E_1$ in cases (i) and (ii), while it is
the other way around in case (iii). For antisymmetric DP states $\psi_n$,
holds the opposite: the $\phi_n^-,E_-$ correspond to $\phi_n^0$ and $E_0$
and the $\phi_n^+,E_+$ to $\phi_n^1$ and $E_1$ in cases (i) and (ii),
and the reverse is valid in case (iii).

The discrete eigenvalue (\ref{fdis}) is dominated by the second term,
since $E_1-E_0$ is a small quantity and $\chi \gg 1$.
For a symmetric DP solution $E_1-E_0$ is positive in cases (i) and (iii)
and negative in case (ii). Therefore, $K_{1,1}$, which has the same sign
as $\chi (E_1-E_0)$, is negative (positive) in cases (i) and (ii) (in case
(iii)), meaning that the symmetric DP solution is unstable (linearly stable)
with a pair of purely imaginary (real) discrete eigenvalues. The situation
is reverse for an antisymmetric DP stationary state since $E_1-E_0$ has the
opposite sign now. Thus, when the symmetric
DP solution is linearly stable (unstable) the antisymmetric one is unstable
(linearly stable). These results are confirmed by the numerical simulations
of sections 4 and 5. The above arguments about the linear stability/instability
of a symmetric or antisymmetric DP solution of DNLS close to the
anti-continuous limit, are generally applied at any dimension. Therefore,
one expects that, for large $|\chi|$, also in 3D an antisymmetric DP
stationary state is linearly stable and a symmetric unstable, except when
$\chi$ is positive and the interpeak distance $S_x+S_y+S_z$ is odd, where
they interchange roles regarding their stability.

The previous discussion can be quantified and using Eqs. (\ref{fdis}) and
(\ref{distb}) analytical expressions are
obtained for the discrete pair of eigenvalues in the limit of $|\chi| \gg 1$.
The wavefunctions $g_n$ in (\ref{distb}) are given by Eq.~(\ref{vap}) for
$\frac{\chi}{2}$. Taking into account that $\zeta=-\frac{2}{\chi}$ one
obtains to leading order: $P=(-sgn\chi)^S (1+S) \zeta^S$ in 1D and
$P=(-sgn\chi)^{S_x+S_y} (1+S_x)(1+S_y) \zeta^{S_x+S_y}$ in 2D [see also Eqs.
(\ref{ov1}) and (\ref{ov2})], the sum $\sum_n(\phi_n^0\phi_n^1)^2=\frac{1}{2}$,
the coupling $\upsilon=(-sgn\chi)^S (-S\zeta^{S-1}+\chi\zeta^S)$ in 1D and
$\upsilon= (-sgn\chi)^{S_x+S_y} [-(S_x+S_y+2S_xS_y)\zeta^{S_x+S_y-1}+\chi
\zeta^{S_x+S_y}]$ in 2D, and the on-site energy $\epsilon=\frac{\chi}{2}$.
Then for symmetric DP states in 1D is
$E_1-E_0=-2(\upsilon-\epsilon P)=\frac{2^S}{|\chi|^{S-1}}$ in case (i), 
$E_1-E_0=-2(\upsilon-\epsilon P)=-\frac{2^S}{\chi^{S-1}}$ in case (ii), and
$E_1-E_0=2(\upsilon-\epsilon P)=\frac{2^S}{\chi^{S-1}}$ in case (iii).
The signs are the same in 2D, but
the magnitudes change to $\frac{(1+S_xS_y)2^{S_x+S_y}}{|\chi|^{S_x+S_y-1}}$.
Therefore Eq.~(\ref{fdis}) yields that in cases (i) and (ii)
$\omega^2=-\frac{2^S}{|\chi|^{S-2}}$ in 1D and 
$\omega^2=-\frac{(1+S_xS_y)2^{S_x+S_y}}{|\chi|^{S_x+S_y-2}}$ in 2D, while
in case (iii) $\omega^2$ is positive with the same magnitude. For the
antisymmetric DP states the $E_1-E_0$ and consequently the $\omega^2$
are the same like previously, but with opposite sign. As a result the
discrete eigenvalues are
\be  \label{unsteig}
\omega=\pm i 2^{S/2} |\chi|^{1-S/2}  \hspace{0.2cm} \mbox{in 1D}
\hspace{0.3cm} \mbox{and} \hspace{0.3cm}   \omega=\pm i \sqrt{1+S_xS_y} \;
2^{(S_x+S_y)/2} |\chi|^{1-\frac{S_x+S_y}{2}}  \hspace{0.2cm} \mbox{in 2D}
\ee
for the (unstable) symmetric DP states in cases (i) and (ii) and the
antisymmetric in case (iii), while
\be  \label{steig}
\omega=\pm  2^{S/2} |\chi|^{1-S/2}  \hspace{0.2cm} \mbox{in 1D}
\hspace{0.3cm} \mbox{and} \hspace{0.3cm}   \omega=\pm  \sqrt{1+S_xS_y} \;
2^{(S_x+S_y)/2} |\chi|^{1-\frac{S_x+S_y}{2}}  \hspace{0.2cm} \mbox{in 2D}
\ee
for the (linearly stable) symmetric states in case (iii) and the
antisymmetric ones in cases (i) and (ii). Eq.~(\ref{unsteig}) provides the
instability magnitudes given in Eqs. (\ref{inst1}) and (\ref{inst2}),
while Eq.~(\ref{steig}) the stable eigenvalue used in Eq.~(\ref{antun}).
By decreasing $|\chi|$, as the stationary state $\psi_n$ and the
potential $U_n$ of (\ref{etb}) become more extended, additional discrete
levels of $H_l$ may appear, resulting in additional discrete eigenvalues
of the linear stability problem.

For the DP states, $E_0 \equiv \omega_0$ equals, to leading order, to
$\epsilon = \frac{\chi}{2}$. The next corrections are
$\pm (\upsilon-\epsilon P)$, which are equal to
$\pm \frac{E_1-E_0}{2}=\pm \frac{2^{S-1}}{|\chi|^{S-1}}$ in 1D and a similar
expression in 2D (see above). These corrections are inverse powers of $\chi$,
apart from the case of $S=1$ (or $S_x+S_y=1$). Then, if $\chi<0$
for example, the symmetric DP states have
$E_0 \equiv \omega_0=\frac{\chi}{2}+(\upsilon-\epsilon P)=\frac{\chi}{2}-1$
and the antisymmetric have
$E_0 \equiv \omega_0=\frac{\chi}{2}-(\upsilon-\epsilon P)=\frac{\chi}{2}+1$
(see the lower and upper branches of DP states in Figs. \ref{figFS1}
and \ref{fFS2}, while similar is the case for the single branches of
the QP states presented in 2D). The knowledge of $E_0$ and the continuous
band of $H_l$ (from $-2d$ to $2d$) provides the bands of stability
eigenvalues $\omega=\pm |E_\nu-E_0|$, Eq.~(\ref{fcon}). The latter, for
positive $\omega$, extends from $\frac{|\chi|}{2}-2d$ to
$\frac{|\chi|}{2}+2d$, apart from the case of $S=1$ (or $S_x+S_y=1$), where
it extends between $\frac{|\chi|}{2}\pm1-2d$ and $\frac{|\chi|}{2}\pm1+2d$
(the upper signs correspond to symmetric DP states in cases i) and ii) and
antisymmetric in case iii), while the minus signs correspond to the
complementary situations). These results are in accordance with the
numerical observations in sections 4 and 5.

\end{document}